\documentclass[
 reprint,
superscriptaddress,
 amsmath,amssymb,
 aps,
prd,
floatfix,
]{revtex4-2}

\usepackage{graphicx}
\usepackage{dcolumn}
\usepackage{bm}
\usepackage[mathlines]{lineno}
\usepackage{hyperref}
\usepackage{xspace}
\usepackage[T1]{fontenc}
\usepackage{subcaption}
\usepackage{xcolor}
\usepackage[normalem]{ulem}

\newcommand{\gevc}{\textrm{GeV}/c}

\newcommand{\gev}{\textrm{GeV}}

\newcommand{\dat}{\delta\alpha_\textrm{T}}
\newcommand{\dphit}{\delta\phi_\textrm{T}}
\newcommand{\pn}{p_\textrm{N}}
\newcommand{\dpt}{\delta p_\textrm{T}}
\newcommand{\vecdpt}{\delta \vec{p}_\textrm{T}}
\newcommand{\dptx}{\delta p_\textrm{Tx}}
\newcommand{\dpty}{\delta p_\textrm{Ty}}
\newcommand{\dptt}{\delta p_\textrm{TT}}

\newcommand{\deltapp}{\Delta^{++}}

\newcommand{\pp}{p_\textrm{p}}
\newcommand{\thetap}{\theta_\textrm{p}}

\newcommand{\vecpp}{\vec{p}_\textrm{p}}
\newcommand{\vecppi}{\vec{p}_{\pi}}

\newcommand{\vecpmu}{\vec{p}_{\mu}}

\newcommand{\degree}{\circ}
\newcommand{\tick}{\checkmark}

\newcommand{\p}{\textrm{p}}
\newcommand{\n}{\textrm{n}}
\newcommand{\pip}{\pi^+}
\newcommand{\pim}{\pi^-}
\newcommand{\piz}{\pi^0}

\newcommand{\ttkzpi}{T2K-$0\pi$\xspace}
\newcommand{\ttkpip}{T2K-$\pi^{+}$\xspace}

\newcommand{\minzpi}{MINERvA-$0\pi$\xspace}
\newcommand{\minpiz}{MINERvA-$\pi^{0}$\xspace}

\newcommand{\cpimfp}{S_\lambda^{\pi^{\pm}}}
\newcommand{\pizmfp}{S_\lambda^{\pi^{0}}}
\newcommand{\picex}{S_\textrm{CEX}^{\pi}}
\newcommand{\piinel}{S_\textrm{INEL}^{\pi}}
\newcommand{\cpiabs}{S_\textrm{ABS}^{\pi^{\pm}}}
\newcommand{\pizabs}{S_\textrm{ABS}^{\pi^0}}
\newcommand{\pipiprod}{S_\textrm{PIPD}^{\pi}}
\newcommand{\nmfp}{S_\lambda^\textrm{N}}
\newcommand{\ncex}{S_\textrm{CEX}^\textrm{N}}
\newcommand{\ninel}{S_\textrm{INEL}^\textrm{N}}
\newcommand{\nabs}{S_\textrm{ABS}^\textrm{N}}
\newcommand{\npiprod}{S_\textrm{PIPD}^\textrm{N}}

\newcommand{\srcfr}{R_\textrm{SRC}}
\newcommand{\nurmec}{E_\textrm{RM}^\textrm{C}}

\newcommand{\geighteen}{\texttt{G18\_10a\_02\_11b}}
\newcommand{\newtune}{\texttt{G24\_20i\_00\_000}\xspace}
\newcommand{\restunefull}{\texttt{G24\_20i\_06\_22c}\xspace}

\newcommand{\alttune}{\texttt{G24\_20i\_fullpara\_alt}\xspace}

\newcommand{\gZero}{\texttt{G24-0}\xspace}
\newcommand{\gC}{\texttt{G24-c}\xspace}
\newcommand{\gT}{\texttt{G24-t}\xspace}
\newcommand{\redpar}{\texttt{RedPar}\xspace}
\newcommand{\allpar}{\texttt{AllPar}\xspace}
\newcommand{\cbRedPar}{\texttt{Combi-Best-\redpar}\xspace}
\newcommand{\cbAllPar}{\texttt{Combi-Best-\allpar}\xspace}

\newcommand{\sfcfg}{SF-CFG\xspace}
\newcommand{\fwid}{0.42}

\newcommand{\genie}{\textsc{genie}\xspace}

\begin{document}

\title{
\boldmath First combined tuning on transverse kinematic imbalance data with and without pion production constraints
}

\author{Weijun Li}
\email{weijun.li@physics.ox.ac.uk}
\affiliation{University of Oxford, Dept. of Physics, Oxford OX1 3RH, UK}

\author{Marco Roda}
\email{mroda@liverpool.ac.uk}
\affiliation{University of Liverpool, Dept. of Physics, Liverpool L69 7ZE, UK}

\author{J\'{u}lia Tena-Vidal}
\email{jtenavidal@tauex.tau.ac.il}
\affiliation{Tel Aviv University, Tel Aviv 69978, Israel}

\author{Costas Andreopoulos}
\email{c.andreopoulos@cern.ch}
\affiliation{University of Liverpool, Dept. of Physics, Liverpool L69 7ZE, UK}

\author{Xianguo Lu}
\email{xianguo.lu@warwick.ac.uk}
\affiliation{University of Warwick, Coventry CV4 7AL, United Kingdom}

\author{Adi Ashkenazi}
\affiliation{Tel Aviv University, Tel Aviv 69978, Israel}

\author{Joshua Barrow}
\affiliation{University of Minnesota Twin Cities, Minneapolis, MN 55455, USA}

\author{Steven Dytman}
\affiliation{University of Pittsburgh, Dept. of Physics and Astronomy, Pittsburgh PA 15260, USA}

\author{Hugh Gallagher}
\affiliation{Tufts University, Dept. of Physics and Astronomy, Medford MA 02155, USA}

\author{Alfonso Andres Garcia Soto}
\affiliation{Instituto de Física Corpuscular, 46980, Paterna, Valéncia, Spain}

\author{Steven Gardiner}
\affiliation{Fermi National Accelerator Laboratory, Batavia, Illinois 60510, USA}

\author{Matan Goldenberg}
\affiliation{Tel Aviv University, Tel Aviv 69978, Israel}

\author{Robert Hatcher}
\affiliation{Fermi National Accelerator Laboratory, Batavia, Illinois 60510, USA}

\author{Or Hen}
\affiliation{Massachusetts Institute of Technology, Dept. of Physics, Cambridge, MA 02139, USA}

\author {Igor D. Kakorin}
\affiliation{Joint Institute for Nuclear Research (JINR), Dubna, Moscow region, 141980, Russia}

\author {Konstantin S. Kuzmin}
\affiliation{Joint Institute for Nuclear Research (JINR), Dubna, Moscow region, 141980, Russia}
\affiliation{Alikhanov Institute for Theoretical and Experimental Physics (ITEP) of NRC ``Kurchatov Institute'', Moscow, 117218, Russia}

\author{Anselmo Meregalia}
\affiliation{CENBG, Universit\'{e} de Bordeaux, CNRS/IN2P3, 33175 Gradignan, France}

\author {Vadim A. Naumov}
\affiliation{Joint Institute for Nuclear Research (JINR), Dubna, Moscow region, 141980, Russia}

\author{Afroditi Papadopoulou}
\affiliation{Argonne National Laboratory, Argonne, IL 60439, USA}

\author{Gabriel Perdue}
\affiliation{Fermi National Accelerator Laboratory, Batavia, Illinois 60510, USA}

\author{Komninos-John Plows}
\affiliation{University of Liverpool, Dept. of Physics, Liverpool L69 7ZE, UK}

\author{Alon Sportes}
\affiliation{Tel Aviv University, Tel Aviv 69978, Israel}

\author{Noah Steinberg}
\affiliation{Fermi National Accelerator Laboratory, Batavia, Illinois 60510, USA}

\author{Vladyslav Syrotenko}
\affiliation{Tufts University, Dept. of Physics and Astronomy, Medford MA 02155, USA}

\author{Jeremy Wolcott}
\affiliation{Tufts University, Dept. of Physics and Astronomy, Medford MA 02155, USA}

\author{Qiyu Yan}
\affiliation{University of the Chinese Academy of Sciences, Beijing 100864, China}
\affiliation{University of Warwick, Coventry CV4 7AL, United Kingdom}

\collaboration{GENIE Collaboration}

\date{\today}

\begin{abstract}
We present the first combined tuning, using \genie, of four transverse kinematic imbalance measurements of neutrino-hydrocarbon scattering, both with and without pion final states, from the T2K and MINERvA experiments. As a proof of concept, we have simultaneously tuned the initial state and final-state interaction models (\sfcfg and hA, respectively), producing a new effective model that more accurately describes the data. 
\end{abstract}

\maketitle

\section{\label{sec:1-intro}Introduction}

Neutrinos play a central role in advancing our understanding of physics and addressing fundamental inquiries in contemporary science. The Hyper-Kamiokande experiment~\cite{Hyper-Kamiokande:2018ofw} aims to conduct precise measurements of charge parity violation within the neutrino sector, a phenomenon believed to be closely linked to the observed matter-antimatter asymmetry in the universe. Similarly, the Deep Underground Neutrino Experiment (DUNE)~\cite{DUNE:2016hlj,DUNE:2015lol,DUNE:2016evb,DUNE:2016rla,DUNE:2021tad}  promises the same, in addition to elucidating the neutrino mass ordering. Whether seeking to ascertain Standard Model parameters or probing for exotic phenomena, significant enhancements in both theoretical frameworks and computational simulations of neutrino-nucleus interactions are imperative. As neutrinos interact mainly with nucleons inside the nucleus, the interaction is subject to nuclear effects, namely the nucleon initial state and final-state interactions (FSIs). These effects are difficult to calculate and can alter the final event topology by changing the number of final state pions with respect to the interactions on free nucleons. Notably, the neutrino flux at DUNE yields a comparable proportion of events with and without pions in the final state, highlighting the pressing need for a generator capable of accurately describing both event topologies.

The large data samples and superb imaging capabilities of modern neutrino experiments offer us a detailed new look at neutrino interaction physics.
Recently, the GENIE~\cite{Andreopoulos:2009rq, GENIE:2021npt} Collaboration has made substantial progress towards a global tuning using neutrino, charged lepton and hadron scattering data, in an attempt to integrate new experimental constraints with state-of-the-art theories and construct robust and comprehensive simulations of neutrino interactions with matter. 
Cross-experiment and cross-topology analyses are challenging tasks as each measurement features its unique selection criteria and various other  aspects, such as the neutrino flux. \genie has built an advanced tuning framework that enables the validation and tuning of comprehensive interaction models using an extensive curated database of measurements of neutrino, charged lepton and hadron scattering off nucleus and nuclei. So far, the non-resonant backgrounds~\cite{GENIE:2021zuu}, hadronization~\cite{GENIE:2021wox} and the quasielastic (QE) and 2-particle-2-hole (2p2h)  components~\cite{GENIE:2022qrc} of the neutrino-nucleus interaction have been tuned with $\nu_\mu$ and $\bar{\nu}_\mu$ charged-current (CC) pionless (0$\pi$) data from MiniBooNE, T2K and MINERvA. A partial tune was performed for each experiment, highlighting the neutrino energy dependence on the QE and 2p2h tuned cross sections. Even though post-tune predictions enhanced the data description for each experiment, the added degrees of freedom were not sufficient to fully describe all CC0$\pi$ data and exhibited tensions with some proton observables~\cite{GENIE:2022qrc}. More exclusive measurements result in additional model constraints. In addition, observables that are sensitive to targeted aspects of the complex dynamics of neutrino interactions are invaluable for model tuning. The transverse kinematic imbalance (TKI)~\cite{Lu:2015hea, Lu:2015tcr}, a final-state correlation between the CC lepton and the handronic system, is a good example since it is sensitive to the initial-state nuclear environment and hadronic FSI. Our next step is to incorporate TKI data from experiments where various exclusive topologies at different energies are considered. This marks the first combined tuning on TKI data with and without pions in the final states and serves as the starting point of a more comprehensive tuning effort in the energy region most relevant for future accelerator-based neutrino experiments.  

This paper is structured as follows: In Sec.~\ref{sec:tki} and Sec.~\ref{sec:genie}, we review the TKI measurements and \genie models respectively. In Sec.~\ref{sec:Tuning}, we detail the tuning considerations and procedures. Results are summarized in Sec.~\ref{sec:results}, highlighting how the data-MC discrepancy in the \minpiz TKI measurement~\cite{MINERvA:2020anu} is resolved while maintaining good data-MC agreement elsewhere. We conclude in Sec.~\ref{sec:summary}. 

\section{TKI measurements}\label{sec:tki}

TKI is a methodology based on the conservation of momentum in neutrino interactions. In essence, it involves quantifying the imbalance between the observed transverse momentum of the final-state particles and the expected transverse momentum from neutrino interactions with free nucleons~\cite{Lu:2015hea, Lu:2015tcr}. This ``kinematic mismatch'' together with its longitudinal and three-dimensional variations~\cite{Furmanski:2016wqo, Lu:2019nmf}, and the derived asymmetry~\cite{Cai:2019jzk}, has been a crucial set of observables, establishing a pathway to extract valuable information about the participating particles and the underlying nuclear processes. Recent experimental results from neutrino experiments such as  T2K~\cite{T2K:2018rnz, T2K:2021naz}, MINERvA~\cite{MINERvA:2018hba, MINERvA:2019ope, MINERvA:2020anu, MINERvA:2021csy}, and MicroBooNE~\cite{MicroBooNE:2022emb, MicroBooNE:2023cmw, MicroBooNE:2023tzj, MicroBooNE:2023wzy, MicroBooNE:2024tmp}, as well as electron scattering experiments such as  CLAS~\cite{CLAS:2021neh}, highlight the efficacy of TKI. 

\begin{figure}[!htb] 	
    \centering 		
    \includegraphics[width=0.35\textwidth]{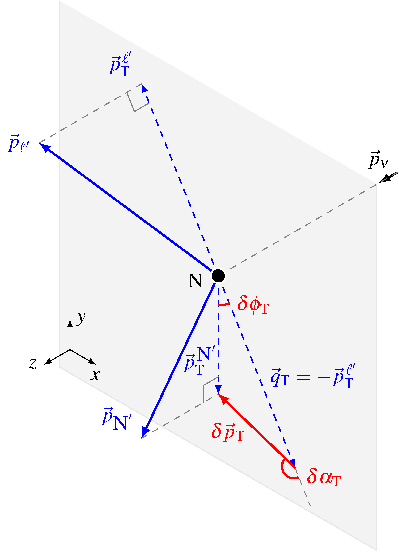}
    \caption{\label{fig:stki} Schematic illustration of the TKI variables. Diagram taken from Ref.~\cite{Lu:2015tcr}.} 
\end{figure}

In neutrino scattering off a free nucleon, the sum of the transverse components of the final products is expected to be zero, visualized through a back-to-back configuration between the final-state lepton and hadronic system in the  plane transverse to the neutrino direction. Hence, in a neutrino interaction with a nucleus, the transverse momentum imbalance, $\dpt$~\cite{Lu:2015tcr}, results from intranuclear dynamics, including Fermi motion and FSIs  as shown in Fig.~\ref{fig:stki}. The deviation from being back-to-back is quantified by the  coplanarity angle $\dphit$~\cite{Lu:2015tcr}, while the transverse boosting angle, $\dat$~\cite{Lu:2015tcr}, represents the direction of $\dpt$ within the transverse plane. Furthermore, analyzing the energy and longitudinal momentum budget~\cite{Furmanski:2016wqo, Lu:2019nmf} enables the conversion of $\dpt$ to the emulated (initial) nucleon momentum, $\pn$,  providing further insight into the Fermi motion; this conversion amounts to a correction on the order of $\mathcal{O}(20\%)$~\cite{Yang:2023dxk}. 
With one-body currents in the absence of FSIs, $\dat$ remains uniform (except for second-order effects, such as variations in the center-of-mass energy), given the isotropic nature of the initial nucleon motion. However, as the final products propagate through the nuclear medium, they experience FSIs, thereby disturbing the isotropy and the Fermi motion peak of the $\dat$ and $\pn$ ($\dpt$) distributions, respectively. Hence, $\dpt$ and $\pn$ elucidate the Fermi motion details, while $\dat$  characterizes the FSI strength---crucial for understanding medium effects in neutrino interactions. A notable advantage of these observables is their minimal dependence on neutrino energy~\cite{Lu:2015tcr}. Moreover, the double TKI variable, $\dptt$~\cite{Lu:2015hea}, is the projection of $\vecdpt$ along the axis perpendicular to the lepton scattering plane (hence ``double''). In addition to its use for extracting neutrino-hydrogen interactions~\cite{Lu:2015hea, Hamacher-Baumann:2020ogq}, it has also been applied to study nuclear effects in neutrino pion productions~\cite{MINERvA:2020anu, T2K:2021naz}. Its equivalent in pionless production, $\dptx$, has been proposed and studied together with its orthogonal companion, $\dpty$, in MINERvA~\cite{MINERvA:2019ope}. 

This work surveyed four TKI data sets: T2K $0\pi$~\cite{T2K:2018rnz}, T2K $\pi^+$~\cite{T2K:2021naz}, MINERvA $0\pi$~\cite{MINERvA:2018hba, MINERvA:2019ope}, and MINERvA $\pi^0$~\cite{MINERvA:2020anu} measurements. All four measurements require the presence of one CC muon and at least one proton in the final state. While the \ttkpip measurement requires exactly one $\pi^+$, and \minpiz requires at least one $\pi^0$, the other two datasets require the absence of any pions. The definitions of the kinematic cuts for these samples are summarized in Table~\ref{tab:data-sets-phase-space-cut}. 

\begin{figure*} 
    \centering 		
    \includegraphics[width=\fwid\textwidth]{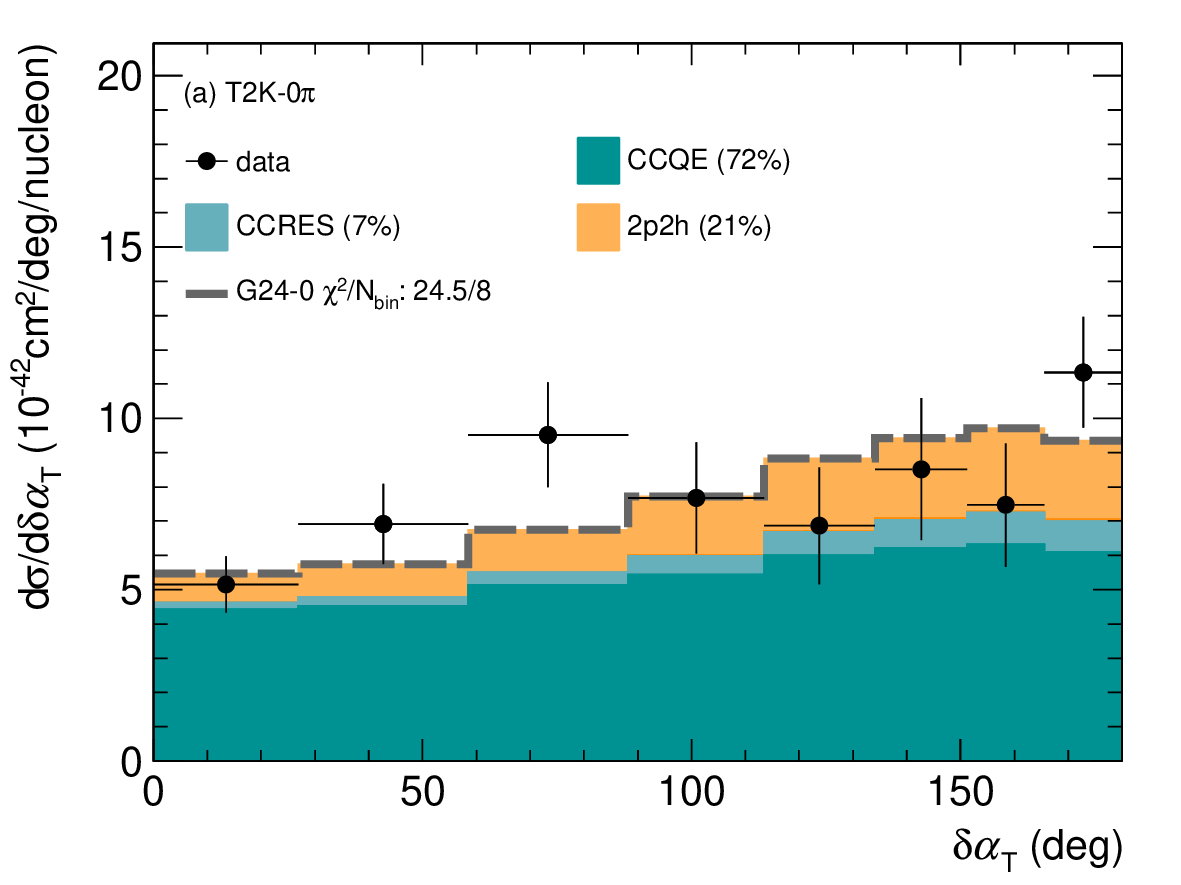} 
    \includegraphics[width=\fwid\textwidth]{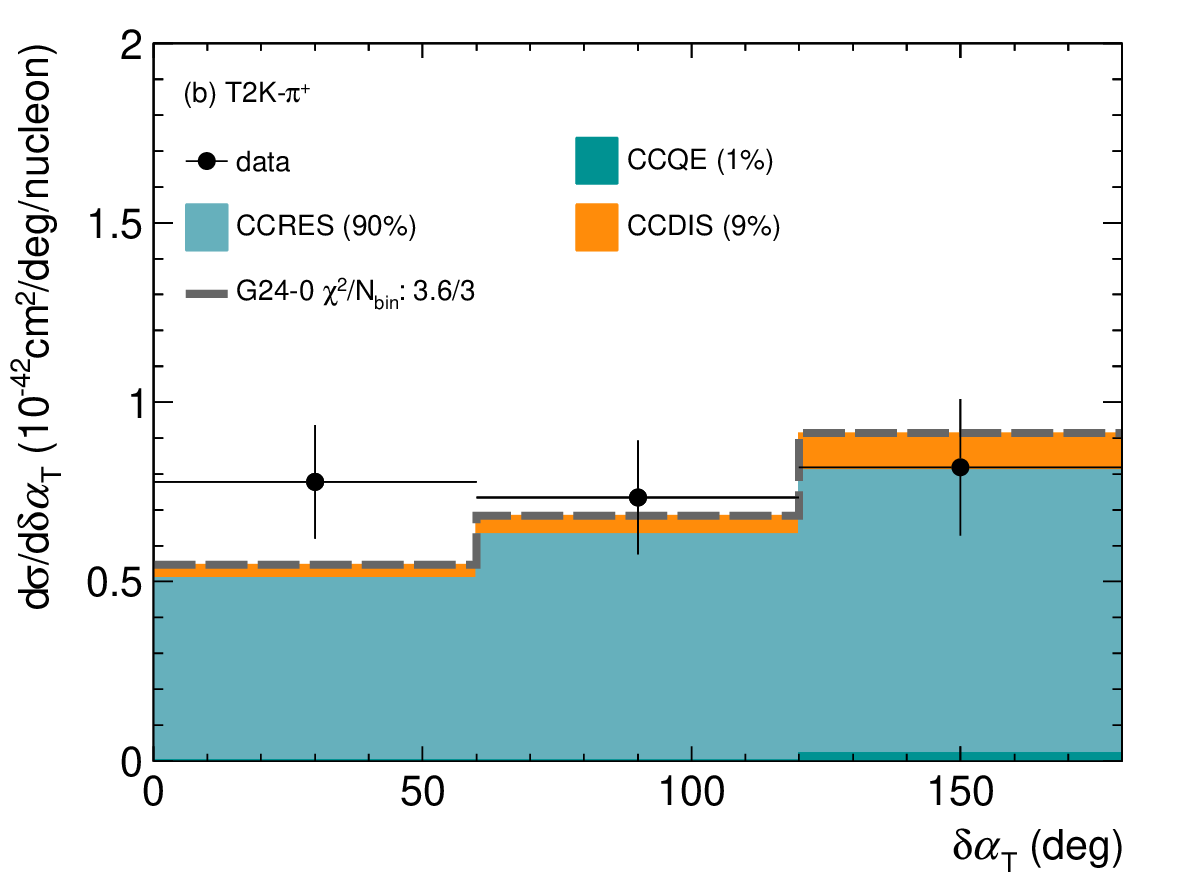} 
    \includegraphics[width=\fwid\textwidth]{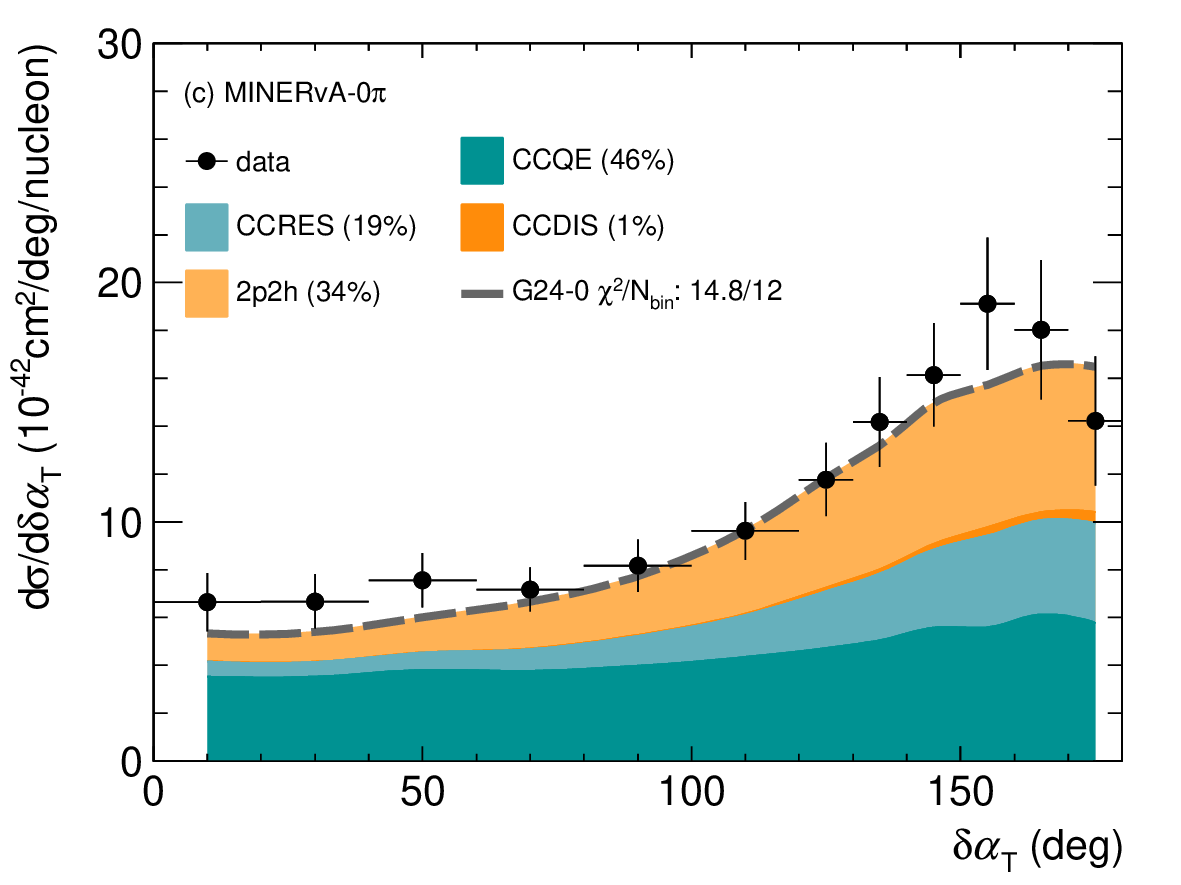} 
    \includegraphics[width=\fwid\textwidth]{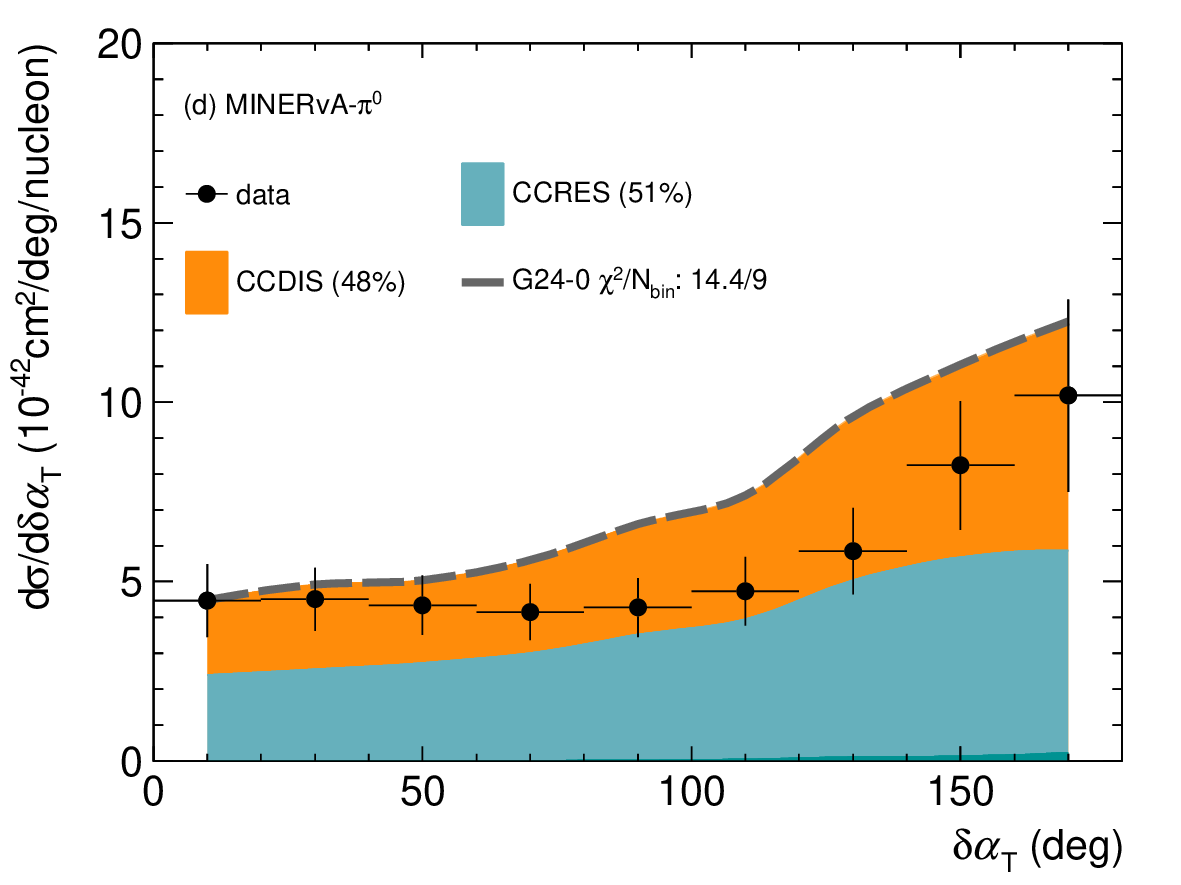}
    \caption{$\dat$ measurements decomposed in interaction types, compared to \gZero prediction.}   \label{fig:g24-0-dat-reac} 
    		
    \includegraphics[width=\fwid\textwidth]{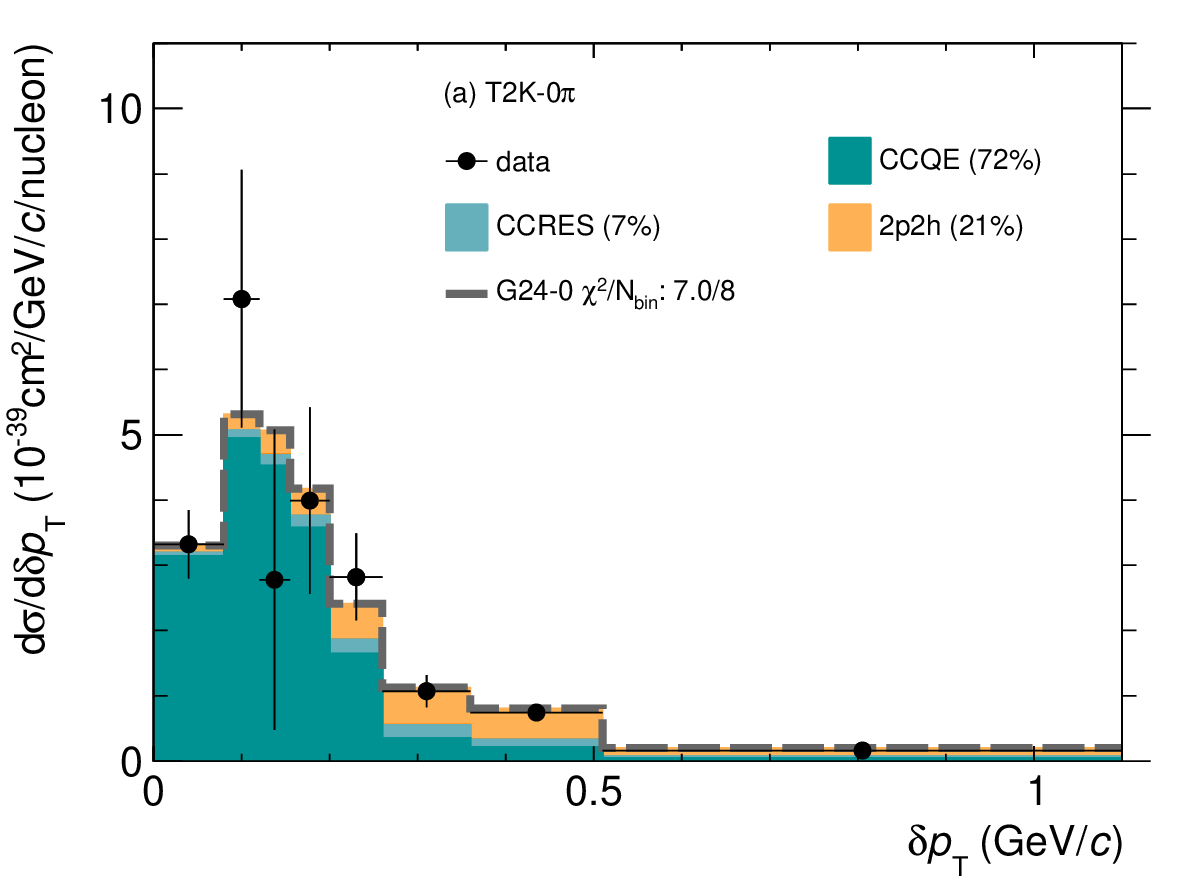}
    \includegraphics[width=\fwid\textwidth]{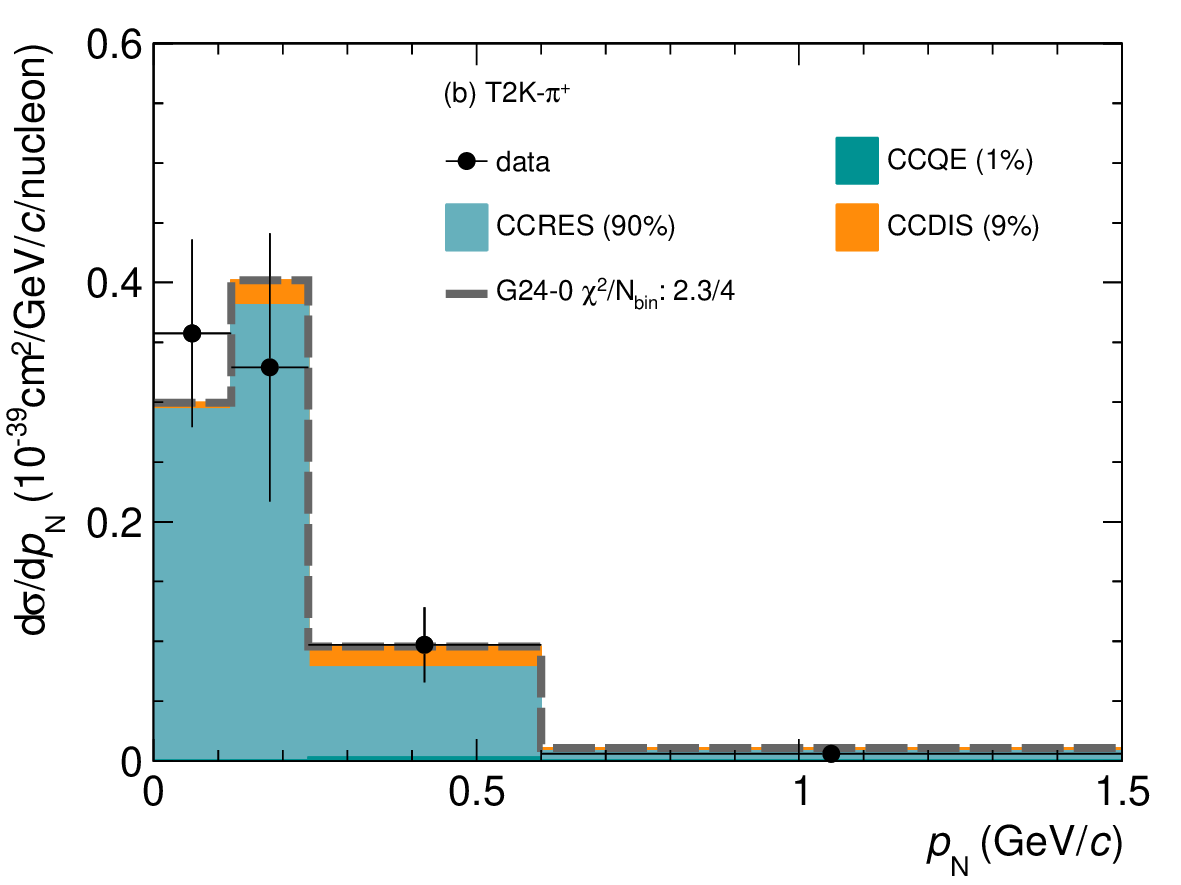}
    \includegraphics[width=\fwid\textwidth]{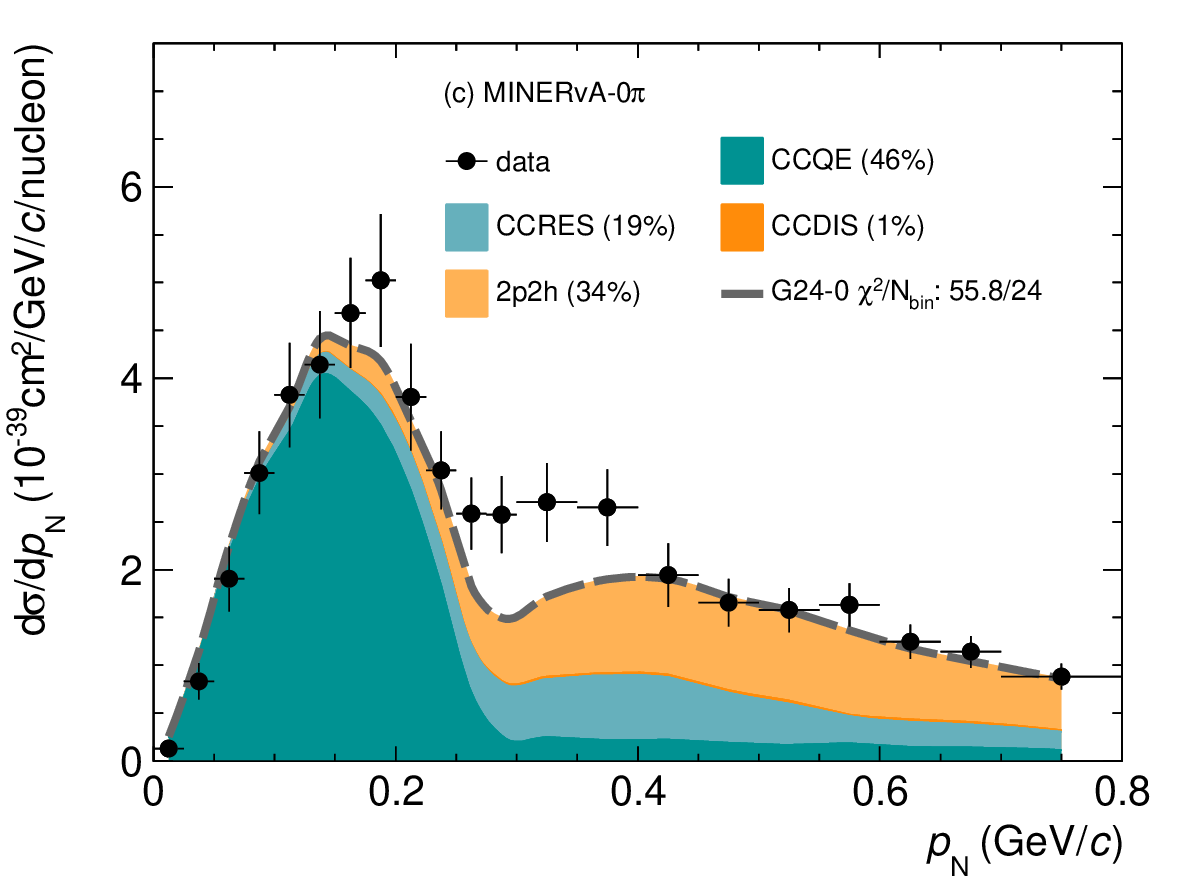}
    \includegraphics[width=\fwid\textwidth]{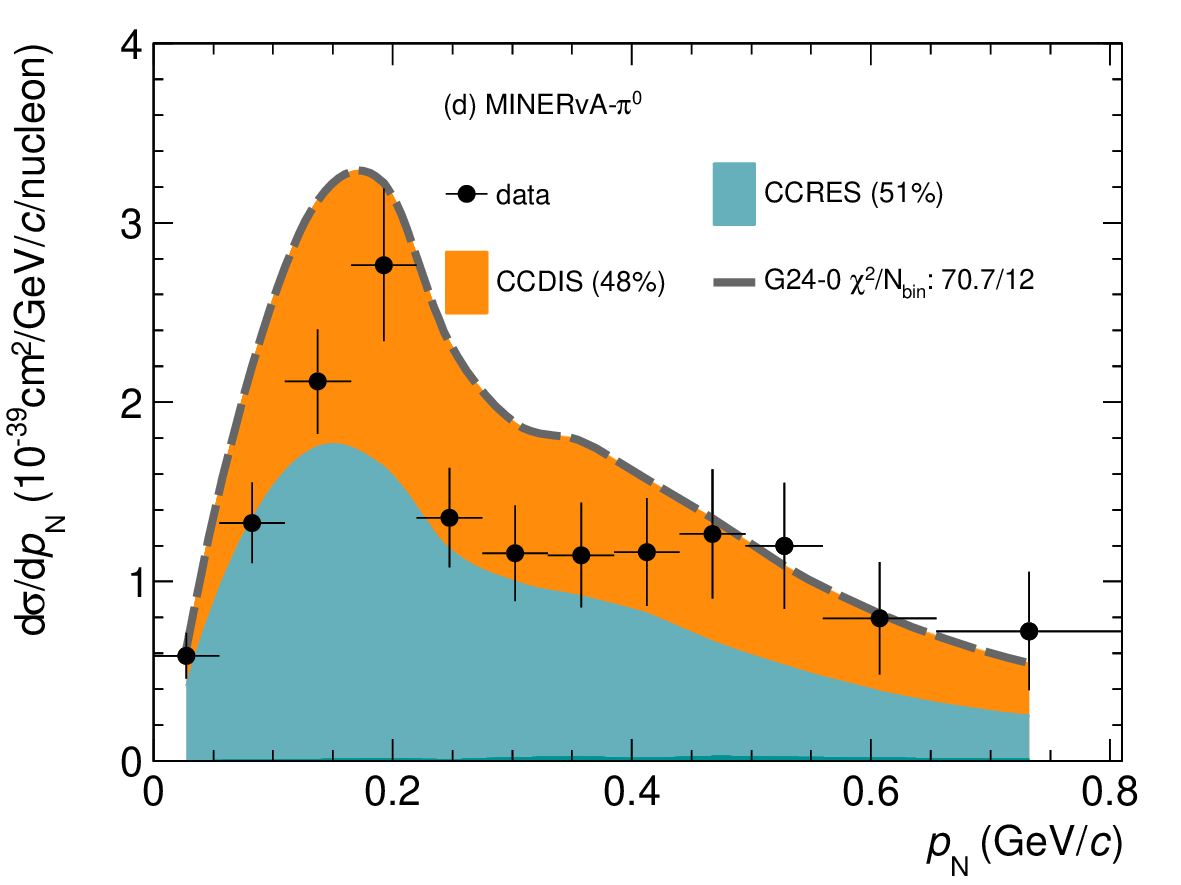}
    \caption{\label{fig:g24-0-pn-reac} Similar to Fig.~\ref{fig:g24-0-dat-reac} but for the $\pn$ (\ttkpip, \minzpi and \minpiz) and $\dpt$ (\ttkzpi) measurements.
    } 
\end{figure*}

\begin{table}[!htb]
    \centering
    \begin{tabular}{cc}
    \hline
    \hline
    Variables & Cuts ($p$ in $\gevc$) \\
    \hline
    \multicolumn{2}{c}{\ttkzpi~\cite{T2K:2018rnz}} \\
    \hline
    $\vecpmu$    &  $0.25 < p_\mu $, $\cos\theta_\mu>-0.6$   \\
    $\vecpp$     & $0.45< p_\text{p} <1.0$ , $\cos\theta_\text{p}>0.4$     \\
    \hline
    \multicolumn{2}{c}{\ttkpip~\cite{T2K:2021naz}} \\
    \hline
    $\vecpmu$    & $0.25 < p_\mu < 7$ , $\theta_\mu < 70^\degree$  \\
    $\vecpp$     & $0.45 < p_\text{p} <1.2$  ,  $\theta_\text{p} < 70^\degree$   \\
    $\vecppi$    & $0.15 < p_\pi <  1.2$, $\theta_\pi < 70^\degree$ \\
    \hline
    \multicolumn{2}{c}{\minzpi~\cite{MINERvA:2018hba, MINERvA:2019ope}} \\
    \hline
    $\vecpmu$     & $1.5< p_\mu < 10$ , $\theta_\mu < 20^\degree $  \\
    $\vecpp$      & $0.45< p_\text{p} <1.2$  , $\theta_\text{p} < 70^\degree$    \\
    \hline
    \multicolumn{2}{c}{\minpiz~\cite{MINERvA:2020anu}} \\
    \hline
    $\vecpmu$   & $1.5< p_\mu < 20$ , $\theta_\mu < 25^\degree$  \\
    $\vecpp$    & $0.45< p_\text{p} $                      \\
    \hline
    \hline
    \end{tabular}
    \caption{\label{tab:data-sets-phase-space-cut}
    Kinematic cuts for the samples of the TKI measurements.
    }
\end{table}

Given the significant physics potential of TKI measurements, a wealth of literature has emerged on the combined analysis of T2K and MINERvA CC0$\pi$ TKI data~\cite{Dolan:2018zye, Bourguille:2020bvw, Franco-Patino:2021yhd, Ershova:2022jah, Franco-Patino:2022tvv, GENIE:2022qrc, Chakrani:2023htw, Ershova:2023dbv}. However, model studies incorporating TKI data with pion production have been notably absent. Since the TKI measurements are sensitive to the initial state and FSI, this work investigates a partial tune of these two processes, in an effort to derive an effective model describing neutrino-nuclear pion production, while maintaining the efficacy with the pionless production.

\section{\genie model selection}\label{sec:genie}
\genie uses formatted strings to uniquely define configurations, which are usually called tunes. 
Regardless of the name, they can refer to configurations either before or after an actual tuning procedure. 

Free nucleon tuning effort in Ref.~\cite{GENIE:2021zuu} has produced a tune, $\geighteen$, with better description of bubble chamber data. 
This is the starting point of the model we want to tune. 
The first effect to apply to this model is adding the proper initial state. 
Among the available \genie initial state models, the best choice is the spectral-function-like Correlated Fermi Gas model (\sfcfg)~\cite{sfcfg-talk,sfcfg-GitHubCommit,GENIE:2021npt}. 
\sfcfg is essentially based on the previously implemented Local Fermi Gas (LFG) model in \genie with a larger number of adjustable degrees of freedom for tuning and improved physics. 
It differs from the previous implementation of LFG in two aspects. 
``Spectral-function-like'' refers to the implementation of removal energy as a varying function rather than a fixed value, while ``Correlated'' highlights the incorporation of the high-momentum tail above the Fermi momentum due to nucleon-nucleon short-range correlations (SRC), as evidenced by electron-scattering data~\cite{PhysRevLett.96.082501}.
These two improvements are considered and modeled in the Valencia Model in Ref.~\cite{Nieves:2004wx}.
Hence, \sfcfg resembles closer the initial state of the Valencia model than previous LFG implementations.
More details on the \sfcfg can be found in Ref.~\cite{GENIE:2021npt}. 
Note that the original Valencia Model in Ref.~\cite{Nieves:2004wx} has its own method of modeling FSI. 
However, since FSI is factorized from other models in implementation in \genie, this method is not used.
Instead, GENIE develops the INTRANUKE hA FSI model (hA for short) for its simplicity and interpretability, which is the FSI candidate for the model to be tuned in this work.

Further improvements to the configuration to be tuned include: 1) using $z$-expansion axial-vector form factor~\cite{Hill:2010yb} rather than the dipole form factor of the Valencia model in QE processes~\cite{Nieves:2004wx}; and 2) replacing the Valencia model~\cite{Nieves:2011pp} in 2p2h processes with the SuSAv2 Model~\cite{Gonzalez-Jimenez:2014eqa} since it covers larger region of the $q^0$, $q^3$ phase space.

The complete list of the model components are given in Table ~\ref{tab:default-gen-list}. 
In the \genie code, this configuration is identified as \newtune and it is currently the default \genie configuration for a number of LAr based experiments. 
As \newtune is widely used in the paper, we will call it \gZero for simplicity. 

Comparing the \genie predictions for \gZero against TKI datasets, Figs.~\ref{fig:g24-0-dat-reac} and~\ref{fig:g24-0-pn-reac},  shows that the model fails to  accurately describe the MINERvA $\pi^0$~\cite{MINERvA:2020anu} TKI measurement. 

\begin{table}[!htb]
    \centering
    \begin{tabular}{p{4cm}c}
    \hline
    \hline
    \textrm{Simulation component} & \textrm{Model} \\
    \hline
    \textrm{Nuclear state}              & \sfcfg~\cite{sfcfg-talk,sfcfg-GitHubCommit,GENIE:2021npt} \\ 
    \textrm{QE}               & Valencia~\cite{Nieves:2004wx} \\
    \textrm{2p2h}               & SuSAv2~\cite{Gonzalez-Jimenez:2014eqa} \\
    \textrm{QE $\Delta S=1$}           & Pais~\cite{Pais:1971er} \\
    \textrm{QE $\Delta C=1$}                  & Kovalenko~\cite{Kovalenko:1990zi} \\
    \textrm{Resonance (RES)}                        & Berger-Sehgal~\cite{Berger:2007rq}\\
    Shallow/Deep inelastic \par scattering (SIS/DIS)                    & Bodek-Yang~\cite{Bodek:2002vp}\\
    \textrm{DIS $\Delta C=1$}           & Aivazis-Tung-Olness~\cite{Aivazis:1991fy}\\
    \textrm{Coherent $\pi$ production}  & Berger-Sehgal~\cite{Berger:2008xs}\\
    \hline
    \textrm{Hadronization}              & AGKY~\cite{Yang:2009zx}\\
    \textrm{FSI}                        & INTRANUKE hA~\cite{Andreopoulos:2015wxa}\\
    \hline
    \hline
    \end{tabular}
    \caption{\label{tab:default-gen-list} Model components of \gZero. Processes with non-trivial $\Delta S$ and $\Delta C$ are those with strangeness and charm production, respectively.}
\end{table}

\section{\label{sec:Tuning}Tuning initial-state and FSI models on TKI data}

\begin{figure}[!htb] 	
    \centering 		
    \includegraphics[width=\fwid\textwidth]{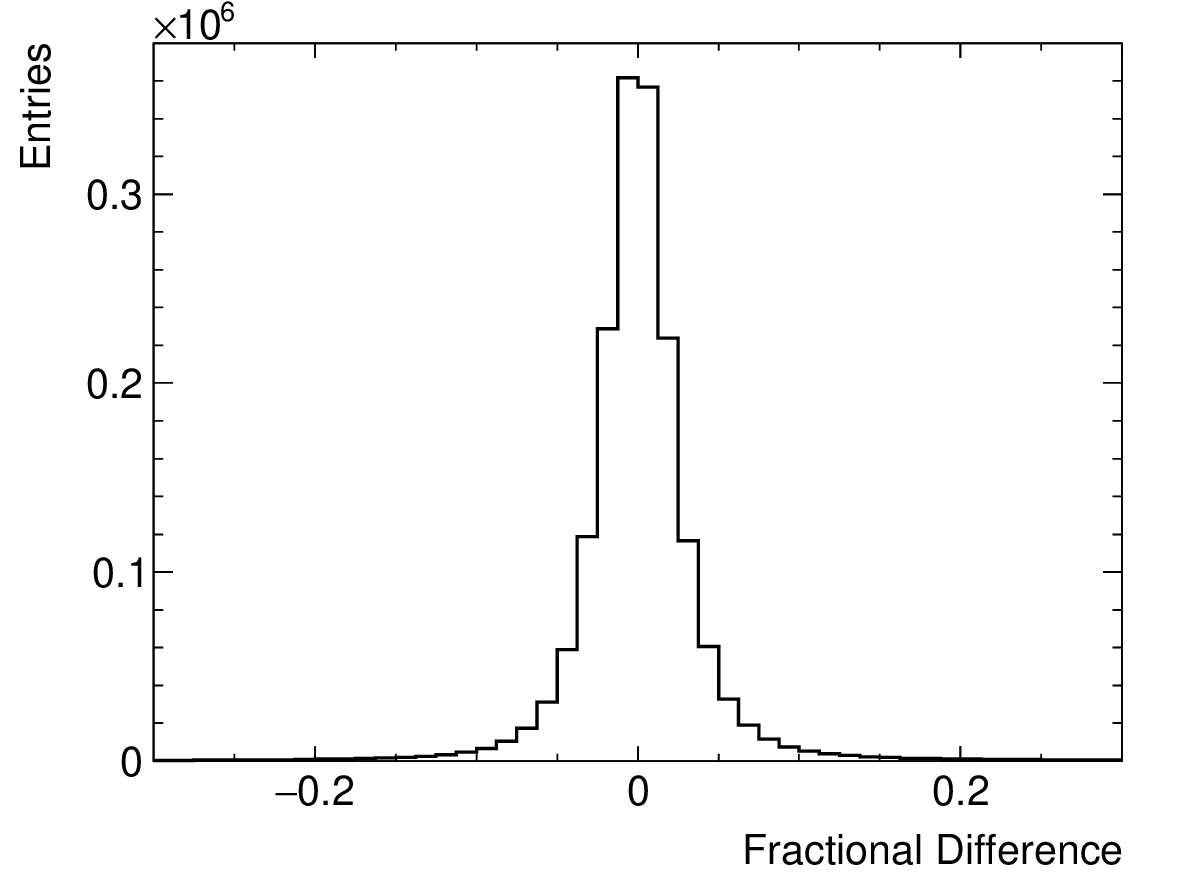}
    \caption{\label{fig:residual} Fractional difference for the bin-by-bin cross sections between MC truth and the parameterized approximation with order-4 polynomials,  both in the Norm-Shape (NS)~\cite{DAgostini:1993arp,Hanson:2005mrg} space with \allpar. See Table~\ref{tab:hALFG-para} for the definition of \allpar. The residual has a mean of $0.003$ and a standard deviation of $0.073$.} 
\end{figure}

This study adopts the tuning procedure outlined in Ref.~\cite{GENIE:2022qrc}, utilizing $N_{\textrm{par}}$ model parameters. The objective is to identify the best fit---that is, the optimal set of parameter values---within the parameter space that minimizes the $\chi^2$ between model predictions and data. Given the high dimensionality of the parameter space, conducting a brute-force, point-by-point scan on a grid is infeasible. Therefore, a sufficient number of points is randomly sampled. Each point is used to run a full simulation. The simulation output for each data observable bin is then parameterized using Professor~\cite{Buckley:2009bj} with a polynomial of the model parameters of a chosen order, $N_{\textrm{ord}}$. The minimal number of points ($N_{\textrm{s}}$) required for tuning $N_{\textrm{par}}$ parameters with an order $N_{\textrm{ord}}$ polynomial is determined by the combination formula,
\begin{equation}
    N_{\textrm{s}} = \binom{N_{\textrm{par}}+N_{\textrm{ord}}}{N_{\textrm{ord}}}.
\end{equation}
Given that $N_\textrm{s}$ increases factorially, caution is essential when augmenting the number of parameters or the polynomial degree. Given that our tuning incorporates all $2$ \sfcfg and $12$ hA parameters (see Sec.~\ref{sec:tuning-para-choice}), a degree-$4$ parameterization  necessitates over 6,000 generations. Hence, only up to order $4$ polynomials are explored, and it has shown to reproduce the MC predictions well, as shown in Fig.~\ref{fig:residual}. 

Furthermore, to circumvent Peelle's Pertinent Puzzle~\cite{PPP_FNL,Chakrani:2023htw}, the Norm-Shape (NS) transformation prescription~\cite{DAgostini:1993arp,Hanson:2005mrg} is applied. The extremal point is then found from a minimization of $\chi^2$ between this NS-transformed polynomial approximation and NS-transformed data. In minimization, priors, usually based on systematic uncertainty, can be imposed on each parameter to penalize it from getting too far from its default value. The following subsections elaborate on the specific choice of measurement observables and model parameters in this work.

\subsection{\label{sec:tuning-obs-choice} Data points}

\begin{figure}[!htb] 
    \centering 		
    \includegraphics[width=\fwid\textwidth]{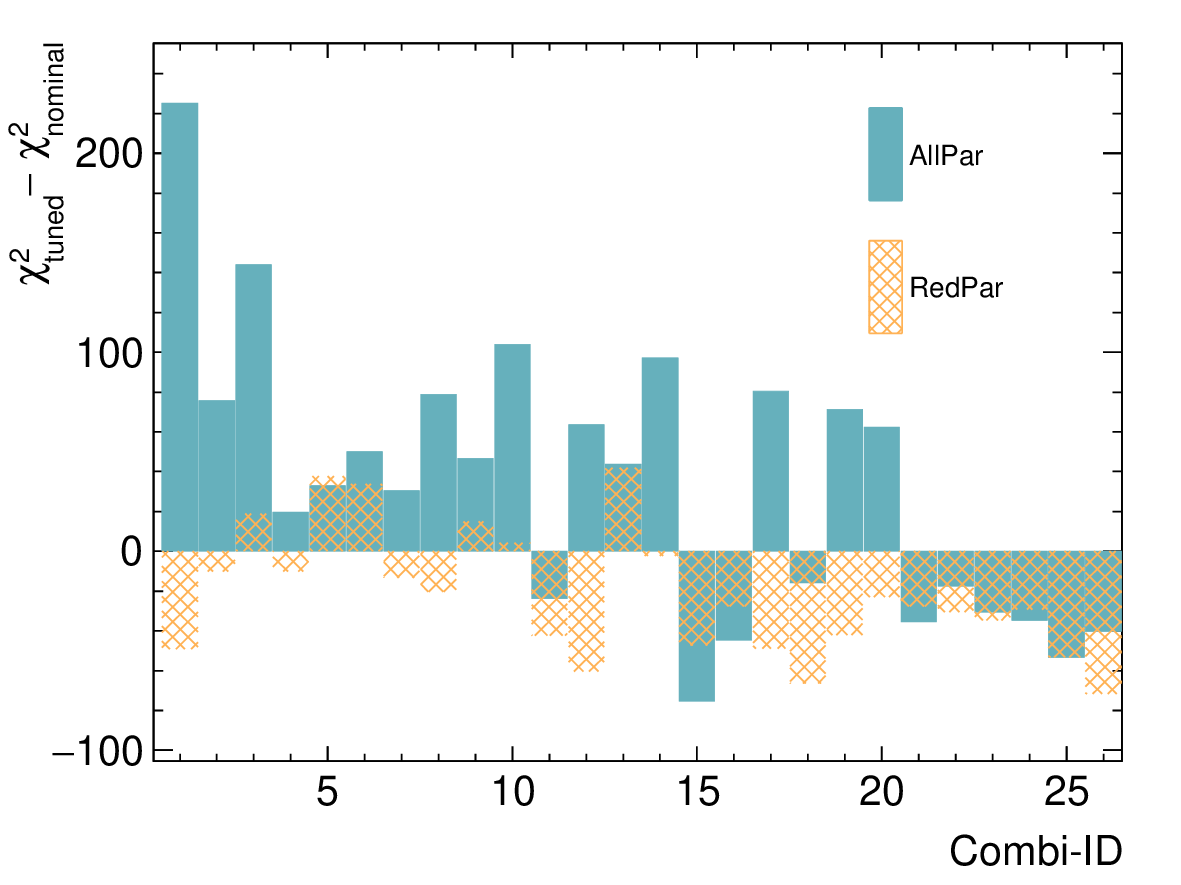} 
    \caption{\label{fig:allchi} Change of $\chi^2$ calculated for the full (i.e., tuned plus validation) observable set as a function of the tuned combination (cf. Table~\ref{tab:fit-var-combo} in Appendix~\ref{sec:appcombi}). The two model parameter sets (\allpar and \redpar, see Table~\ref{tab:hALFG-para} for definitions) are compared and it can be seen that the respective minima happen at \texttt{Combi-15} and 26. }   
\end{figure}

The observables for the reported differential cross sections across the four TKI measurements---\ttkzpi, \ttkpip, \minzpi, and \minpiz---are detailed in Table~\ref{tab:data-sets}. To identify the most sensitive variables, various combinations were evaluated  for tuning purposes. A total of 26 combinations were assessed,  with the superset comprising $\dat$, $\dpt$, $\dphit$, $\pn$, and $\dptt$ (see Table~\ref{tab:fit-var-combo} in Appendix~\ref{sec:appcombi} for all combinations). Each combination was evaluated such that non-selected variables---including proton momentum and angle ($p_\text{p}$ and $\theta_\text{p}$)~\cite{MINERvA:2018hba}, as well as $\dptx$ and $\dpty$~\cite{MINERvA:2019ope} in \minzpi---were used solely for validation; the model is expected to accurately predict these variables post-tuning. When constructing the combinations, it was observed that $\dphit$ is strongly dependent  on neutrino beam energy~\cite{Lu:2015tcr}, which suppresses its sensitivity to nuclear effects. Additionally, correlations exist between various observables, notably between $\dpt$ and $\pn$. Therefore, determining the optimal combination is not straightforward; this study employs a criterion based on  the $\chi^2$ with the complete (tuned plus validation) observable set (Fig.~\ref{fig:allchi}).

\begin{table}[!htb]
    \centering
    \begin{tabular}{cp{1.1cm}p{1.5cm}p{1.5cm}p{1.5cm}}
    \hline
    \hline
    Observables & No. of \par bins & \texttt{Combi-} \par \texttt{Superset}  & \texttt{Combi-} \par \texttt{Best-} \par \allpar& \texttt{Combi-} \par \texttt{Best-} \par \redpar\\
    \hline
    \multicolumn{5}{c}{\ttkzpi} \\
    \hline
       $\dat$            & $8$                & $\tick$     &  & $\tick$  \\ 
       $\dpt$            & $8$                & $\tick$     & $\tick$  & $\tick$ \\ 
       $\dphit$          & $8$                & $\tick$     &  &  \\      
    \hline
    \multicolumn{5}{c}{\ttkpip} \\
    \hline
      $\dat$            & $3$                & $\tick$      &  & $\tick$  \\
      $\pn$             & $4$                & $\tick$      & $\tick$  & $\tick$ \\ 
      $\dptt$           & $5$                & $\tick$      &  & $\tick$  \\ 
    \hline
    \multicolumn{5}{c}{\minzpi} \\
    \hline  
      $\dat$            & $12$               & $\tick$      & & $\tick$  \\ 
      $\pn$             & $24$               & $\tick$      & $\tick$  & $\tick$ \\
      $\dpt$            & $24$               & $\tick$      & $\tick$ &  \\     
      $\dphit$          & $23$               & $\tick$      &  &  \\     
      $\pp$             & $25$               &      &  &  \\     
      $\thetap$         & $26$               &      &  &  \\     
      $\dptx$           & $32$               &      &  &  \\  
      $\dpty$           & $33$               &      &  & \\     
    \hline
    \multicolumn{5}{c}{\minpiz} \\
    \hline
      $\dat$            & $9$               & $\tick$      & & $\tick$      \\  
      $\pn$             & $12$               & $\tick$     & $\tick$  & $\tick$ \\ 
      $\dptt$           & $13$               & $\tick$     &  & $\tick$  \\
    \hline
    \hline
    \end{tabular}
    \caption{\label{tab:data-sets}
	Observables of the TKI measurements and their binning. Those with ``$\tick$'''s are used for tuning, while those without  are for the respective validation. See Table~\ref{tab:hALFG-para} for definitions of \cbRedPar and \cbAllPar.
    }
\end{table}

\subsection{\label{sec:tuning-para-choice} Model parameters}
As outlined in Sec.~\ref{sec:genie}, we focus exclusively on the currently tunable parameters within the \sfcfg and hA models. Therefore, $14$ parameters (collectively referred to as the \allpar set) are utilized for tuning, as detailed in Table~\ref{tab:hALFG-para}. Nominal values and associated uncertainties for the hA model are taken from Table 17.3 in Ref.~\cite{Andreopoulos:2015wxa}. 

\begin{table}[!htb]
    \centering
    \begin{tabular}{cp{1.5cm}p{1.5cm}p{1.2cm}p{1.2cm}}
    \hline
    \hline
    \textrm{Parameter} & \textrm{Nominal} (\gZero)     & \textrm{Range In} \par \textrm{Tuning} & \allpar \par (\gT)  & \redpar \par (\gC) \\ 
    \hline
    \multicolumn{5}{c}{\sfcfg} \\
    \hline
    \textrm{$\srcfr$} & 0.12 & (0.0, 0.5)  & \tick & \tick\\
    \textrm{$\nurmec$} & 0.01 & (0.0, 0.2) & \tick & \\
    \hline
    \multicolumn{5}{c}{hA} \\
    \hline
    \textrm{$\cpimfp$} & $1.0\pm0.2$ & (0.0, 3.0) & \tick & \\
    \textrm{$\pizmfp$} & $1.0\pm0.2$ & (0.0, 3.0) & \tick & \tick\\
    \textrm{$\nmfp$} & $1.0\pm0.2$ & (0.0, 3.0) & \tick &\\
    \hline
    \textrm{$\picex$} &  $1.0\pm0.5$ & (0.0, 3.0) & \tick & \tick \\
    \textrm{$\ncex$} & $1.0\pm0.5$ & (0.0, 3.0)  & \tick & \tick\\
    \hline
    \textrm{$\piinel$} & $1.0\pm0.4$ & (0.0, 3.0) & \tick & \\
    \textrm{$\ninel$} & $1.0\pm0.4$ & (0.0, 3.0)  & \tick &\\
    \hline
    \textrm{$\cpiabs$} & $1.0\pm0.2$ & (0.0, 3.0) & \tick &\\
    \textrm{$\pizabs$} & $1.0\pm0.2$ & (0.0, 3.0) & \tick &\\
    \textrm{$\nabs$} & $1.0\pm0.2$ & (0.0, 3.0)  & \tick & \tick\\
    \hline
    \textrm{$\pipiprod$} & $1.0\pm0.2$ & (0.0, 3.0) & \tick &\\
    \textrm{$\npiprod$} & $1.0\pm0.2$ & (0.0, 3.0)  & \tick & \tick\\
    \hline
    \hline
    \end{tabular}
    \caption{\label{tab:hALFG-para}
    Tuneable parameters and their ranges in the  \sfcfg (\textit{uppermost} group) and hA (\textit{lower} groups, uncertainties from Ref.~\cite{Andreopoulos:2015wxa}) models. Parameters to be tuned in the two sets are marked with ``\tick'''s. See later text for definitions of \gT and \gC.
    }
\end{table}

In our analysis of the \sfcfg model, the two most relevant parameters are $\srcfr$, representing the fraction of the high-momentum tail from SRC, notably above the Fermi surface, and $\nurmec$, marking the commencement of the nuclear removal energy distribution for carbon. A larger $\srcfr$ indicates the presence of more energetic initial nucleons, while an increased $\nurmec$ implies that greater energy is necessary to liberate a nucleon from the carbon nucleus, so the product particles will possess lower energy. Given the novelty and limited constraints of the spectral-function-like implementation, this study employs relatively relaxed priors for $\srcfr$ and $\nurmec$, at 0.12$\pm$0.12 and 0.01$\pm$0.005 \gev, respectively.

\genie users employ the hA model~\cite{Andreopoulos:2015wxa} for the majority of neutrino analysis, owing to its straightforward reweighting capability. Unlike cascade models such as the hN model~\cite{Andreopoulos:2015wxa}, this model determines the FSI type for each hadron exactly once, without considering further propagation of rescattered hadrons within the nucleus. Its model parameters are explained as follows.

\begin{enumerate}
    \item 
The model evaluates a mean free path (MFP), $\lambda$, which varies with the hadron's energy, $E$, and its radial distance, $r$, from the nucleus center,  as follows:
\begin{equation}
    \lambda(E,r) = \frac{1}{\sigma_\textrm{hN,tot}(E)\rho(r)},
\end{equation}
where $\sigma_\textrm{hN,tot}$ is the total hadron-nucleon cross section and $\rho$ is the number density of the nucleons inside the nucleus. Once a hadron is generated inside the nucleus, it travels a distance $\lambda$ before a probabilistic determination of rescattering occurs, inversely related to  $\lambda$. If no rescattering happens, the hadron will be propagated for another $\lambda$ and the dice will be thrown again. The cycle repeats until a rescattering happens or the hadron is propagated far enough to escape the nucleus without any interactions. If a rescattering happens, the type of rescattering is evaluated according to their cross section. The relative probabilities of each type are extracted from stored hadron scattering data~\cite{LADS:1999dyv,Navon:1983xj,Carroll:1976hj,Clough:1974qt,BAUHOFF1986429,Mashnik:2000up,Ishibashi:1997gbe}. After a specific rescattering type is chosen, the rescattered product will be generated. These rescattered products will not undergo further rescattering and will be considered out of the nucleus as final state particles.  
The MFP values for pions and nucleons can be modified using scaling factors  ($\cpimfp$, $\pizmfp$,  and $\nmfp$) detailed in Table~\ref{tab:hALFG-para}. Both changed pions, $\pip$ and $\pim$, are scaled by the same factor, $\cpimfp$, due to isospin symmetry. The $\lambda$ for $\piz$ is calculated from those of charged pions based on isospin symmetry as well and can be adjusted separately by $\pizmfp$. If the factor is larger than $1$, $\lambda$ will increase and hence the total number of steps required for the hadron to escape the nucleus reduces, thereby reducing the average probability of rescattering. 
Given the high precision with which total hadron-nucleus cross sections are determined~\cite{LADS:1999dyv,Navon:1983xj,Carroll:1976hj,Clough:1974qt,BAUHOFF1986429}, strict Gaussian priors (error---$\sigma$, not to be confused with the various cross sections---of $0.2$, same as the systematic uncertainties shown in Table~\ref{tab:hALFG-para}) are applied to the MFP scaling factors. 
Varying $\pizmfp$ by $\pm2.5~\sigma$ significantly affects the \minpiz $\textrm{d}\sigma/\textrm{d}\pn$, as depicted in Fig.~\ref{fig:minpiz-pn-pi0mfp}. Decreasing MFP naturally leads to increased rescattering. Hence, fewer $\piz$s can escape the nucleus and fewer events will have $\piz$ in the final states, thereby reducing the cross section. 

\begin{figure}[!htb]
    \centering
    \includegraphics[width=\fwid\textwidth]{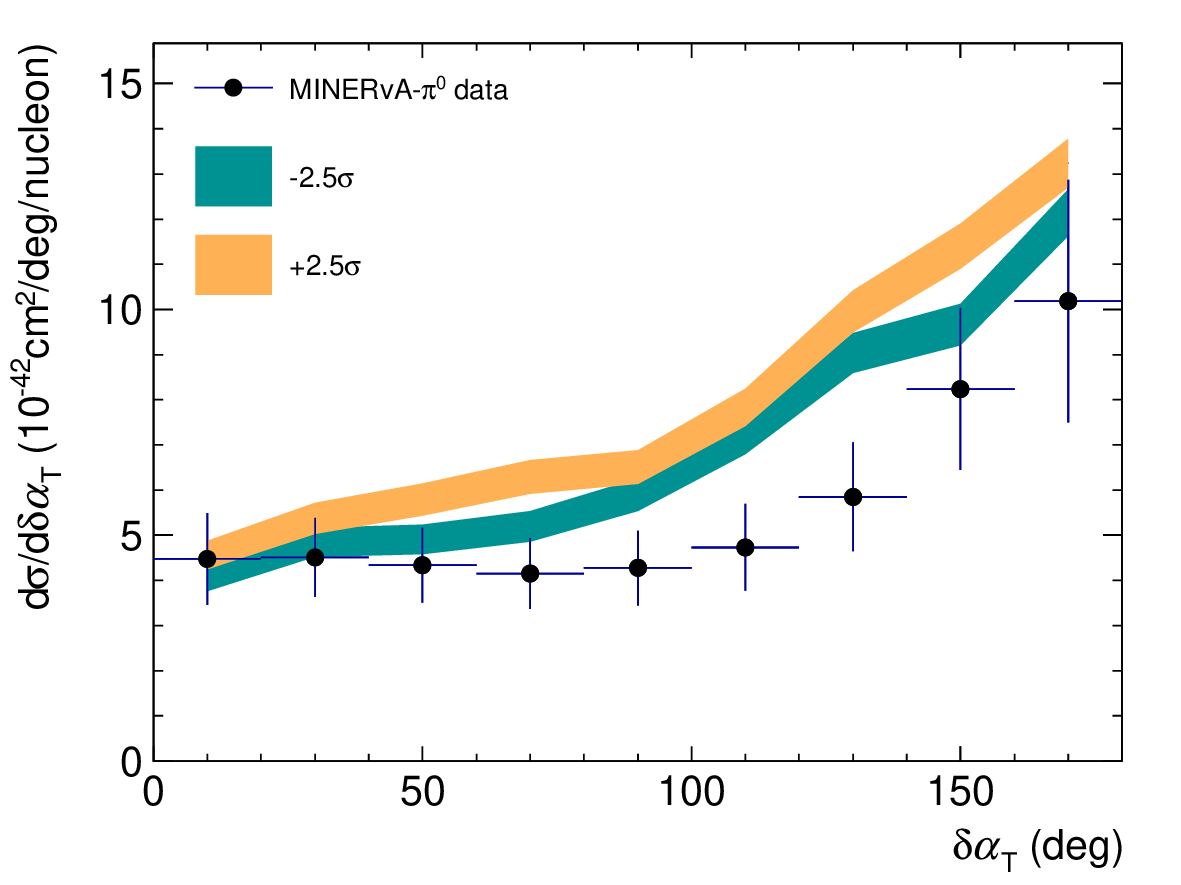}
    \caption{Effect of varying $\pizmfp$ by $\pm2.5~\sigma$ compared to the MINERvA-$\pi^0$ measurement. Each band's width indicates the \genie prediction's statistical uncertainty from $10^5$ events.}
    \label{fig:minpiz-pn-pi0mfp}
\end{figure}

\item 
The relative probability of each rescattering type can be adjusted by a scale factor. There are four rescattering types for both nucleons and pions available for tuning. They are charge exchange (CEX), inelastic scattering (INEL), absorption (ABS), and pion production (PIPD). The default energy-dependent probability distributions come from hadron data tuning~\cite{LADS:1999dyv,Navon:1983xj,Carroll:1976hj,Clough:1974qt,BAUHOFF1986429} and can be adjusted by respective scaling factors, like $\picex$ for CEX of all pions, listed in Table~\ref{tab:hALFG-para}. The relative probabilities of all four rescattering types will be normalized such that they sum up to one. Scaling factors directly multiply  the probability of each type, with new relative probabilities subsequently renormalized. More specifically, suppose the initial fraction of absorption, $f_\textrm{ABS}$, is $0.5$, which means that the sum of the other fractions is  $f_\textrm{other}=f_\textrm{INEL}+f_\textrm{CEX}+f_\textrm{PIPD}=1-0.5=0.5$. If $f_\textrm{ABS}$ is scaled up with $S_\textrm{ABS}=2$, the new absorption fraction is 
\begin{equation}
    f^\prime_\textrm{ABS} = \frac{f_\textrm{ABS}*S_\textrm{ABS}}{f_\textrm{ABS}*S_\textrm{ABS}+f_\textrm{oth}} = 2/3,
\end{equation}
which is not equal to simply multiplying $f_\textrm{ABS}$ by $S_\textrm{ABS}$ due to the presence of the other rescattering types. Hence, it is unavoidable that a change of the scaling factor for one type would affect others. Nonetheless, when the cross section is plotted with a breakdown according to rescattering type, the interpretation will be clear regarding the increase or decrease of a particular FSI type. This correlation implies that the scaling parameters' effective impact on FSI fate is often less pronounced than what the raw numbers suggest. As illustrated in the above example, $f_\textrm{ABS}$ is only scaled up by $ f^\prime_\textrm{ABS}/f_\textrm{ABS}=4/3=1.3$ instead of $S_\textrm{ABS}=2$. Hence, a relatively more relaxed Gaussian prior, with a $\sigma$ of $0.5$, is placed on the FSI fate scales, slightly larger than the systematic uncertainties shown in Table~\ref{tab:hALFG-para}. 

Understanding each rescattering type is crucial for interpreting tuning results. A detailed elaboration of the four tunable FSI types is given as follows. 

\begin{enumerate}
    \item 
CEX involves changing the charge of the participating particles; for example,
\begin{equation}
    \pip + \n \rightarrow \piz + \p,
\end{equation}
or vice versa. This rescattering type is crucial for event topologies requiring the presence of a pion;  depending on the signal pion charge, CEX could migrate events between signal and background. 

\item 
INEL is the case where the nucleus is left in an excited state after the rescattering. This category only contains the situation where a single additional nucleon is emitted/knocked-out after rescattering. Since it does not affect the number of pions produced, it will not convert an event from a pionless topology to a pion-production topology. The effects on nucleons are two-fold. Firstly, it can alter the number of signal events within each event topology. If the inelastic rescattering leads to two low-momentum protons below the detection threshold as opposed to a high-momentum proton, this signal event will be discarded as no protons are observed. Secondly, INEL invariably changes the kinematics of the rescattering particle. Be it the leading proton or the leading pion, based on which the TKI observables are calculated, the TKI distribution shape will be affected. Hence, while $\ninel$ would affect all four data sets, $\piinel$ would only affect \ttkpip and \minpiz. 

\item 
ABS refers to the case where the particle undergoes an interaction so that it does not emerge as a final particle. For example, $\pi^+$ can interact with two or more nucleons, initially forming a baryon resonance that subsequently interacts with other nucleons, emitting multiple nucleons rather than pions. Hence, the $\pi^+$ would not emerge from the nucleus anymore.

\item 
PIPD happens for energetic particles where an extra pion emerges as a result of the rescattering, such as
\begin{equation}
    \p + \p \rightarrow \p + \n + \pip.
\end{equation}
Such an interaction significantly alters the event topology. 

\end{enumerate}
\end{enumerate}

\section{\label{sec:results}The first TKI-driven \genie tunes}

\begin{figure}[!htb] 	
    \centering 		
    \setkeys{Gin}{width=\linewidth}
    \begin{subfigure}{\fwid\textwidth}
    \includegraphics{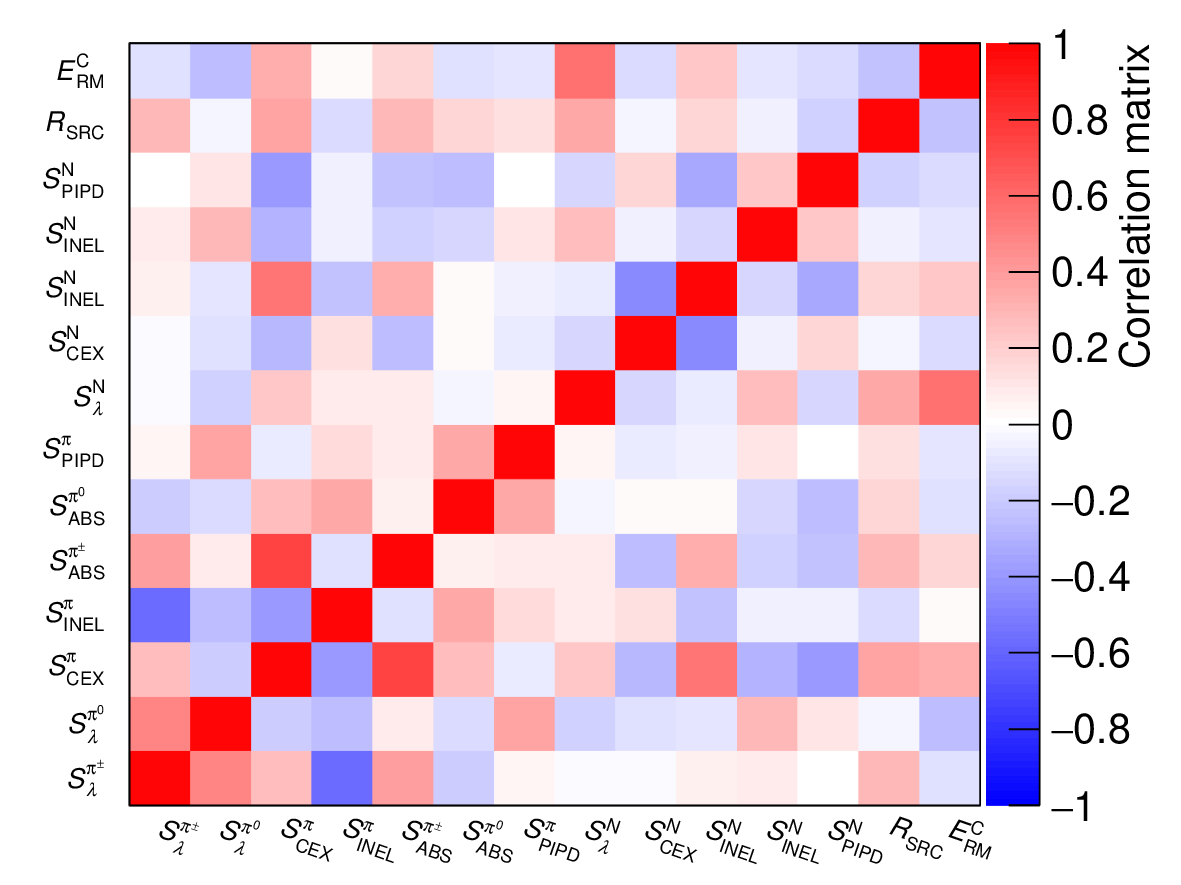}
    \subcaption{\label{fig:comb_26_cor_allpar}\allpar}
    \end{subfigure}
    \begin{subfigure}{\fwid\textwidth}
    \includegraphics{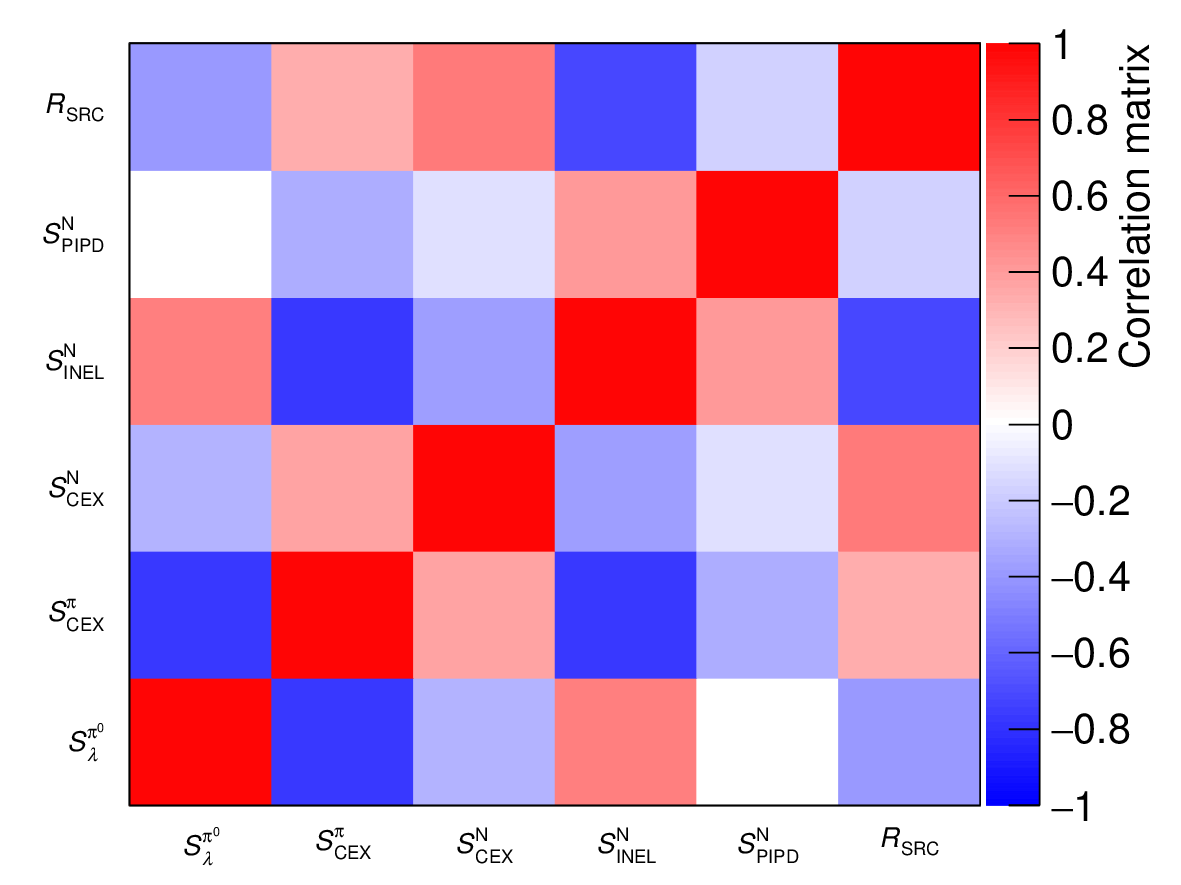}
    \subcaption{\label{fig:comb_26_cor_redpar}\redpar}
    \end{subfigure}
    \caption{ Post-fit Correlation coefficient on \texttt{Combi-26}. } 
\end{figure}

\begin{figure*} 
    \centering 		
    \includegraphics[width=\fwid\textwidth]{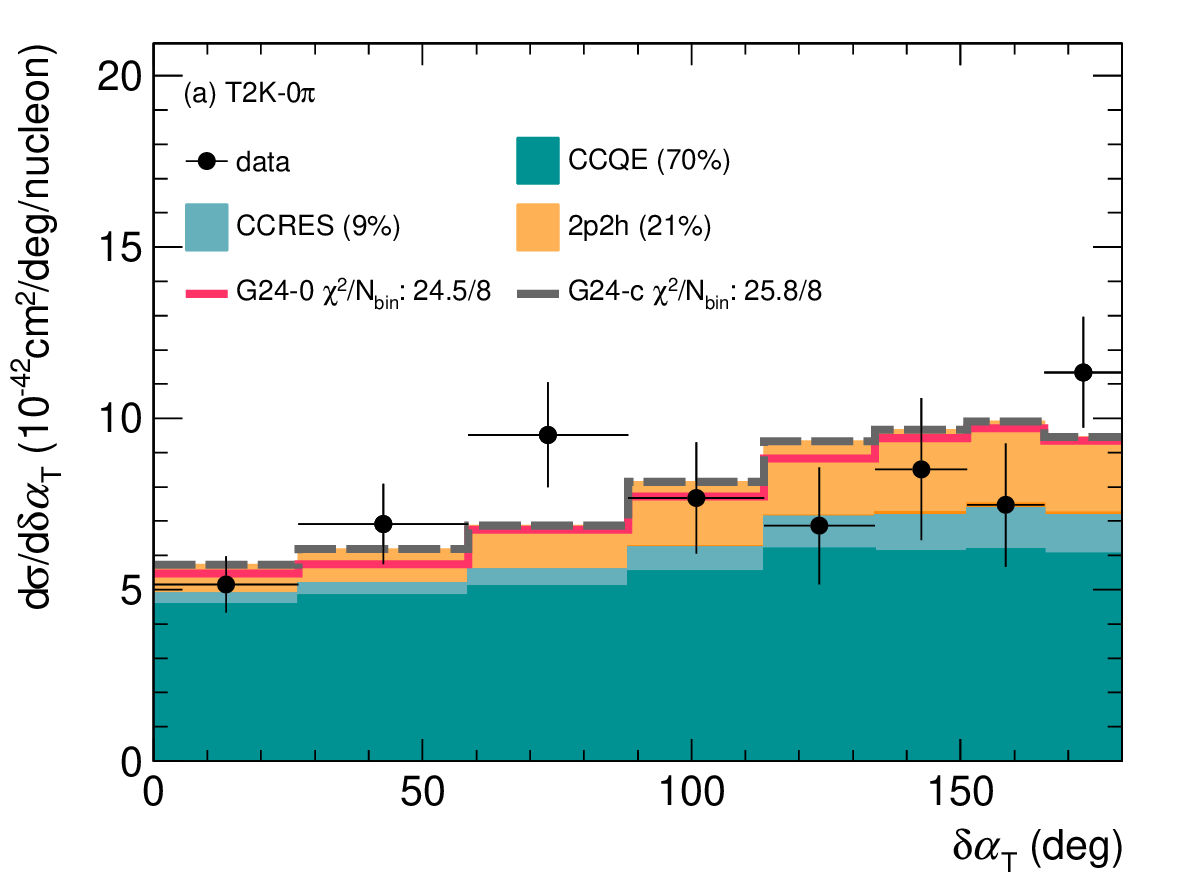}
    \includegraphics[width=\fwid\textwidth]{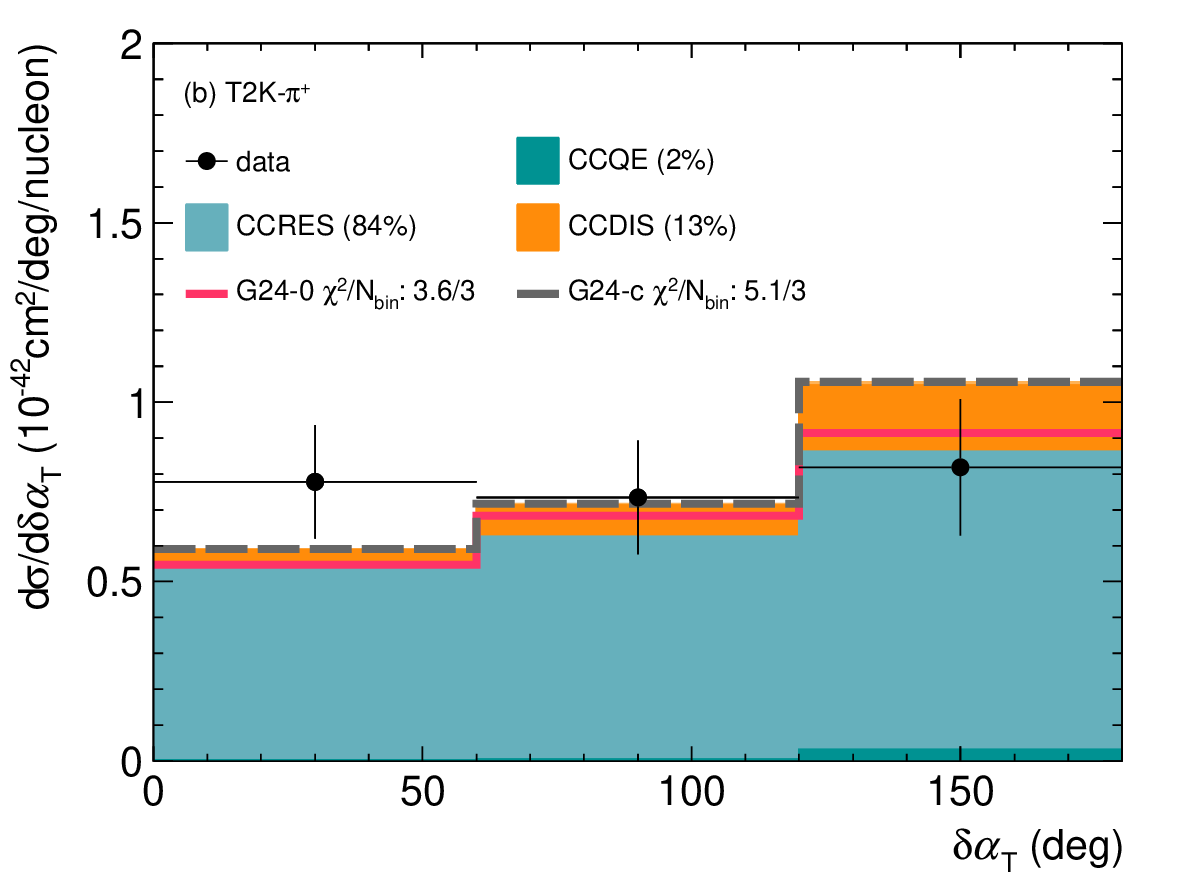}
    \includegraphics[width=\fwid\textwidth]{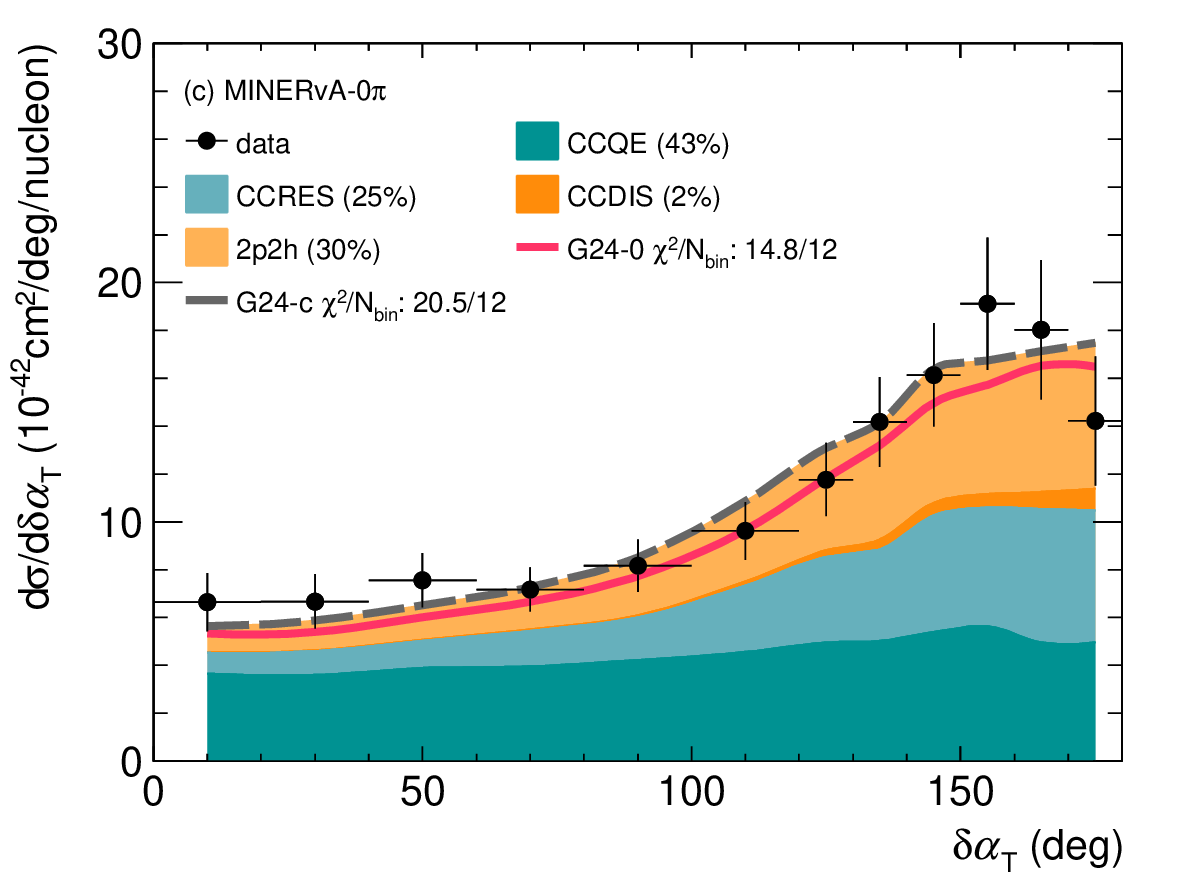}
    \includegraphics[width=\fwid\textwidth]{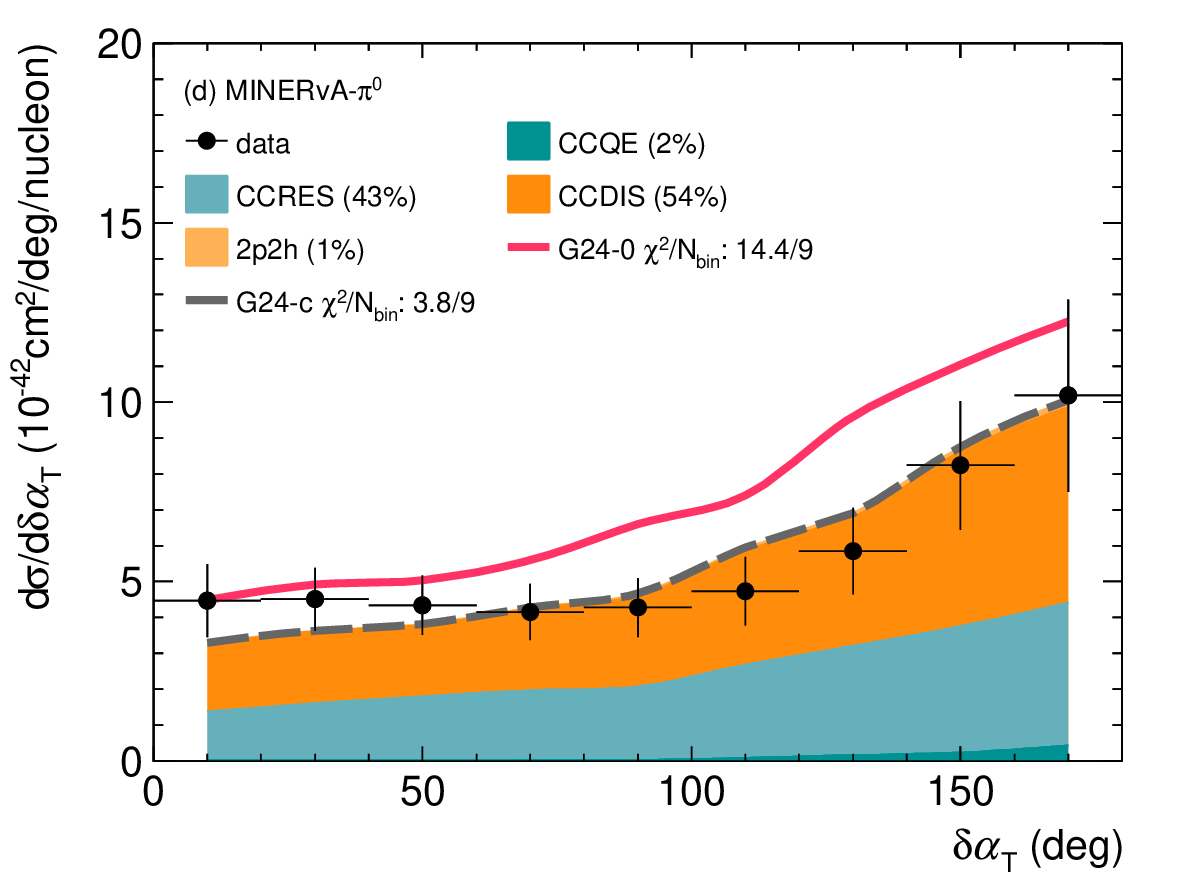}
    \caption{\label{fig:g24-c-dat-reac} 
    Similar to Fig.~\ref{fig:g24-0-dat-reac} but with \gC.  The \gZero prediction is also plotted for comparison. 
    } 

    \includegraphics[width=\fwid\textwidth]{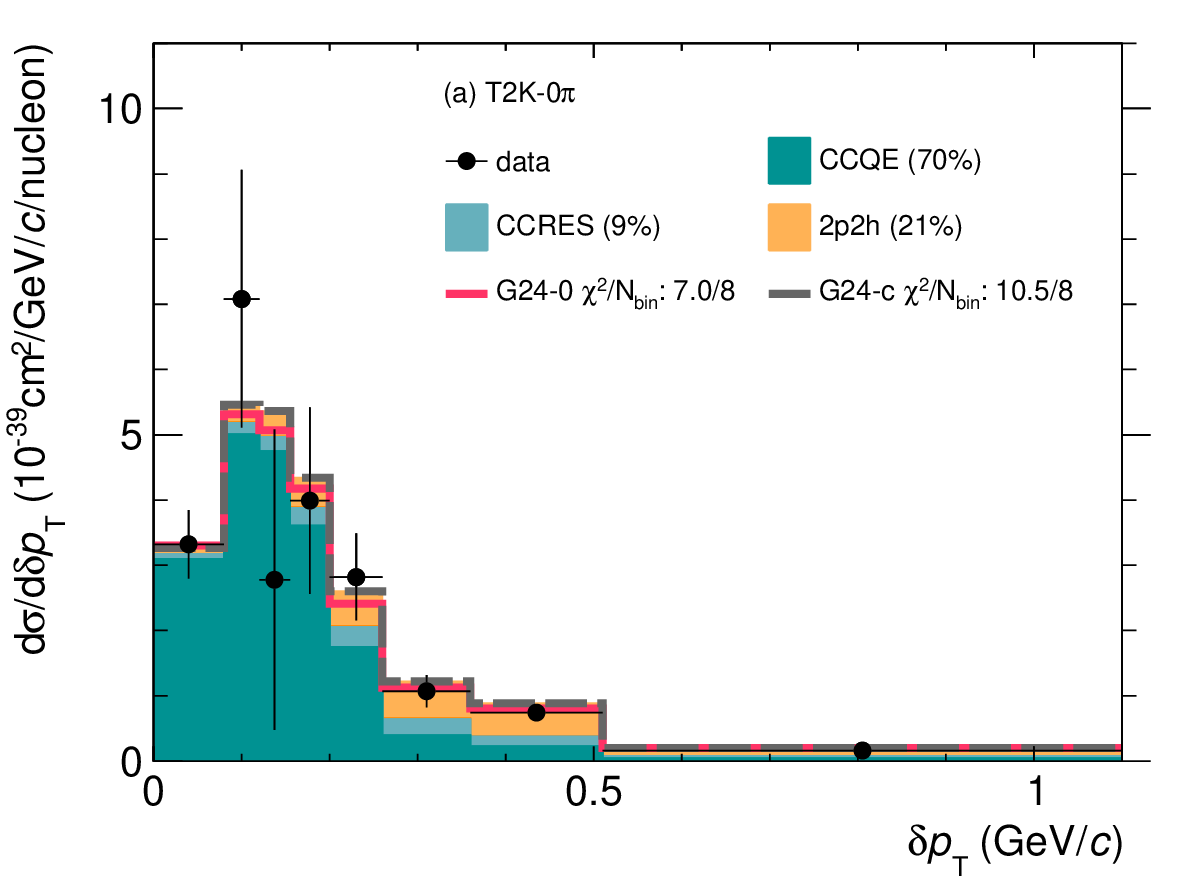}
    \includegraphics[width=\fwid\textwidth]{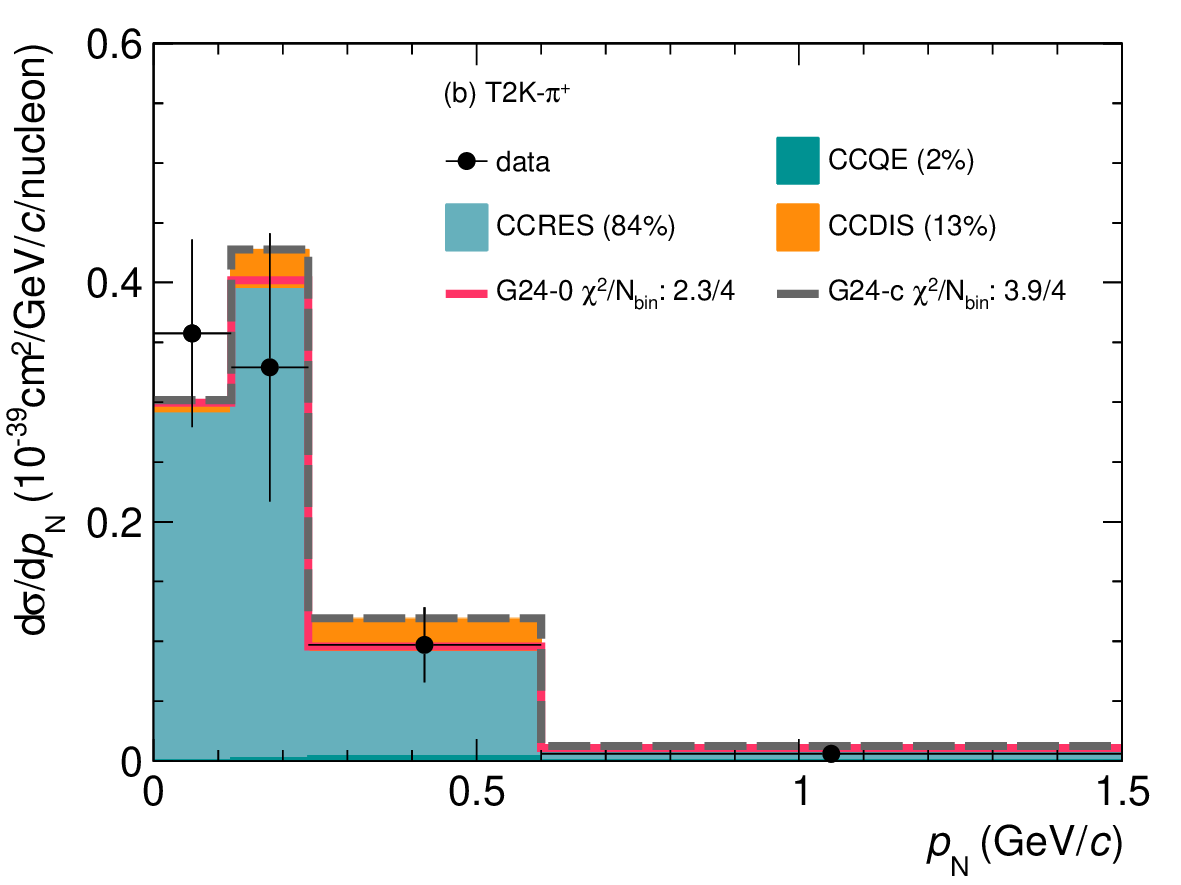}	
    \includegraphics[width=\fwid\textwidth]{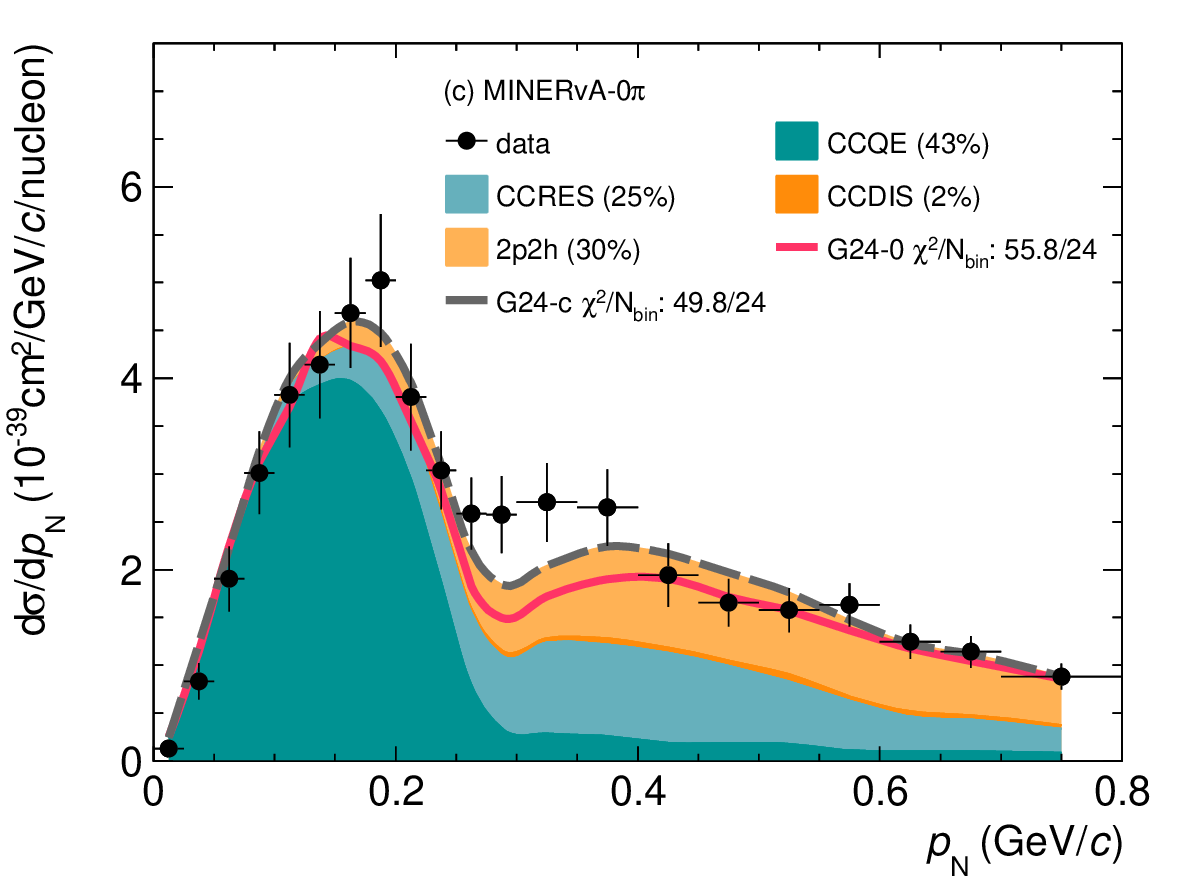}
    \includegraphics[width=\fwid\textwidth]{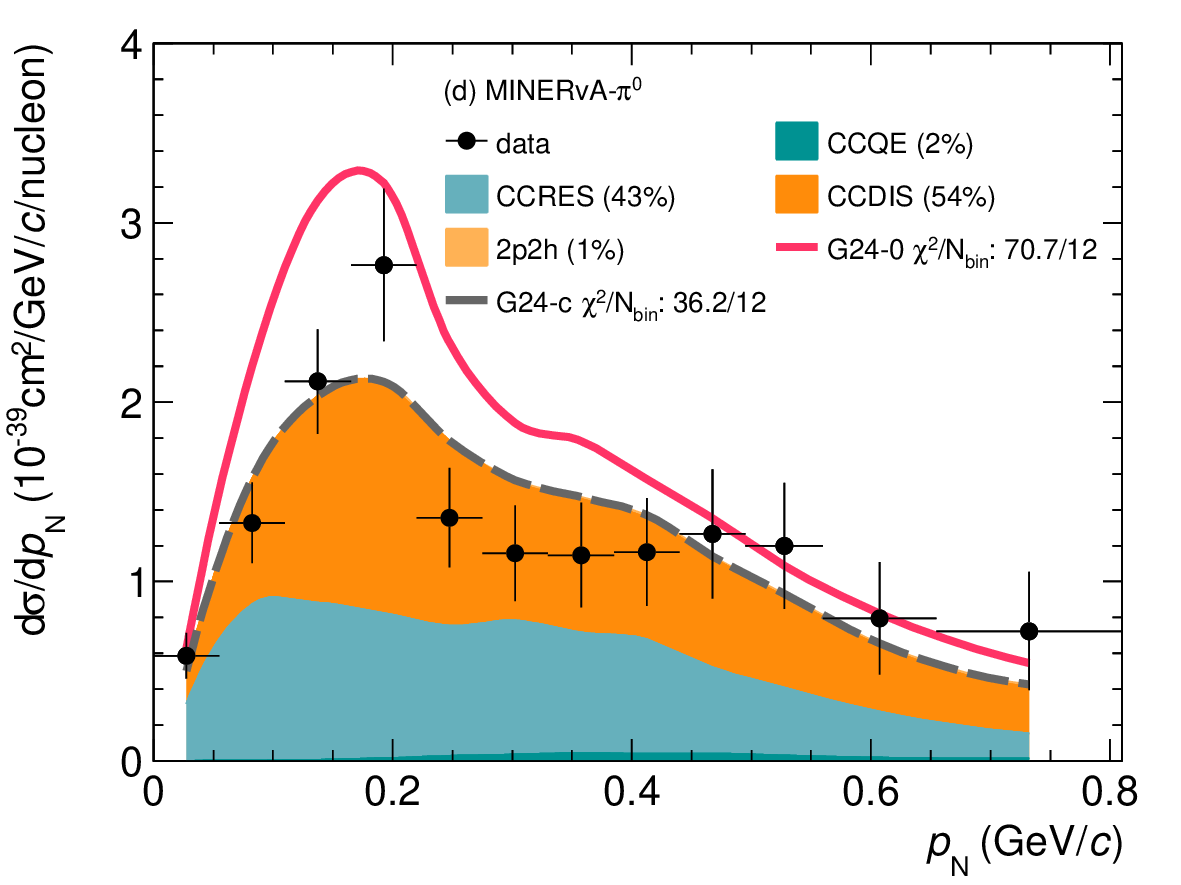}
    \caption{\label{fig:g24-c-pn-reac}  
    Similar to Fig.~\ref{fig:g24-c-dat-reac} but for the $\pn$ (\ttkpip, \minzpi and \minpiz) and $\dpt$ (\ttkzpi) measurements. 
    } 
\end{figure*}

Given the intricate correlations among model parameters, determining which ones are most sensitive to the data proves challenging. Considerable correlation and anti-correlation are observed between different FSI fates for the same particle, e.g. $\cpiabs$ and $\picex$, as shown in Figs.~\ref{fig:comb_26_cor_allpar} and~\ref{fig:comb_26_cor_redpar}. 
Our tuning began with the \allpar parameter set, achieving the best result with \texttt{Combi-15}, which is denoted as \cbAllPar in Table~\ref{tab:data-sets} (cf. Fig.~\ref{fig:allchi} for the $\chi^2$ distribution of all combinations).
Observations showed that for most combinations, certain parameters remained close to their default values, within $1$ $\sigma$ of the imposed priors, such as $\pipiprod$ since none of the data sets has a significant contribution from PIPD of pions---Removing these parameters from the tuning process does not significantly affect the outcome.
Therefore, we then constructed a reduced set comprising only $\srcfr$, $\pizmfp$, $\picex$, $\ncex$, $\nabs$, and $\npiprod$ (denoted as \redpar in Table~\ref{tab:hALFG-para}) and ran the tuning on the 26 combinations with it. Note that the parameters most correlated with $\picex$ and $\pizmfp$ are not present in \redpar. The \redpar tuning proved more stable, with nearly all combinations showing a more negative $\chi^2$ change than the \allpar tuning, as shown in Fig.~\ref{fig:allchi}. Moreover, the single-observable (\texttt{Combi-}1 to 5), single-measurement (\texttt{Combi-}6 to 9), single-experiment (\texttt{Combi-}10 and 11), and single-topology (\texttt{Combi-}12 and 13) combinations all systematically yielded no or limited overall (tuned plus validation) improvement post-tuning. The best tuning then occurred with \texttt{Combi-26} (\cbRedPar in Table~\ref{tab:data-sets}), that is, all TKI variables excluding $\dphit$ and $\dpt$ (unless no $\pn$ is available)---$\dat$, $\pn$ (or $\dpt$ if $\pn$ is unavailable), and $\dptt$---from pionless and pion-production measurements in both T2K and MINERvA. 

Table~\ref{tab:restunes} summarizes the primary results of our tuning process. The tuned \redpar, \restunefull (\gC), is derived from the observable set \cbRedPar, and the tuned \allpar, $\alttune$~(\gT), is from \cbAllPar. The upper part of Table~\ref{tab:restunes} contains the parameter values. For the  \gC tune, changes to the \sfcfg model were moderately sized ($\srcfr$ increases from $0.12$ to $0.15$), while in the hA model, $\pizmfp$ and $\picex$ are highly suppressed and the nucleon FSI has larger CEX and PIPD and lower ABS.  The tune \gT differs most significantly from \gC in $\srcfr$ and $\picex$. The table's lower section illustrates $\chi^2$ improvements for selected observables and their corresponding validation sets. Crucially, tuning should also enhance the accuracy of validation set descriptions. Overall, both tunes have substantial reduction in the total $\chi^2$. In the following, we will discuss the two tunes.

\begin{table}[!htb]
    \centering
    \begin{tabular}{p{1.5cm}p{1.5cm}p{2.1cm}p{2.3cm}}
    \hline
    \hline
    Parameter              & Nominal \par (\gZero) &  \redpar \par (\gC)     & \allpar \par (\gT)   \\
    \hline
    \multicolumn{4}{c}{\sfcfg} \\
    \hline
    $\srcfr$   & 0.12 & 0.15 $\pm$ 0.08         & 0.30  $\pm$ 0.05       \\
    $\nurmec$  & 0.01 & 0.01                    & 0.011 $\pm$ 0.003     \\
    \hline
    \multicolumn{4}{c}{hA} \\
    \hline
    $\cpimfp$  & $1.0\pm0.2$ & 1.0    & 1.11   $\pm$ 0.16       \\
    $\pizmfp$  & $1.0\pm0.2$ & 0.22 $\pm$ 0.07          & 0.17   $\pm$ 0.06        \\
    $\nmfp$    & $1.0\pm0.2$& 1.0                      & 1.20   $\pm$ 0.12     \\
    \hline
    $\picex$   & $1.0\pm0.5$ & 0.26 $\pm$ 0.12          & 1.53   $\pm$ 0.37        \\
    $\ncex$    & $1.0\pm0.4$ & 1.43 $\pm$ 0.34          & 1.41   $\pm$ 0.38       \\
    \hline
    $\piinel$  & $1.0\pm0.4$ & 1.0                      & 0.67   $\pm$ 0.30        \\
    $\ninel$   & $1.0\pm0.4$ & 1.0                     & 1.26   $\pm$ 0.48       \\
    \hline
    $\cpiabs$  & $1.0\pm0.2$ & 1.0                      & 1.59   $\pm$ 0.31         \\
    $\pizabs$  & $1.0\pm0.2$ & 1.0                      & 0.90   $\pm$ 0.28         \\
    $\nabs$    & $1.0\pm0.2$ & 0.25 $\pm$ 0.28          & 0.28   $\pm$ 0.27       \\
    \hline
    $\pipiprod$& $1.0\pm0.2$ & 1.0                      & 1.12   $\pm$ 0.30     \\
    $\npiprod$ & $1.0\pm0.2$ & 2.05 $\pm$ 0.48          & 1.27   $\pm$ 0.48       \\ 
    \hline
    \hline
    \multicolumn{4}{c}{$\chi^2$ \textrm{for} \texttt{combi}} \\
    \hline
    \texttt{untuned}         & & 231.75         & 161.26        \\
    \texttt{tuned}           & & 174.84         & 122.53        \\
    \texttt{diff}            & & -56.91         & -38.73        \\
    \hline
    \multicolumn{4}{c}{$\chi^2$ \textrm{for} \texttt{vald}} \\
    \hline
    \texttt{untuned}    & & 229.5          & 299.99        \\
    \texttt{tuned}            & & 214.7          & 263.41        \\
    \texttt{diff}       & & -14.8          & -36.58        \\
    \hline
    \multicolumn{4}{c}{$\chi^2$ \textrm{for} \texttt{combi+vald}} \\
    \hline
    \texttt{untuned}    & &  461.25          &  461.25        \\
    \texttt{tuned}          &  & 389.54 & 385.94        \\
    \texttt{diff} & & -71.71         & -75.31  \\     
    \hline
    \hline
\end{tabular}
\caption{\label{tab:restunes}
	Parameters in \gZero, \gC, and \gT. Those without errors are not tuned. Lower section: \texttt{combi} indicates that the following $\chi^2$ sums are calculated for the tuned measurements (i.e. \cbRedPar (98 bins) and \cbAllPar (72 bins) for \gC and \gT, respectively), while \texttt{vald} indicates that the respective validation sets are used; \texttt{untuned} means that the $\chi^2$ calculation uses the nominal values of the parameters, while \texttt{tuned} denotes the tuned ones. \texttt{diff} displays the improvement achieved by the respective tuning.  
}
\end{table}

\subsection{Reduced tune: \gC}

To assess the tune's quality, we plot predictions using \gC for $\dat$ and $\pn$ in Figs.~\ref{fig:g24-c-dat-reac} and~\ref{fig:g24-c-pn-reac},  respectively. For \ttkzpi, \ttkpip and \minzpi, the new $\chi^2$ values are comparable to those obtained with \gZero, changing less than the number of bins. For \minpiz, \gC is distinctly better than \gZero, thereby demonstrating the possibility of simultaneous good descriptions of both pionless and pion production samples with constrained parameters from cross-topology TKI tuning. Note that in the region of $\pn\sim0.3~\gevc$, the model deficit previously reported in Ref.~\cite{MINERvA:2018hba} (and also shown in Fig.~\ref{fig:g24-0-pn-reac}c) persists even after the fit. The physical origin of this deficit might be attributed to the strength of the 2p2h contribution~\cite{MINERvA:2018hba}, but it is still a topic of significant discussion within the community.

\begin{figure}[!htb] 	
    \centering 		
    \includegraphics[width=\fwid\textwidth]{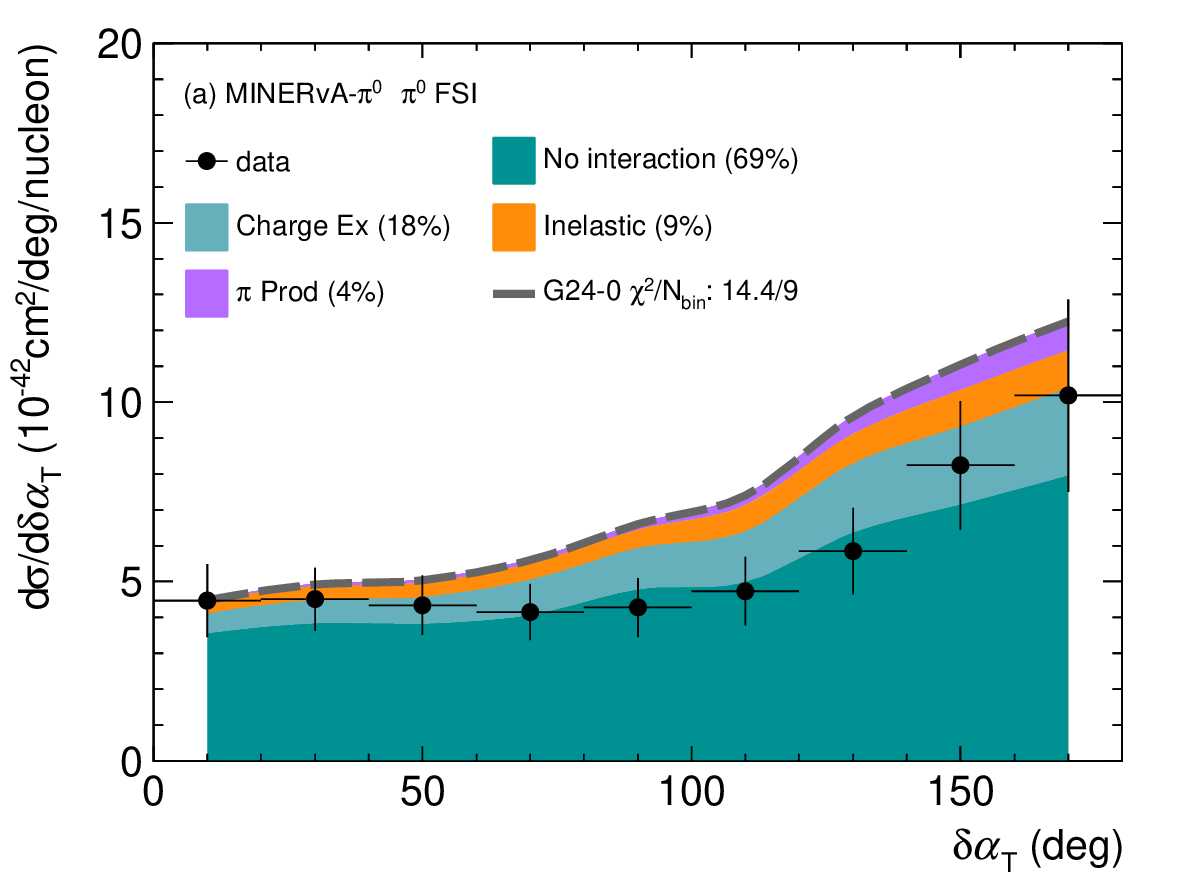}
    \includegraphics[width=\fwid\textwidth]{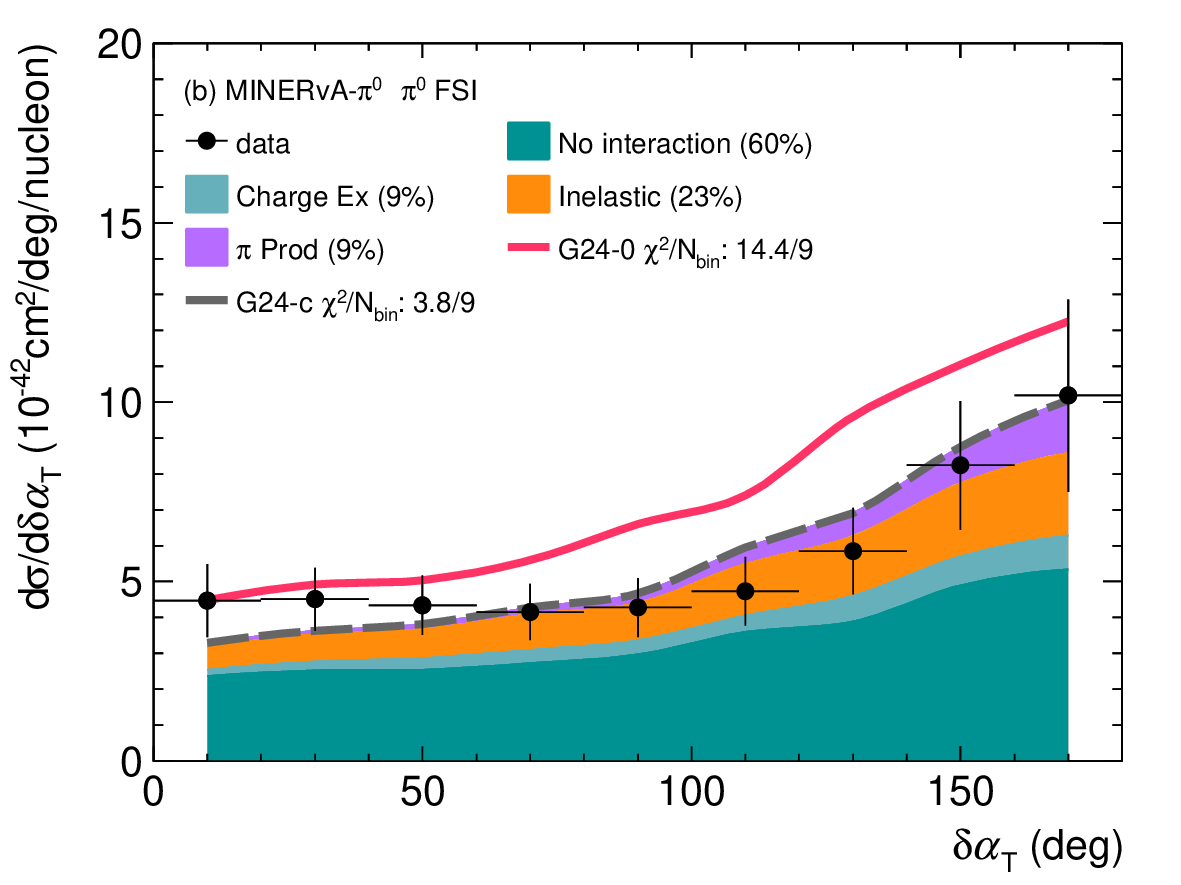}	
    \caption{\label{fig:CEX-minpiz-dat-pi0} \minpiz $\dat$ measurement compared to \genie predictions decomposed in $\piz$ FSI fates with (a) \gZero and (b) \gC.} 
\end{figure}

Examining individual datasets closely helps to understand the origin of the improvement. The decomposition of the cross section for $\dat$ according to the FSI fates of the $\piz$ is presented in Fig.~\ref{fig:CEX-minpiz-dat-pi0}. The hA model rescatters primary interaction products only once, and records this event with a rescattering code for these particles. The rescattered products, which are the final outputs, are stored as daughters of the primary interaction products. Hence, the FSI fate that produces the final products can be checked by the rescattering code of their first parent. More specifically, we loop through all final-state particles to look for the leading $\piz$ and check the rescattering code of its first parent. 

In \minpiz, the number of $\piz$s undergoing no FSI (``No Interaction'', as shown in the figures) is adjustable via the $\pizmfp$ parameter. A decrease in $\pizmfp$ reduces the size of the ``No Interaction'' events,  as is shown in Fig.~\ref{fig:CEX-minpiz-dat-pi0}b when compared to Fig.~\ref{fig:CEX-minpiz-dat-pi0}a. Meanwhile, an increase in $\piz$ rescattering only manifests in pionless measurements through ABS. There is indeed an increase of ABS for $\piz$ in \ttkzpi and \minzpi, but the fraction is small to begin with, so the impact is small. The increase of $\piz$ rescattering can increase \ttkpip via CEX as discussed below. However, due to the significant suppression of CEX, its impact on \ttkpip is also minimal (for detailed breakdown, see Figs.~\ref{fig:g24-0-dat-pi0}-\ref{fig:g24-c-pn-pi0} in Appendix~\ref{sec:appfate}). 

As shown in Fig.~\ref{fig:CEX-minpiz-dat-pi0}a, the \minpiz prediction from the nominal tune has considerable contributions from CEX controlled by $\picex$. It does not impact \ttkzpi and \minzpi as CEX only changes the pion type without removing them. Hence, events with pions in the final state would be rejected in these two data sets regardless of the pion charge. In principle, CEX will also migrate events between the signal and background definitions for \ttkpip when a $\piz$ is converted to a $\pip$ and vice verse. However, \ttkpip does not have a considerable CEX fraction to begin with, so changing CEX will have a minimal impact on its prediction. Hence, this is one of the most effective paths to decrease the \minpiz cross section prediction without affecting other measurements, and indeed the $\picex$ is heavily suppressed as shown in Table~\ref{tab:restunes} and illustrated in Fig.~\ref{fig:CEX-minpiz-dat-pi0}b compared to Fig.~\ref{fig:CEX-minpiz-dat-pi0}a. However, due to the intricate correlation between the FSI fates in the hA implementation discussed in Sec.~\ref{sec:tuning-para-choice}, although $\piinel$ and $\pipiprod$ are not modified explicitly, a considerable increase is observed in both INEL and PIPD for \minpiz. 

\begin{figure}[!htb] 	
    \centering 		
    \includegraphics[width=\fwid\textwidth]{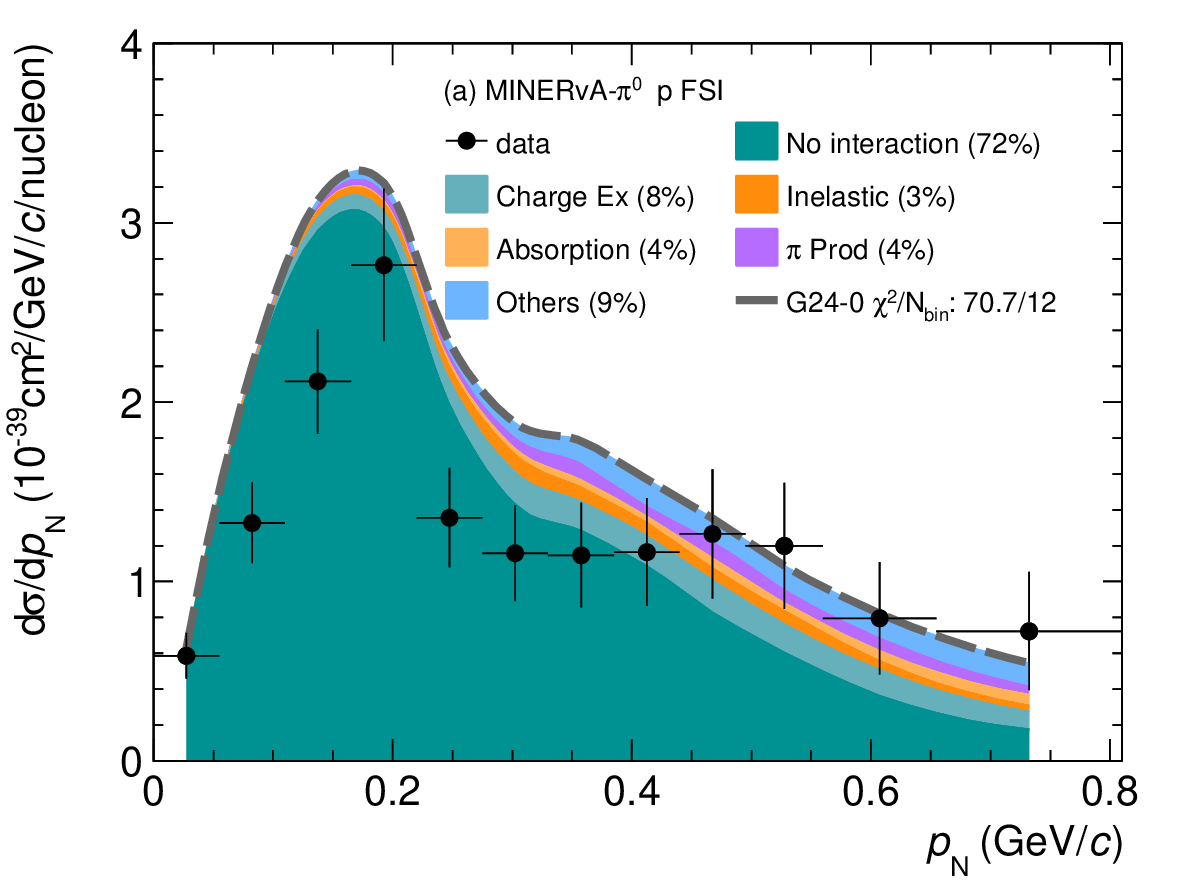}
    \includegraphics[width=\fwid\textwidth]{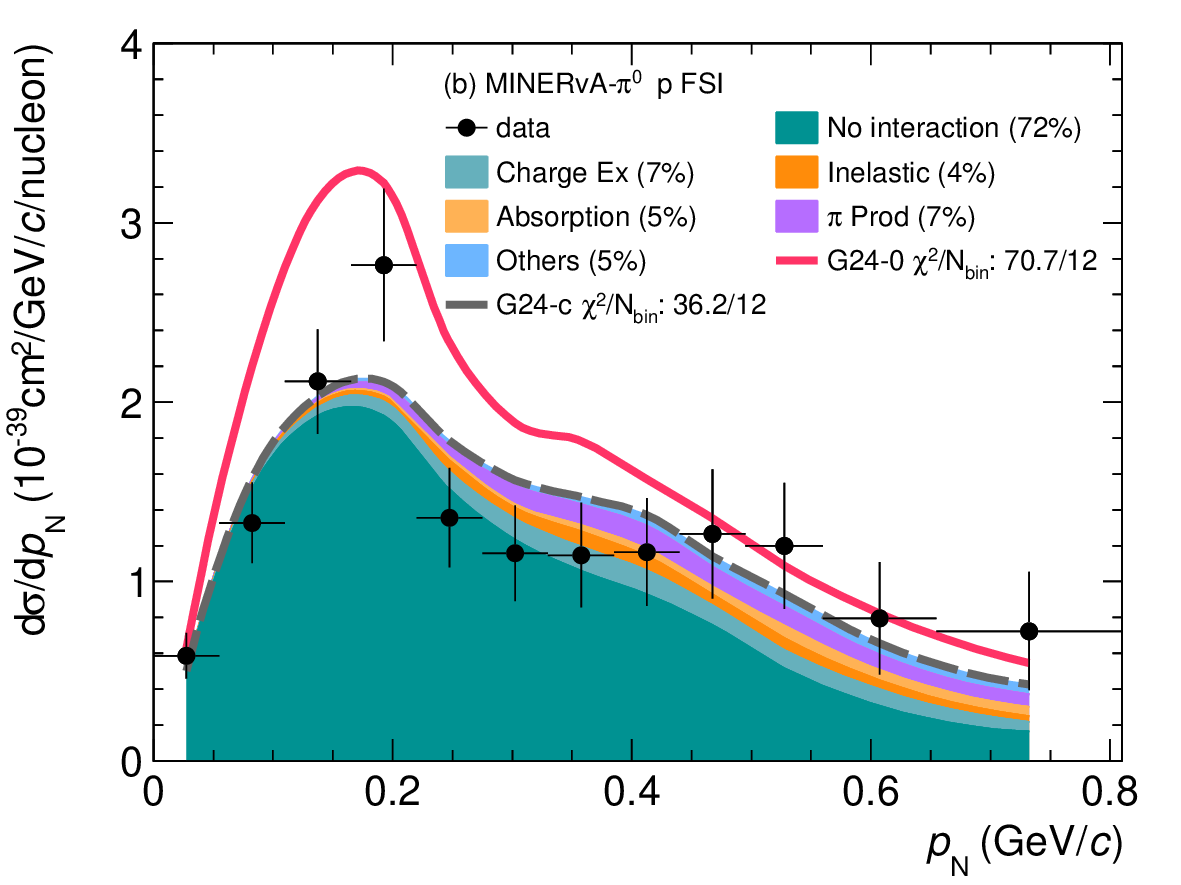}	
    \caption{\label{fig:minpiz-pn-pr} Similar to Fig.~\ref{fig:CEX-minpiz-dat-pi0} but for $\pn$ with proton FSI fate decomposition. Note that the ``Absorption'' of the proton is referring to the absorption of the $\pip$ from the decay of $\deltapp$, which could lead to emission of nucleons as discussed in Sec.~\ref{sec:tuning-para-choice}.     
    } 
\end{figure}

Suppressing both ``No Interaction'' and CEX for $\piz$ adjusts the \minpiz prediction to the appropriate magnitude with small impacts on other data sets.  The effects of the other parameter changes are more transparent in the comparison of the $\pn$ distribution of \minpiz between Figs.~\ref{fig:minpiz-pn-pr}a and~\ref{fig:minpiz-pn-pr}b. A relatively large increase in $\ncex$ and $\npiprod$ moves events away from the Fermi motion peak at $\pn\leq0.25~\textrm{GeV}/c$; the same effect can be achieved by an increased $\srcfr$. 

Another impact of the larger $\npiprod$ manifests in the appearance of 2p2h ($1\%$) and CCQE ($2 \%$) in the \minpiz cross section, as shown in Fig.~\ref{fig:g24-c-pn-reac}d. These neutrino-nucleon interactions do not produce pions, but the product nucleons can produce pions via re-scattering in the nucleus, leading to a contribution to topologies with pions in the final states. However, such contribution remains small.

Although the interaction models, e.g. 2p2h, are not directly tuned, the respective fractions have undergone small changes in Fig.~\ref{fig:g24-c-dat-reac} and Fig.~\ref{fig:g24-c-pn-reac}, compared to Fig.~\ref{fig:g24-0-dat-reac} and Fig.~\ref{fig:g24-0-pn-reac}. 
This is a result of the increased $\srcfr$ which leads to more events with higher initial nucleon momenta such that interactions requiring higher initial state energy, such as RES and DIS, will be more frequent. 
This is reflected by the increase in the ratios of these interactions: For example, RES increases from $19\%$ (Fig.~\ref{fig:g24-0-pn-reac}c) to $25\%$ (Fig.~\ref{fig:g24-c-pn-reac}c) for \minzpi. 
Hence, the other interactions, although not directly tuned, will experience a decrease in the fraction of total cross section: For example, 2p2h drops from $34\%$ (Fig.~\ref{fig:g24-0-pn-reac}c) to $30\%$ (Fig.~\ref{fig:g24-c-pn-reac}c) for \minzpi.

\subsection{Alternative tune: \gT}

\begin{figure}[!htb] 	
    \centering 		
    \includegraphics[width=\fwid\textwidth]{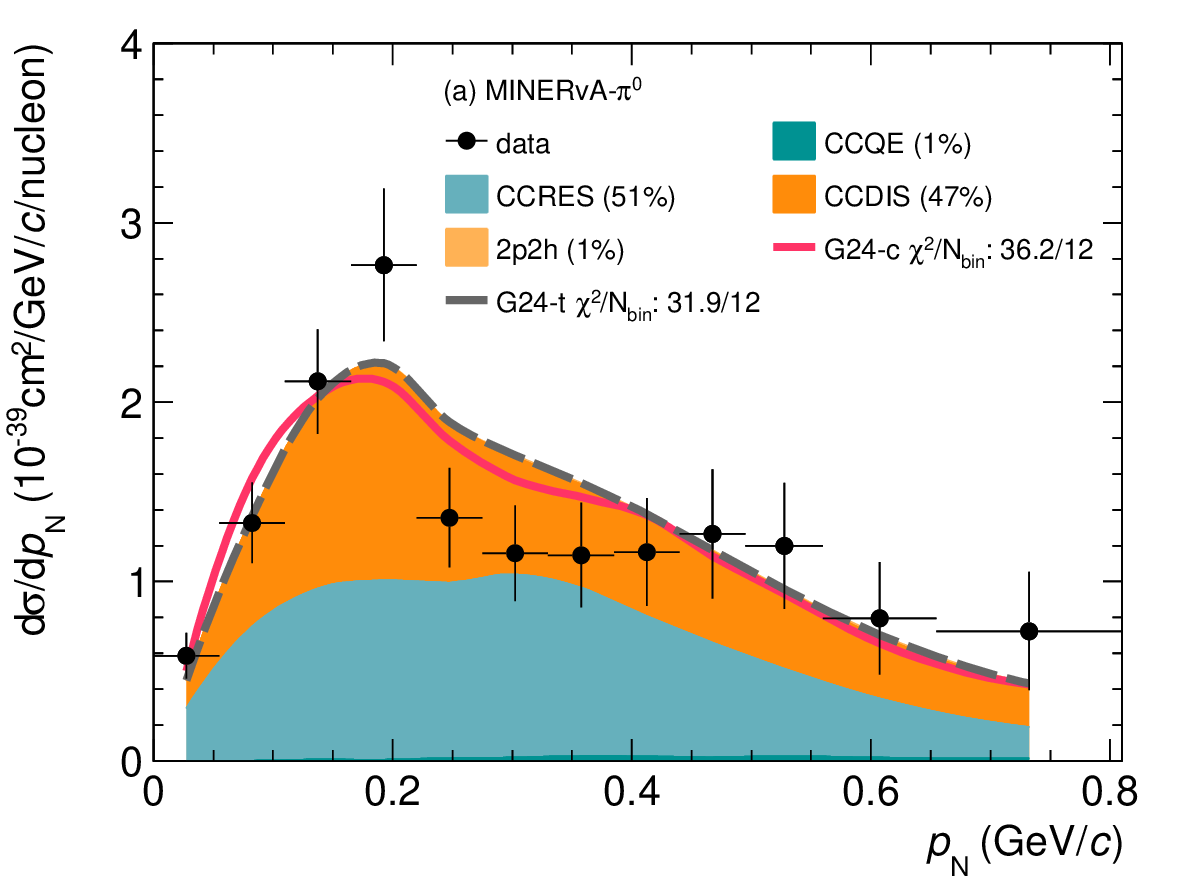}
    \includegraphics[width=\fwid\textwidth]{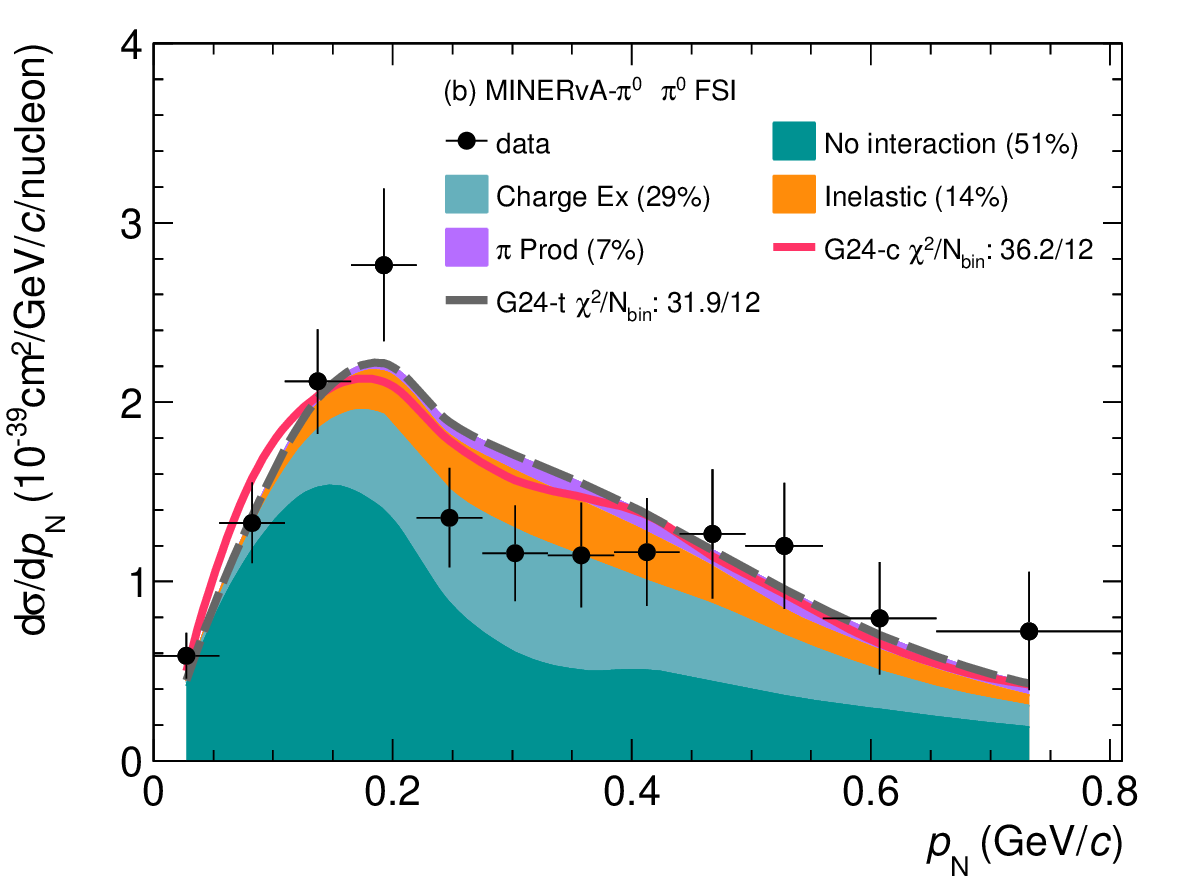}
    \includegraphics[width=\fwid\textwidth]{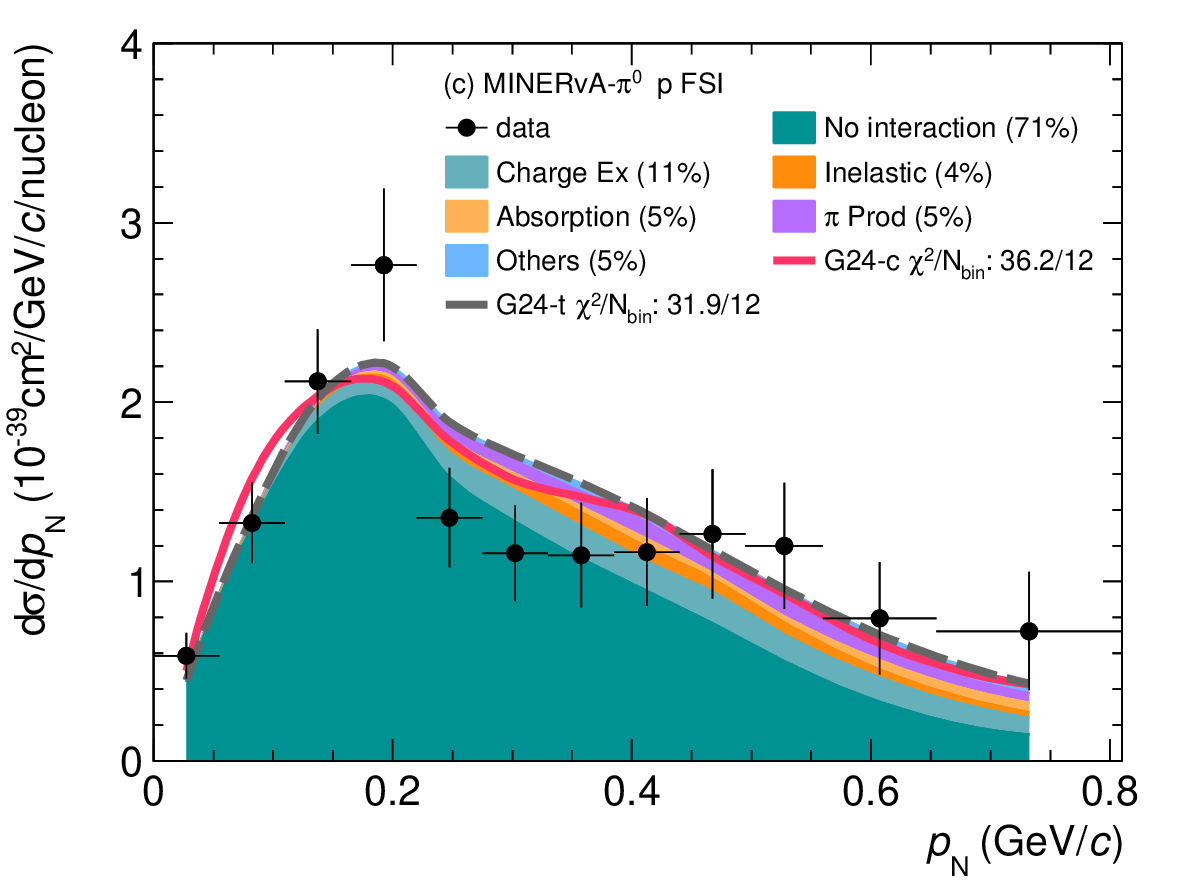}
    \caption{\label{fig:minpiz-alttune} \minpiz $\pn$ measurement compared to \genie predictions decomposed in  (a) $\nu$-N interaction, (b) $\piz$ FSI, and (c) proton FSI for the alternative tune, \gT.} 
\end{figure}

A similar improvement is achieved by the intermediate tune, \gT, illustrated in Fig.~\ref{fig:minpiz-alttune}. Instead of significantly increasing proton PIPD to elevate the $\pn$ tail, this tune notably enhances $\srcfr$---more-energetic nucleons are present and therefore the RES $\pn$ peak is shifted to the right in Fig.~\ref{fig:minpiz-alttune}a with respect to Fig.~\ref{fig:g24-c-pn-reac}d. Rather than suppressing $\picex$ (cf. Fig.~\ref{fig:g24-c-pn-pi0}d in Appendix~\ref{sec:appfate}), this tune significantly increases it (Fig.~\ref{fig:minpiz-alttune}b) to offset the further reduction of ``No Interaction'' $\piz$ events (smaller $\pizmfp$ in \allpar, cf. Table~\ref{tab:restunes}).  Overall, this leads to a reduction of $75.31$ in $\chi^2$ (Table~\ref{tab:restunes}), similar to \gC, as well as a good data-MC agreement across all data sets.

\subsection{Discussion}
\begin{figure}[!htb] 	
    \centering 		
    \includegraphics[width=\fwid\textwidth]{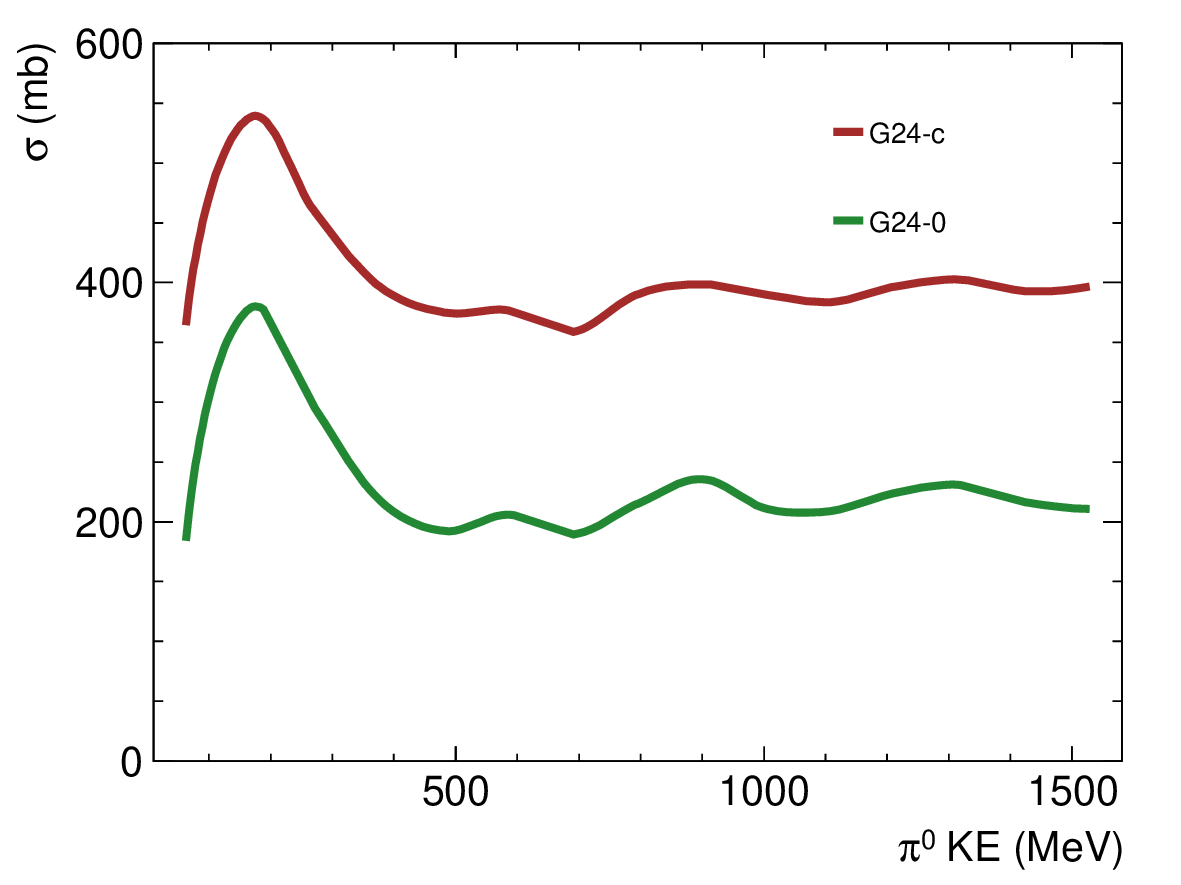}
    \caption{\label{fig:pizmfp_change} Change in MC prediction for $\piz$ cross section between \gZero and \gC . } 
\end{figure}

For both \gC and \gT the fit results suggest extreme values, but the effect of the parameter change is less effective than the change in the parameter suggests. 
For the individual FSI fate cross section, due to the re-normalization step after scaling, the effective change for one particular fate is smaller than the scaling parameter magnitude as explained in Sec.~\ref{sec:tuning-para-choice}. 
As for the total FSI cross sections, only that of $\piz$ is changed. The total $\piz$-C scattering cross section for pion kinetic energy is plotted in Fig.~\ref{fig:pizmfp_change} to assess the overall impact. 
The shape remains similar and consistent with Fig. 2.23 ($\pip$-C reactions) in Ref.~\cite{Andreopoulos:2015wxa} while the peak magnitude increases from $380$ to $538$ mb, a $48\%$ increase, considerably less than the scaling factor, $\pizmfp=0.22$, seemingly suggests. 
Moreover, the default $\piz$ parameter values are calculated from charged pion data assuming isospin symmetry rather than extracted directly from experimental data. 
Hence, this modification does not violate any existing agreement with hadron scattering data.
In addition, agreement of this tune with most of the non-TKI neutrino datasets available is unchanged before and after the tune,  thereby demonstrating the physicality of the tune.

Exclusive electron scattering data is better suited to constrain FSI models.
Regrettably, most electron scattering data to date is inclusive~\cite{electronsforneutrinos:2020tbf} and it is not well suited for the tuning of FSI. 
The ``Electrons for Neutrinos'' (e4nu) collaboration exploits data from the CLAS6 spectrometer to measure exclusive final states, such as a $1\textrm{p}0\pi$ cross-section measurement on carbon as a function of TKI variables~\cite{CLAS:2021neh}. But for electron-scattering pion production, 
currently, the \genie predictions over-predict data due to the lack of tuning of the \genie models on free nucleon data.
Such big uncertainties in the pion production model mask possible nuclear model or FSI dependence.
In order to be sensitive to these effects, one need to first tune electron-scattering to free nucleon data. The current tuning study could be extended in the future when data from dedicated electro-pion production measurements are available. 

In conclusion, this tune is a valid and effective model that can be used as a starting point for an analysis. 
Further refinement will be available once more data will be included in the fit, including non-TKI observables and electron-scattering data.

\section{\label{sec:summary} Summary and outlook}
This work represents the first global tuning effort on TKI data. Our partial tune of the \sfcfg and hA models, $\restunefull$ (\gC), provides an effective theory to better describe both the neutrino-hydrocarbon pionless and pion production.
The largest change in the model is demanded by the \minpiz TKI measurement~\cite{MINERvA:2020anu} that was significantly overestimated in \genie. 
The improvement is crucial for future precision GeV neutrino experiments.  This tuning configuration has been integrated into the \texttt{master} branch of \genie and is slated for inclusion in the upcoming release.  

To develop an effective model, we focused on the most sensitive parameters, $\srcfr$, $\pizmfp$, $\picex$, $\ncex$, $\nabs$, and $\npiprod$, of the \sfcfg and hA models, reducing the total number of parameters from $14$ to $6$. The corresponding best combination of observables is $\dat$, $\pn$ (or $\dpt$ if no $\pn$ is available), and $\dptt$, from pionless and pion-production measurements in both T2K and MINERvA. 
In doing so, the pion production model is held fixed such that the tuned values of $\pizmfp$, $\picex$, $\nabs$, and $\npiprod$ depart from the established constraints from hadron measurements. 
A different pion production model might compensate part of these model tuning effects~\cite{Yan:2024kkg}. 
We also chose the hA FSI model for practical reasons. Its sequential treatment of MFP and rescattering should be considered as a numerical advantage (as was also pointed out in Ref.~\cite{GENIE:2022qrc}). Other more sophisticated models could be considered in future tunes.  With the existing four TKI measurements from T2K and MINERvA, we have also derived other effective tunes like $\alttune$ (\gT). The degeneracy can be resolved with additional data, particularly from argon-based measurements~\cite{MicroBooNE:2022emb, MicroBooNE:2023cmw, MicroBooNE:2023tzj, MicroBooNE:2023wzy, MicroBooNE:2024tmp, MicroBooNE:2015bmn} and e4nu data~\cite{CLAS:2021neh}.

\section*{Acknowledgment}

We thank the CC-IN2P3 Computing Center providing computing resources and for their support. 
X.L. is supported by the STFC (UK) Grant No. ST/S003533/2. 
J.T.-V. acknowledges the support of the Raymond and Beverly Sackler scholarship.

\bibliographystyle{apsrev4-1}
\bibliography{main}

\appendix


\section*{appendix}
\subsection{Combinations of Observables}\label{sec:appcombi}

Table~\ref{tab:fit-var-combo} displays the tested combinations of observables:
\begin{itemize}
    \item \textbf{\texttt{Combi-}1 through 5} are cross-experiment selections of a single observable. 
For example, \texttt{Combi-}$3$ uses only $\dat$ from all four data sets. If a chosen observable is absent from a dataset, that dataset is excluded from the specific combination. 
For example, \ttkzpi is not used in \texttt{Combi-}1 due to the absence of $\pn$.
\item \textbf{\texttt{Combi-}6 through 9} incorporate all variables from a single measurement. For example, \texttt{Combi-}$9$ uses \minpiz only.
\item \textbf{\texttt{Combi-}10 to 13} uses two out of four measurements according to the experiment or topology. For example, \texttt{Combi-}$10$ uses only the two data sets from T2K, and \texttt{Combi-}$13$  only with pion production. 
\item \textbf{\texttt{Combi-}14 to 17} combine two separate combinations:
\begin{align*}
\texttt{Combi-14} &= \texttt{Combi-3} \cup \texttt{Combi-5},\\
\texttt{Combi-15} &= \texttt{Combi-1} \cup \texttt{Combi-2},\\
\texttt{Combi-16} &= \texttt{Combi-1} \cup \texttt{Combi-3},\\
\texttt{Combi-17} &= \texttt{Combi-2} \cup \texttt{Combi-3}.
\end{align*}
\item \textbf{\texttt{Combi-}18 through 22} result from merging three combinations:
\begin{align*}
\texttt{Combi-18} &= \texttt{Combi-1} \cup \texttt{Combi-3} \cup \texttt{Combi-5},\\
\texttt{Combi-19} &= \texttt{Combi-2} \cup \texttt{Combi-3} \cup \texttt{Combi-5},\\
\texttt{Combi-20} &= \texttt{Combi-3} \cup \texttt{Combi-4} \cup \texttt{Combi-5},\\
\texttt{Combi-21} &= \texttt{Combi-1} \cup \texttt{Combi-2} \cup \texttt{Combi-3},\\
\texttt{Combi-22} &= \texttt{Combi-1} \cup \texttt{Combi-2} \cup \texttt{Combi-4}.
\end{align*}
\item \textbf{\texttt{Combi-}23} merges four different combinations:
\begin{align*}
\texttt{Combi-23} &= \texttt{Combi-1} \cup \texttt{Combi-2} \cup \\
& \texttt{Combi-3} \cup \texttt{Combi-5}.
\end{align*}
\item \textbf{\texttt{Combi-}24} encompasses all variables, acting as the superset.
\item \textbf{\texttt{Combi-}25 and 26} are the same as \texttt{Combi-}$21$ and $23$, respectively, except that $\dpt$ in \minzpi is removed to avoid correlation with $\pn$ in the same data set.
\end{itemize}

Figure~\ref{fig:allchi} shows the change in $\chi^2$ for the complete observable set (tuned plus validation) as a function of the tuning combination, where the two model parameter sets, \allpar and \redpar, are compared and it can be seen that the respective minima happen at \texttt{Combi-15} and 26.

\subsection{$\piz$ FSI fate decomposition for \gZero and \gC}\label{sec:appfate}

\begin{figure*}[!htb] 
\centering 		
    \includegraphics[width=\fwid\textwidth]{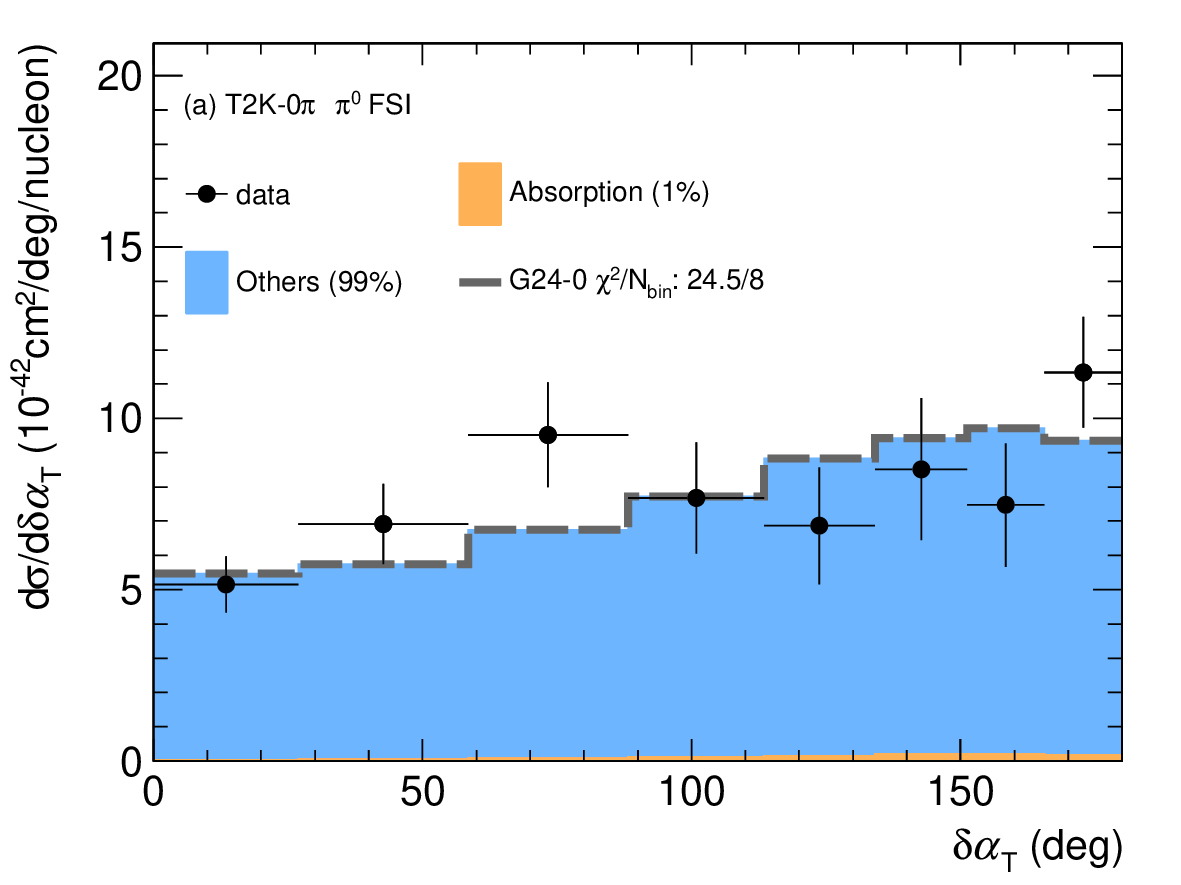} 
    \includegraphics[width=\fwid\textwidth]{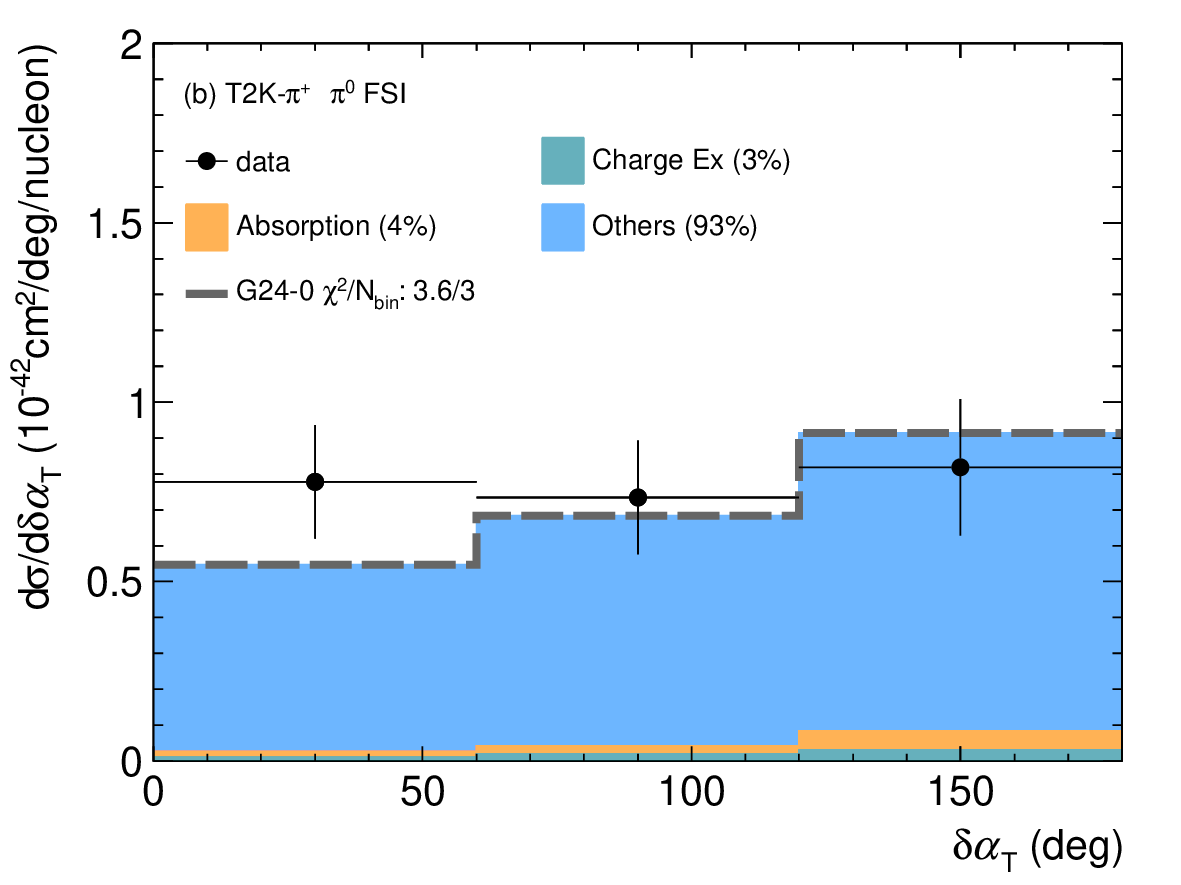} 
    \includegraphics[width=\fwid\textwidth]{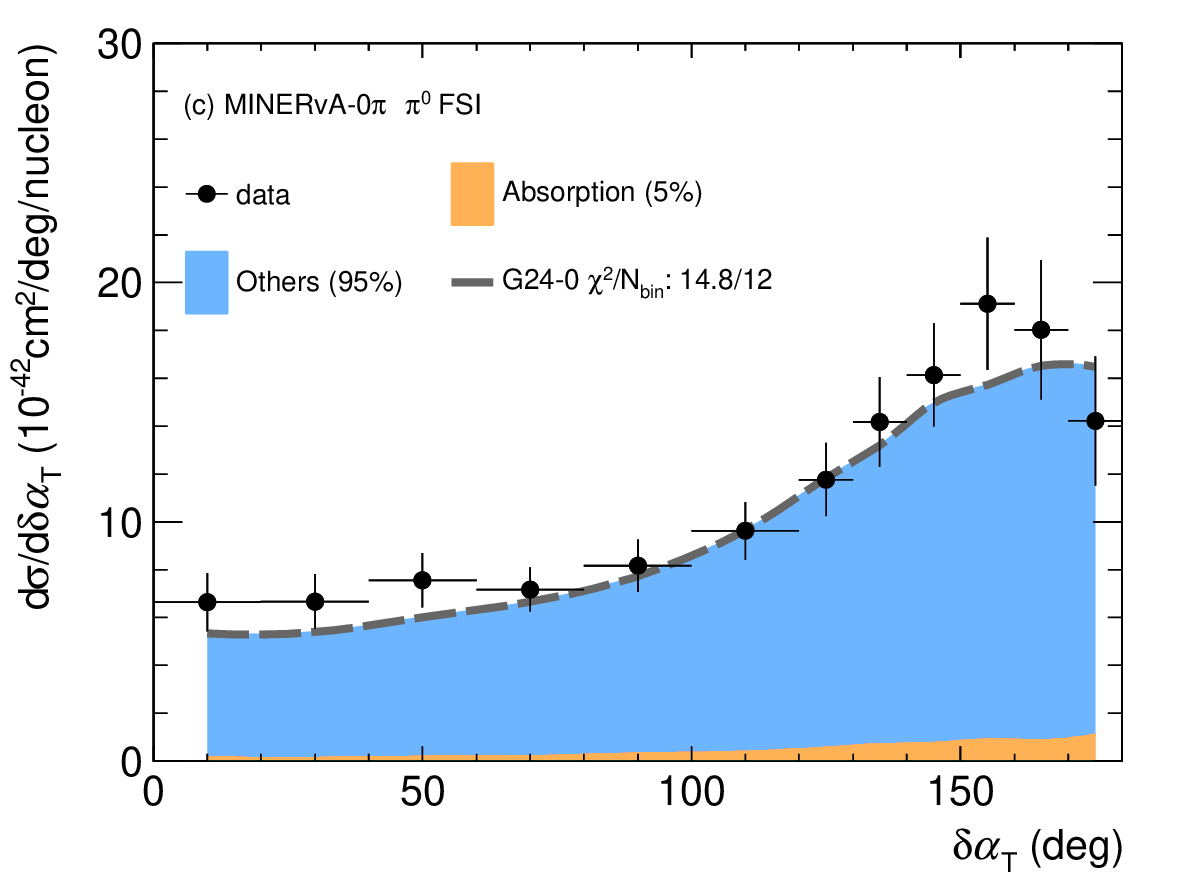} 
    \includegraphics[width=\fwid\textwidth]{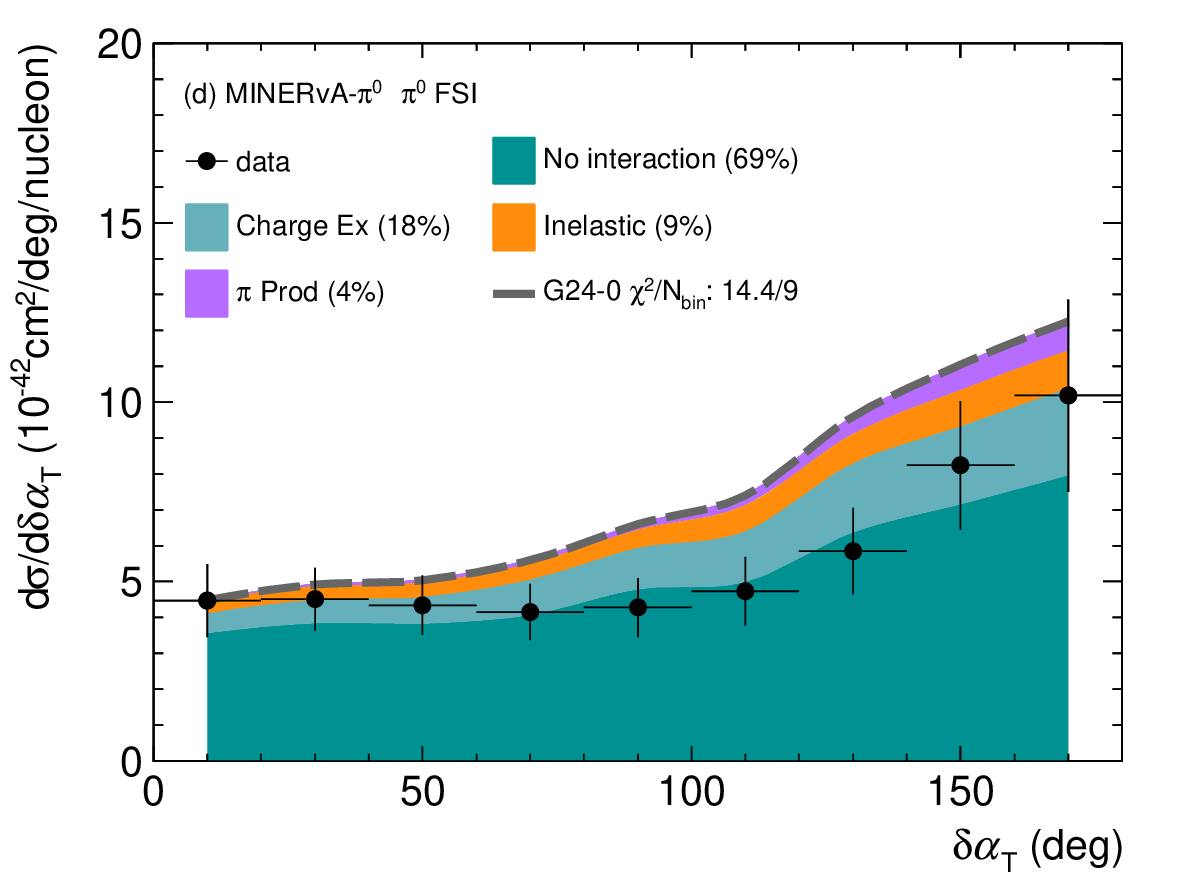}
    \caption{\label{fig:g24-0-dat-pi0} Extension of figure~\ref{fig:CEX-minpiz-dat-pi0}a to all four $\dat$ measurements. }   		
    \includegraphics[width=\fwid\textwidth]{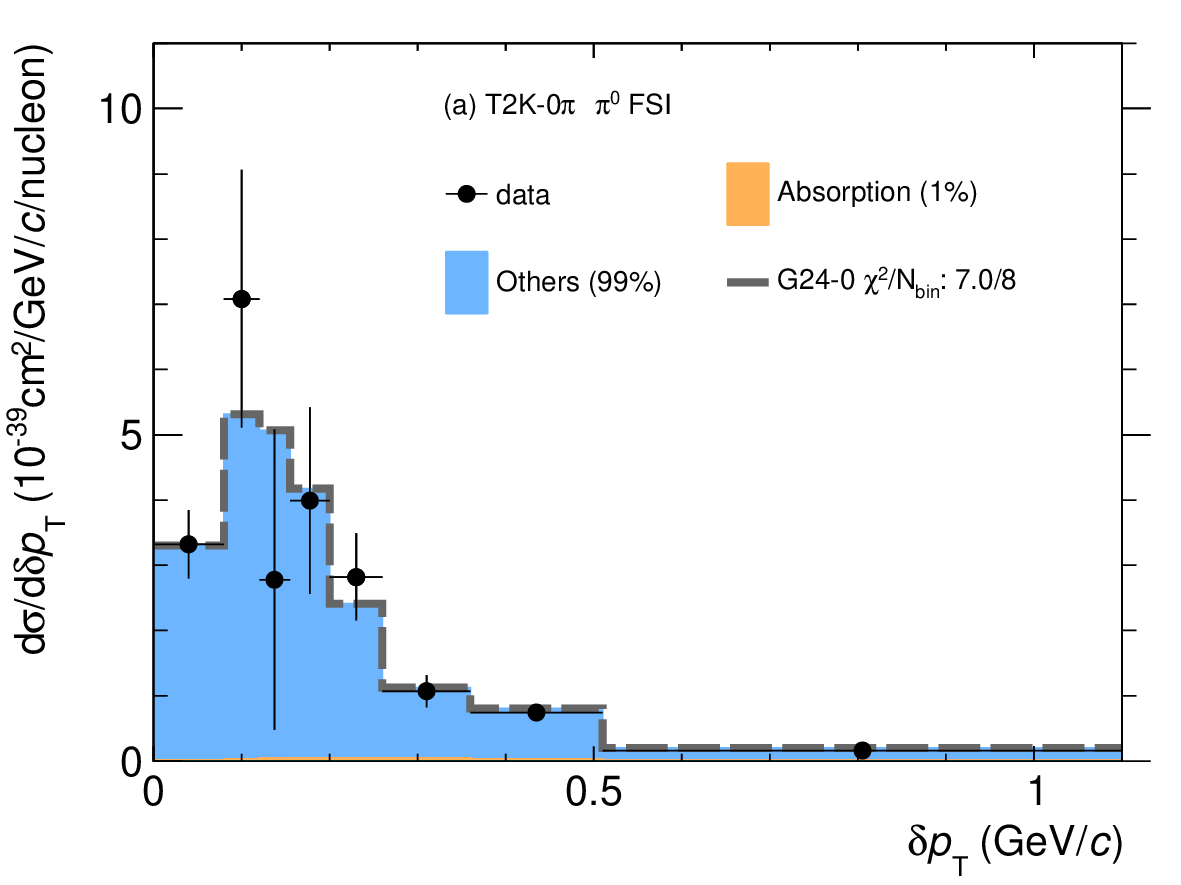}
    \includegraphics[width=\fwid\textwidth]{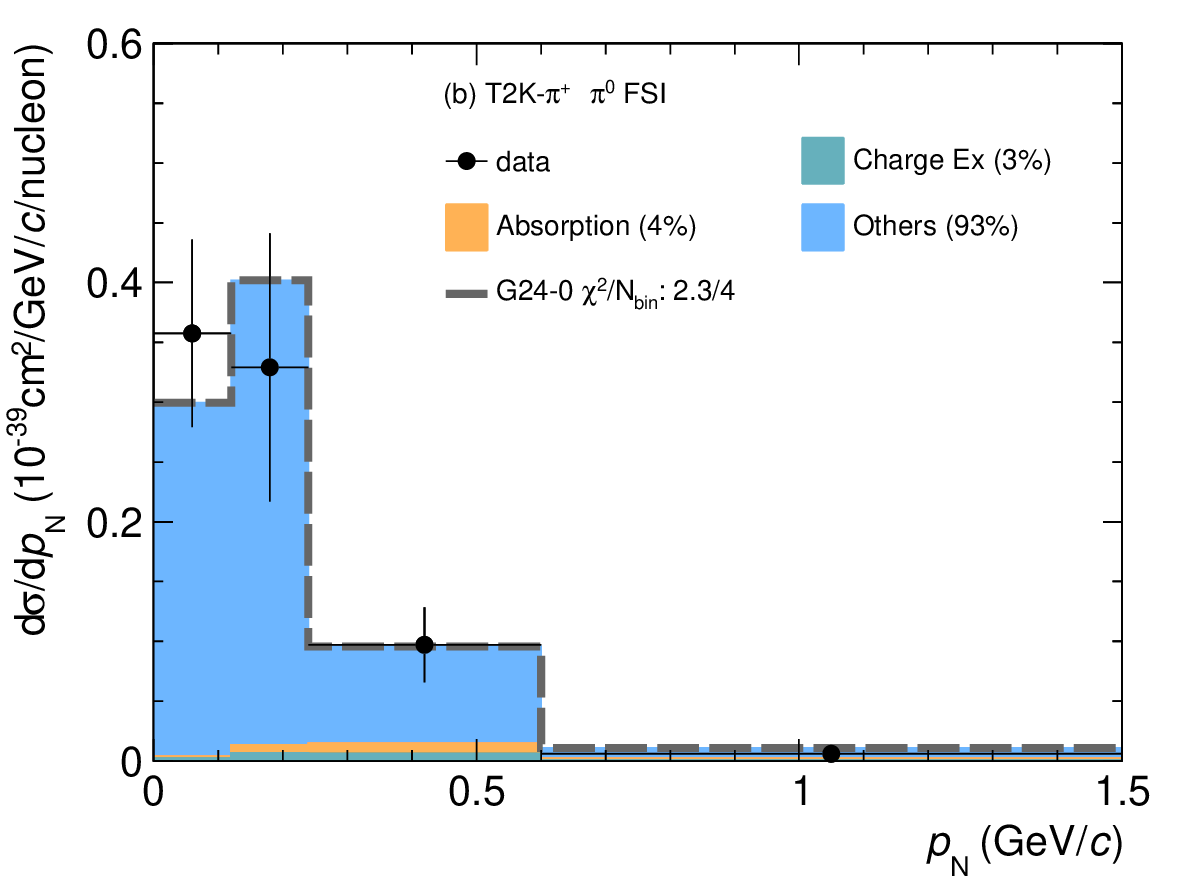}
    \includegraphics[width=\fwid\textwidth]{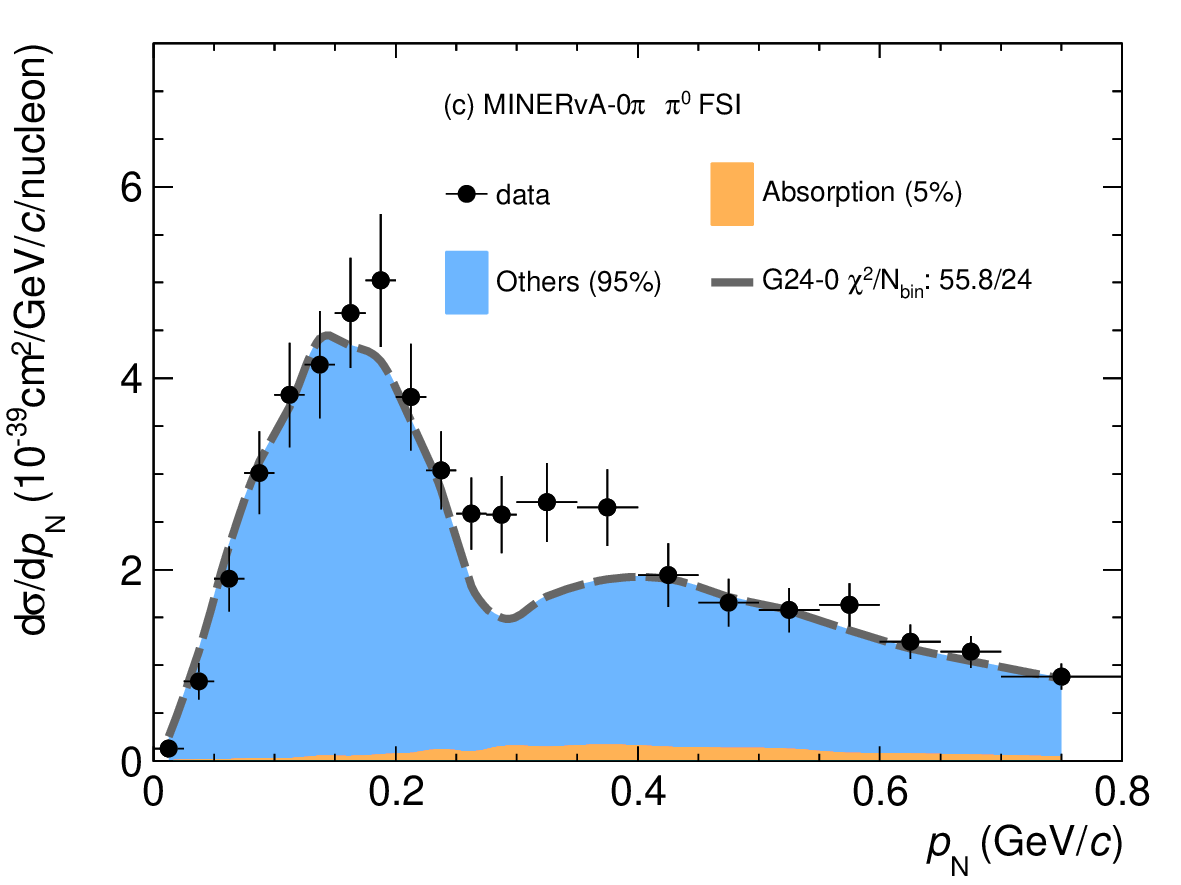}
    \includegraphics[width=\fwid\textwidth]{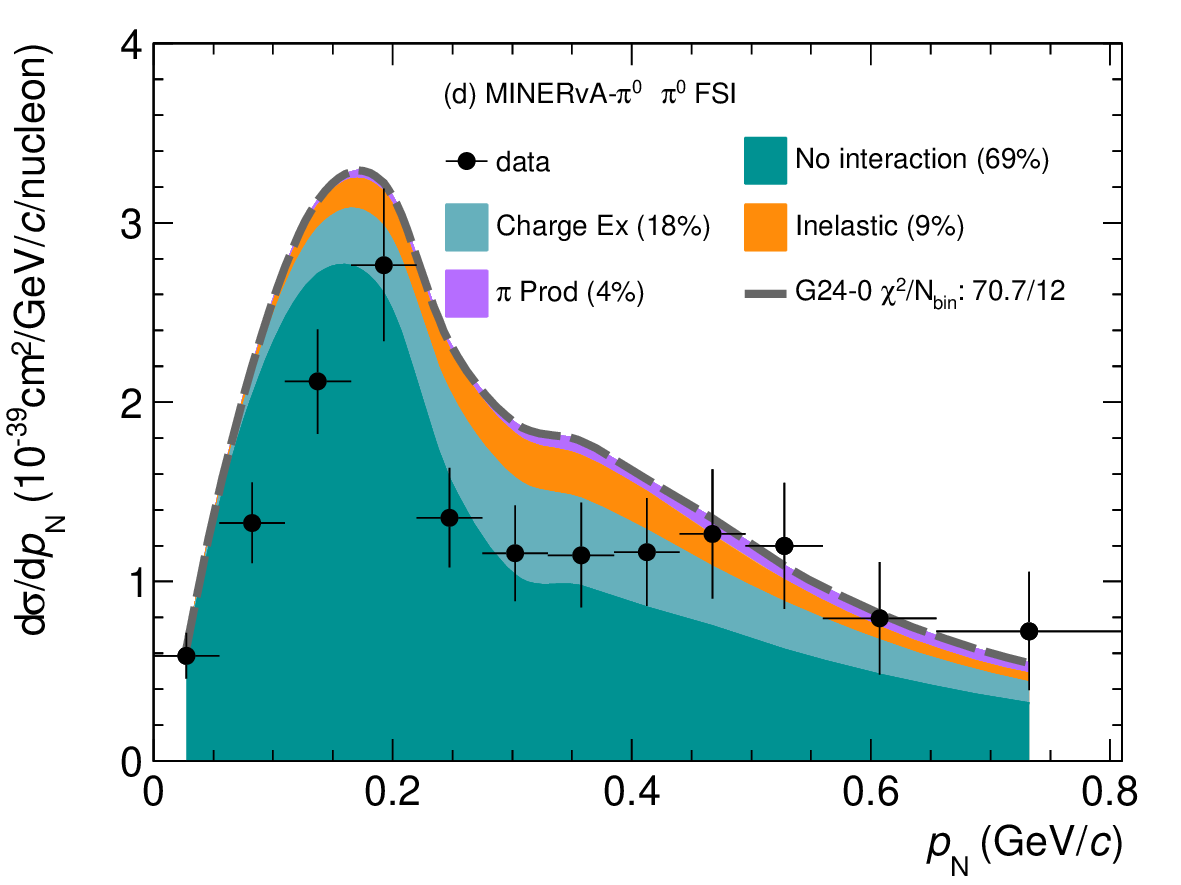}
    \caption{\label{fig:g24-0-pn-pi0}  Extension of figure~\ref{fig:CEX-minpiz-dat-pi0}a but to all four $\pn$ measurements. } 
\end{figure*}

\begin{figure*}[!htb] 
    \centering 		
    \includegraphics[width=\fwid\textwidth]{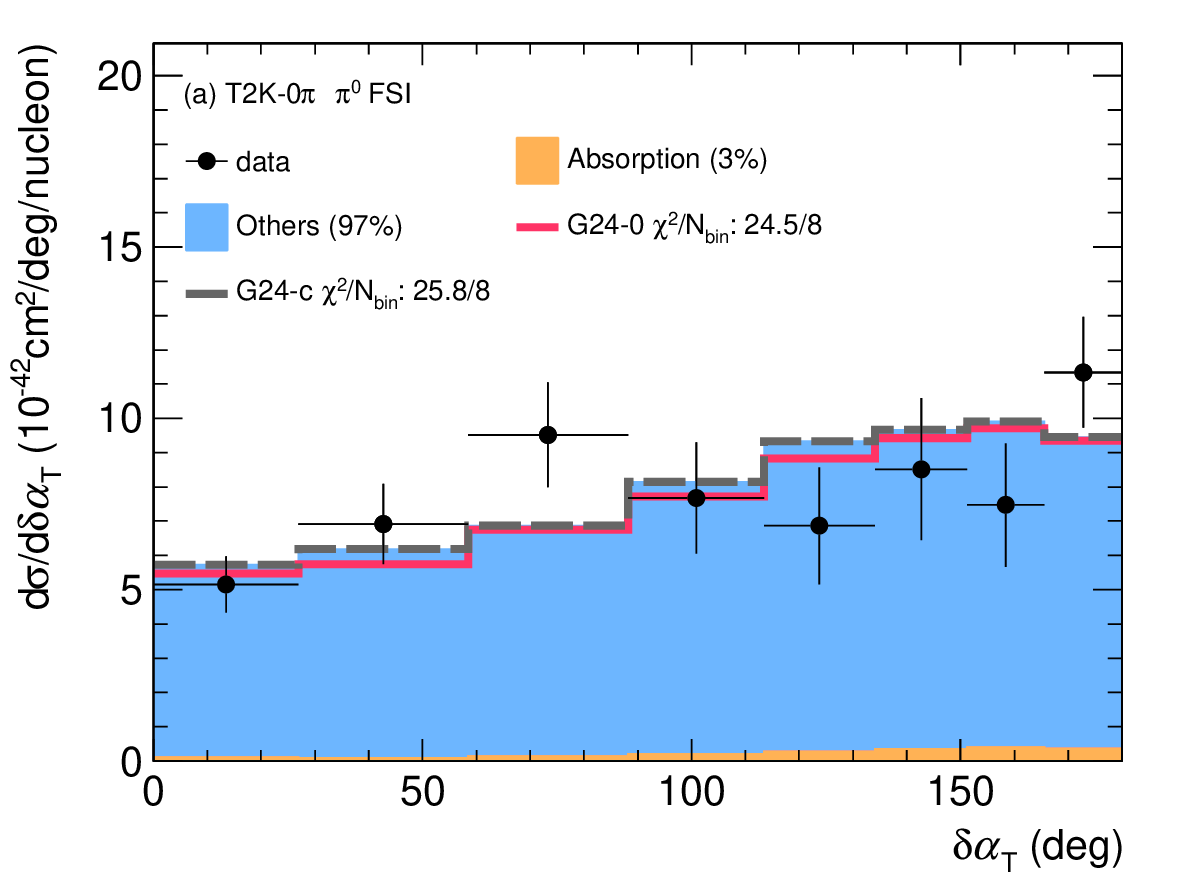}
    \includegraphics[width=\fwid\textwidth]{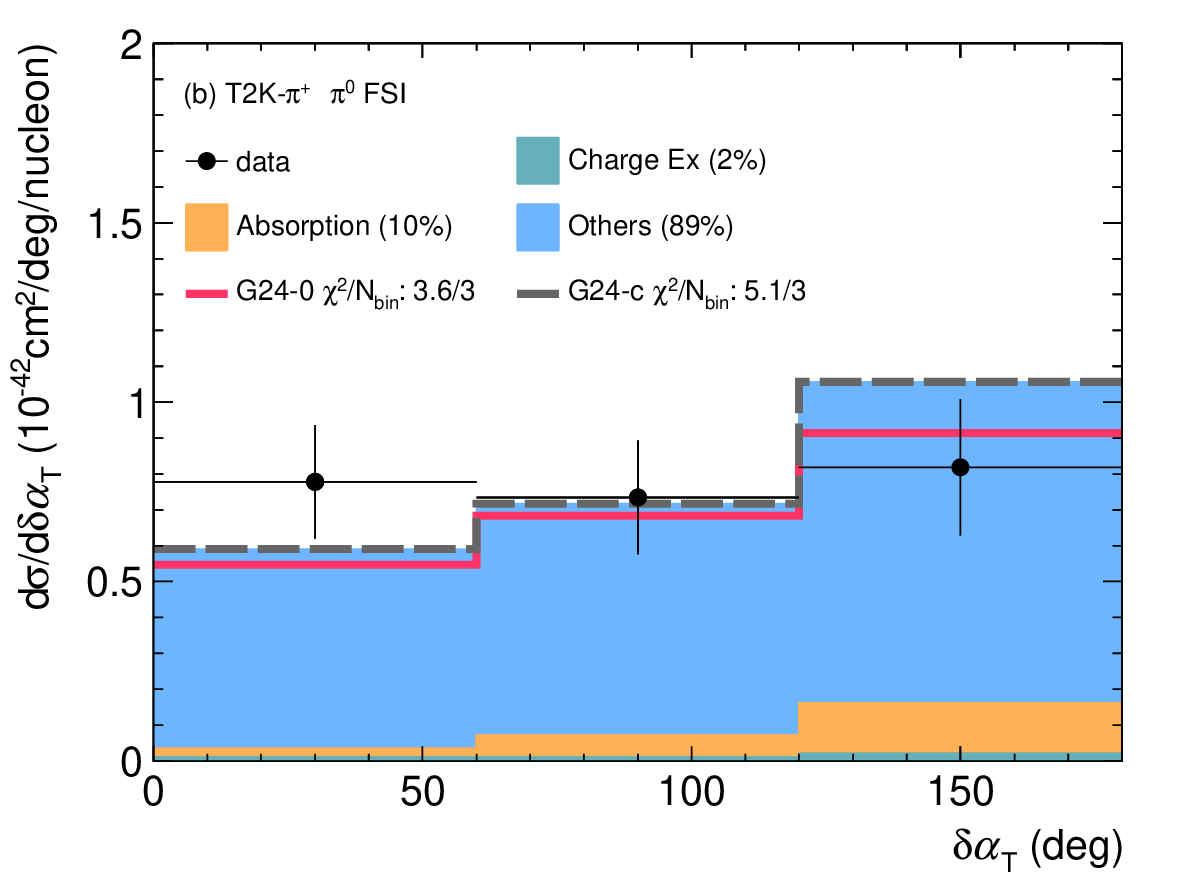}
    \includegraphics[width=\fwid\textwidth]{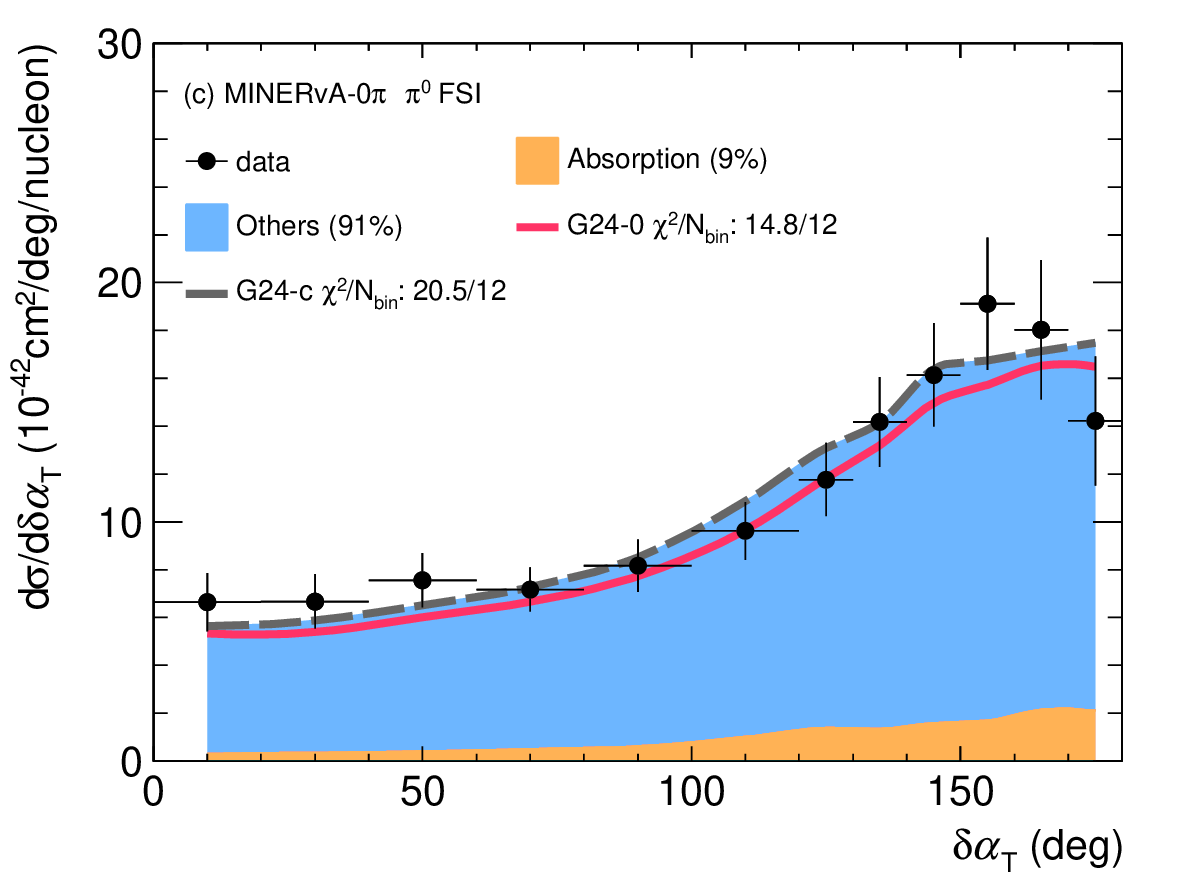}
    \includegraphics[width=\fwid\textwidth]{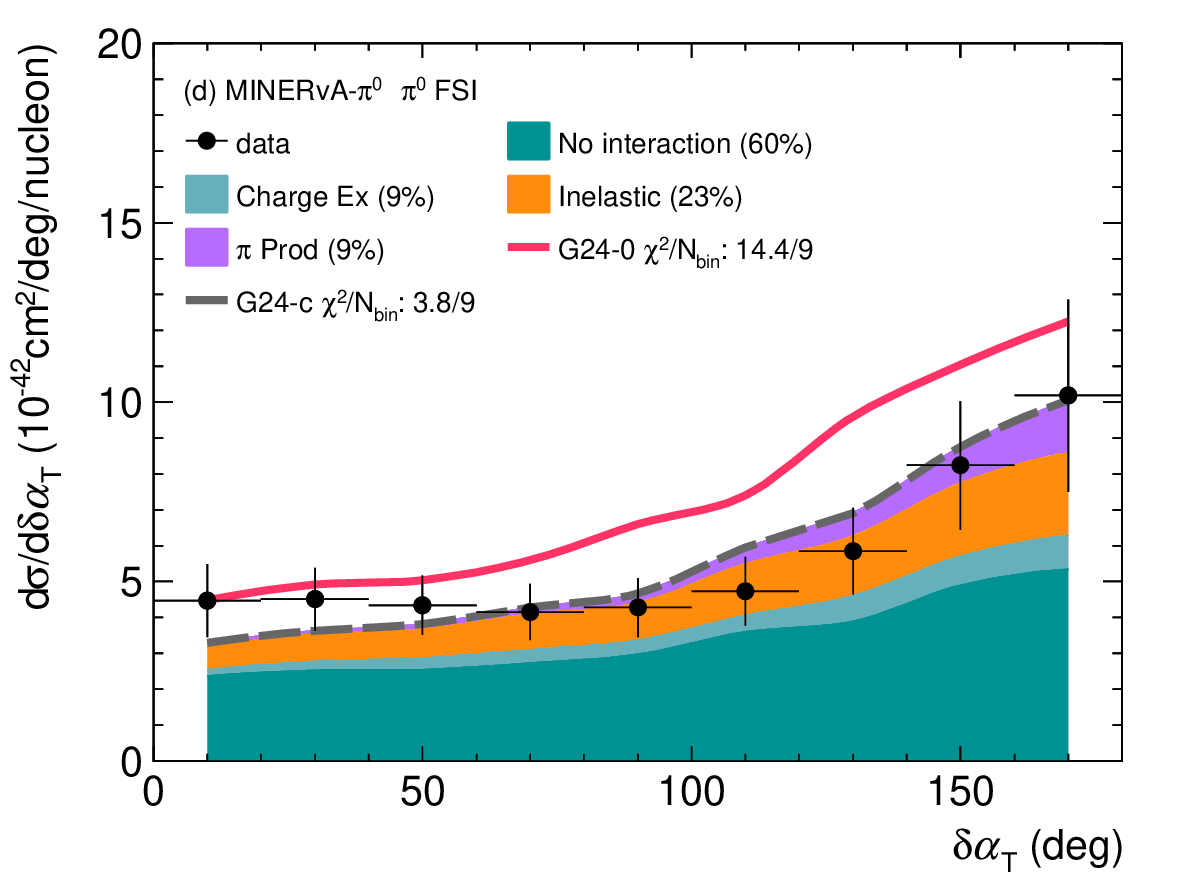}
    \caption{\label{fig:g24-c-dat-pi0} Extension of figure~\ref{fig:CEX-minpiz-dat-pi0}b to all four $\dat$ measurements. } 
		
    \includegraphics[width=\fwid\textwidth]{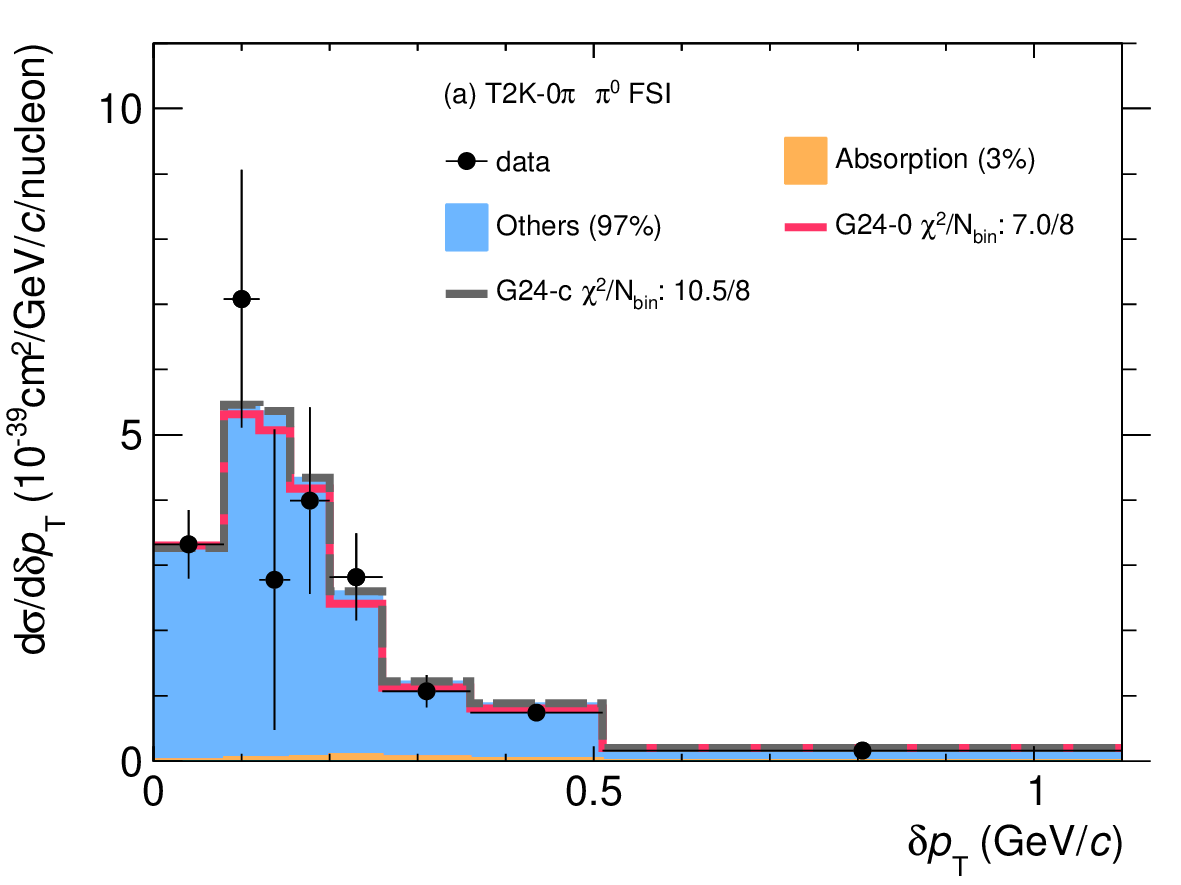}
    \includegraphics[width=\fwid\textwidth]{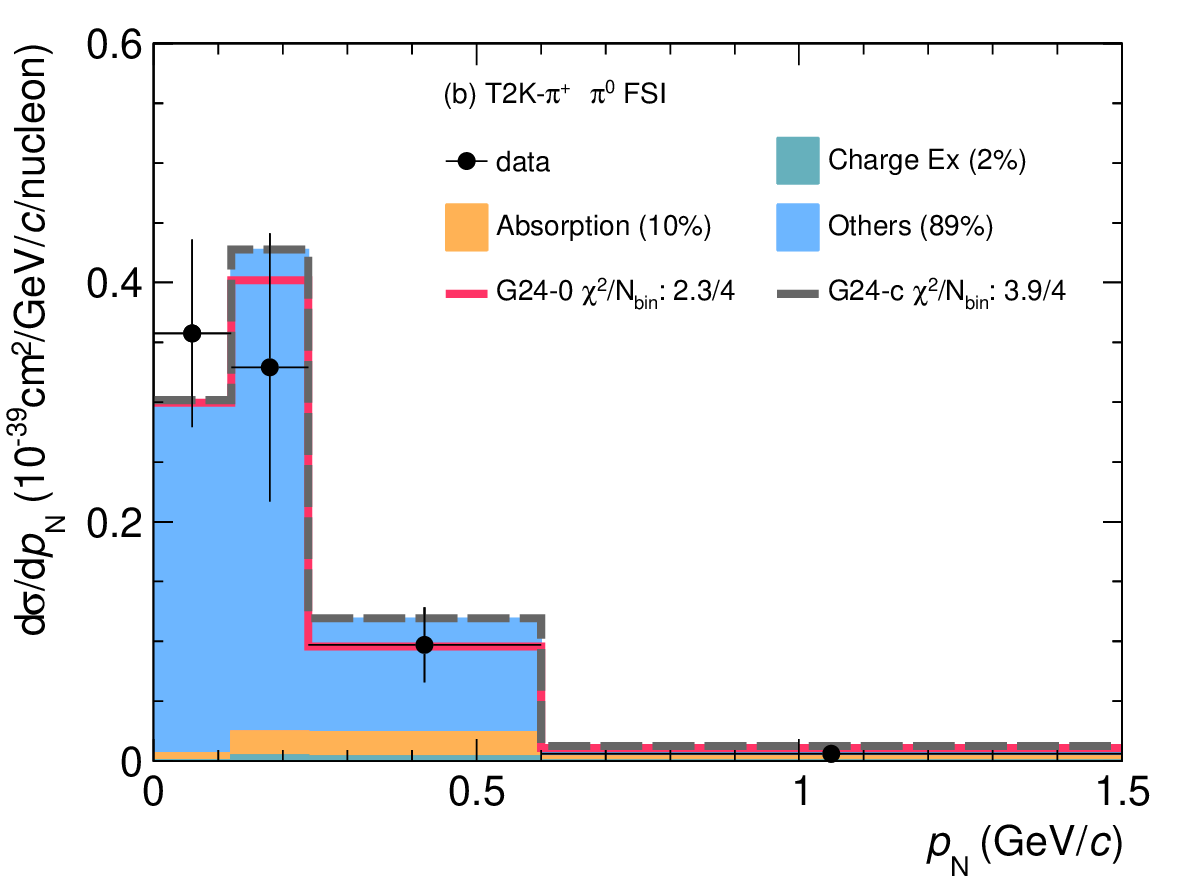}	
    \includegraphics[width=\fwid\textwidth]{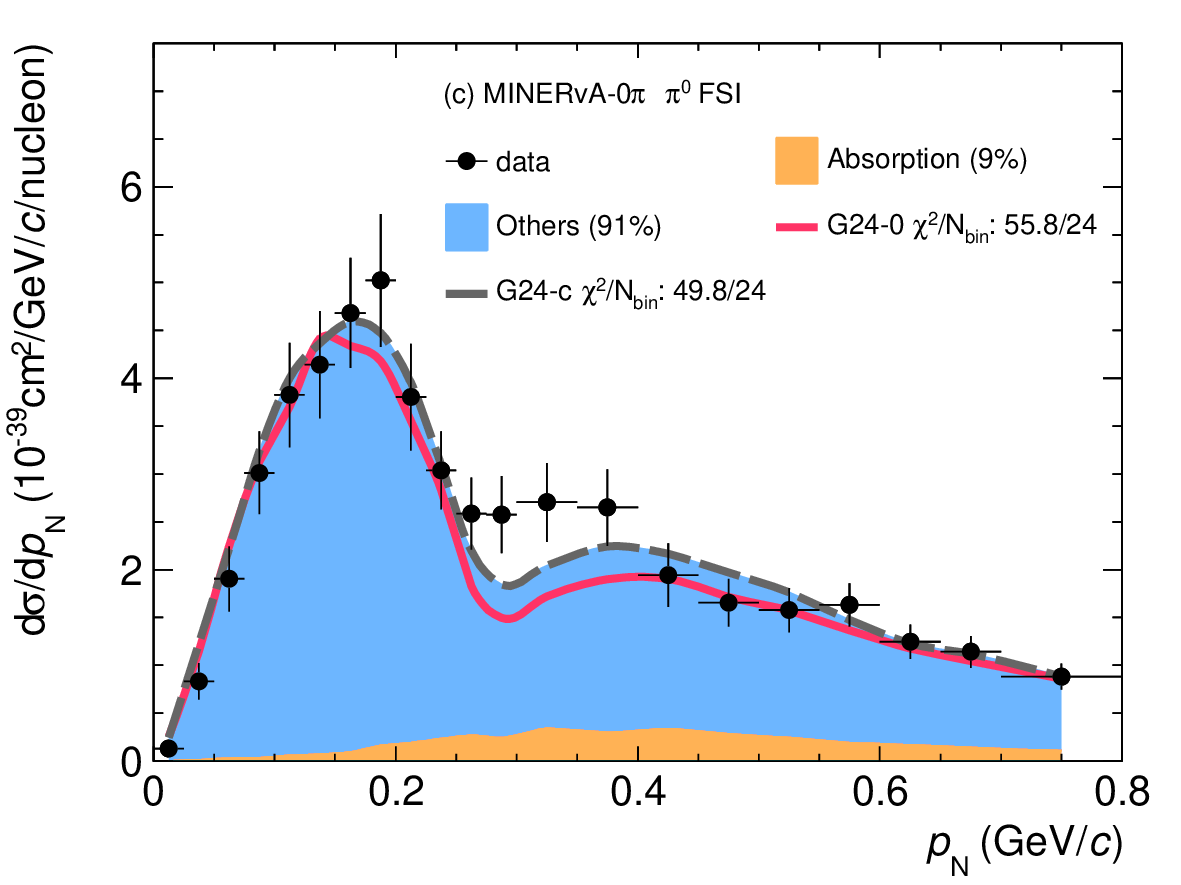}
    \includegraphics[width=\fwid\textwidth]{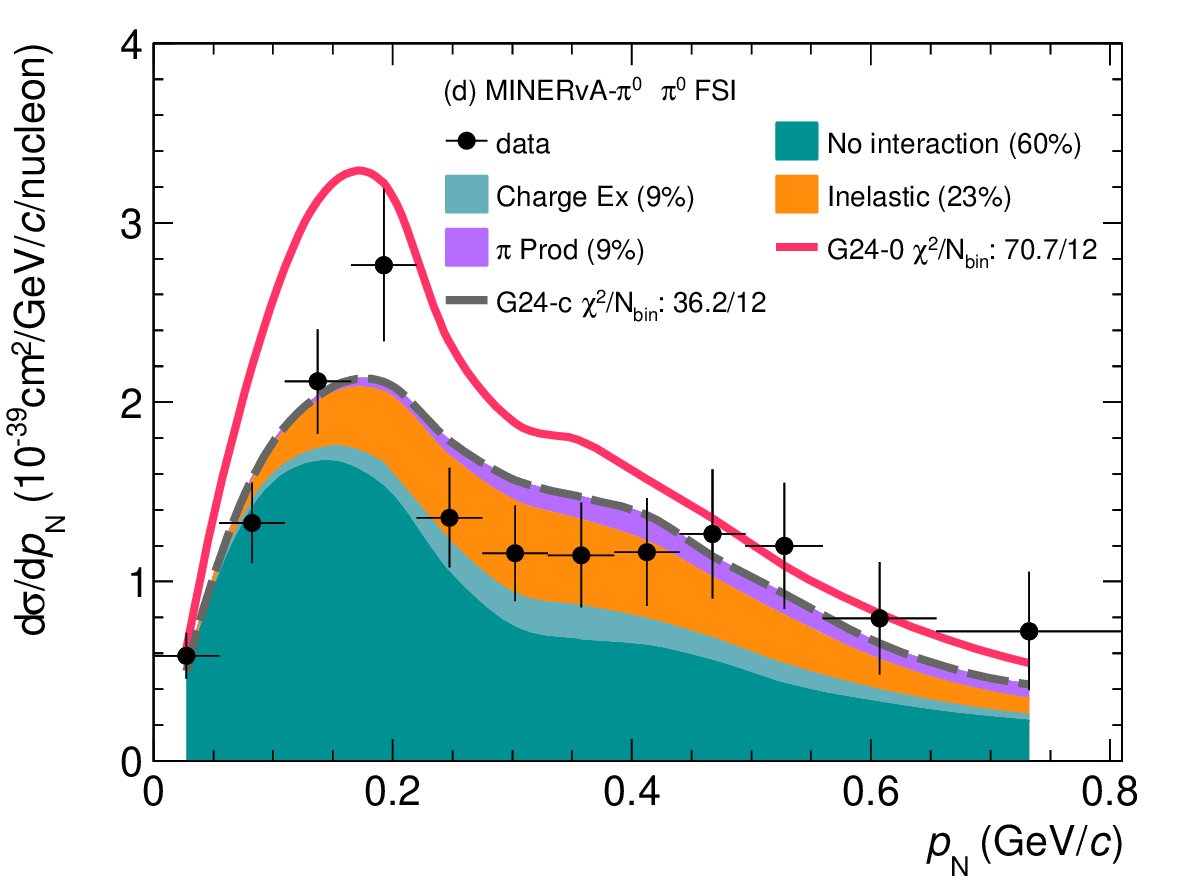}
    \caption{\label{fig:g24-c-pn-pi0} Extension of figure~\ref{fig:CEX-minpiz-dat-pi0}b but to all four $\pn$ measurements.} 
\end{figure*}

Figures~\ref{fig:g24-0-dat-pi0}-\ref{fig:g24-c-pn-pi0} display comparisons of  \gZero and \gC predictions to data, detailing the composition by $\piz$ FSI fate.

\newpage

\begin{table*}[h]
\begin{tabular}{l|lllll|llll|llll|lp{1cm}ll|lllll|l|p{1cm}|lp{1cm}}
\hline
\hline
 \texttt{Combi-}        & 1 & 2 & 3 & 4 & 5 & 6 & 7 & 8 & 9 & 10 & 11 & 12 & 13 & 14 & 15 \par (\texttt{Best-}\par\allpar) & 16 & 17 & 18 & 19 & 20 & 21 & 22 & 23 & 24 \par (\texttt{Super-}\par\texttt{set}) & 25 & 26 \par (\texttt{Best-}\par\redpar)\\
         \hline
\multicolumn{27}{c}{\ttkzpi} \\
\hline
$\dat$      &   &   & $\tick$ &   &   & $\tick$ &   &   &   & $\tick$  &    & $\tick$  &    & $\tick$  &    & $\tick$  & $\tick$  & $\tick$  & $\tick$  & $\tick$  & $\tick$  &    & $\tick$  & $\tick$  & $\tick$  & $\tick$  \\
$\dpt$   &   & $\tick$ &   &   &   & $\tick$ &   &   &   & $\tick$  &    & $\tick$  &    &    & $\tick$  &    & $\tick$  &    & $\tick$  &    & $\tick$  & $\tick$  & $\tick$  & $\tick$  & $\tick$  & $\tick$  \\
$\dphit$ &   &   &   & $\tick$ &   & $\tick$ &   &   &   & $\tick$  &    & $\tick$  &    &    &    &    &    &    &    & $\tick$  &    & $\tick$  &    & $\tick$  &    &    \\
\hline
\multicolumn{27}{c}{\ttkpip} \\
\hline
$\dat$      &   &   & $\tick$ &   &   &   & $\tick$ &   &   & $\tick$  &    &    & $\tick$  & $\tick$  &    & $\tick$  & $\tick$  & $\tick$  & $\tick$  & $\tick$  & $\tick$  &    & $\tick$  & $\tick$  & $\tick$  & $\tick$  \\
$\pn$       & $\tick$ &   &   &   &   &   & $\tick$ &   &   & $\tick$  &    &    & $\tick$  &    & $\tick$  & $\tick$  &    & $\tick$  &    &    & $\tick$  & $\tick$  & $\tick$  & $\tick$  & $\tick$  & $\tick$  \\
$\dptt$     &   &   &   &   & $\tick$ &   & $\tick$ &   &   & $\tick$  &    &    & $\tick$  & $\tick$  &    &    &    & $\tick$  & $\tick$  & $\tick$  &    &    & $\tick$  & $\tick$  &    & $\tick$  \\
\hline
\multicolumn{27}{c}{\minzpi} \\
\hline
$\dat$      &   &   & $\tick$ &   &   &   &   & $\tick$ &   &    & $\tick$  & $\tick$  &    & $\tick$  &    & $\tick$  & $\tick$  & $\tick$  & $\tick$  & $\tick$  & $\tick$  &    & $\tick$  & $\tick$  & $\tick$  & $\tick$  \\
$\pn$       & $\tick$ &   &   &   &   &   &   & $\tick$ &   &    & $\tick$  & $\tick$  &    &    & $\tick$  & $\tick$  &    & $\tick$  &    &    & $\tick$  & $\tick$  & $\tick$  & $\tick$  & $\tick$  & $\tick$  \\
$\dpt$      &   & $\tick$ &   &   &   &   &   & $\tick$ &   &    & $\tick$  & $\tick$  &    &    & $\tick$  &    & $\tick$  &    & $\tick$  &    & $\tick$  & $\tick$  & $\tick$  & $\tick$  &    &    \\
$\dphit$    &   &   &   & $\tick$ &   &   &   & $\tick$ &   &    & $\tick$  & $\tick$  &    &    &    &    &    &    &    & $\tick$  &    & $\tick$  &    & $\tick$  &    &    \\
\hline
\multicolumn{27}{c}{\minpiz} \\
\hline
$\dat$      &   &   & $\tick$ &   &   &   &   &   & $\tick$ &    & $\tick$  &    & $\tick$  & $\tick$  &    & $\tick$  & $\tick$  & $\tick$  & $\tick$  & $\tick$  & $\tick$  &    & $\tick$  & $\tick$  & $\tick$  & $\tick$  \\
$\pn$       & $\tick$ &   &   &   &   &   &   &   & $\tick$ &    & $\tick$  &    & $\tick$  &    & $\tick$  & $\tick$  &    & $\tick$  &    &    & $\tick$  & $\tick$  & $\tick$  & $\tick$  & $\tick$  & $\tick$  \\
$\dptt$     &   &   &   &   & $\tick$ &   &   &   & $\tick$ &    & $\tick$  &    & $\tick$  & $\tick$  &    &    &    & $\tick$  & $\tick$  & $\tick$  &    &    & $\tick$  & $\tick$  &    & $\tick$ \\
\hline
\hline    
\end{tabular}
\caption{\label{tab:fit-var-combo}
Specifications of observable combinations within the tuning superset in Table~\ref{tab:data-sets}. \texttt{Combi-15} is \texttt{Best-}\allpar, \texttt{Combi-24} is \texttt{Superset}, and \texttt{Combi-26} is \texttt{Best-}\redpar. 
}
\end{table*}

\end{document}